\documentclass[%
 reprint,
superscriptaddress,
showpacs,preprintnumbers,
 amsmath,amssymb,
 aps,
 pra,
nobalancelastpage,
letterpaper
]{revtex4-1}

\usepackage{graphicx}
\usepackage{dcolumn}
\usepackage{bm}

\usepackage{amsmath,amsfonts,amssymb,amsthm}
\usepackage{txfonts}
\usepackage{enumerate}
\usepackage{color}
\usepackage{braket}
\usepackage{cases}
\usepackage{comment}
\usepackage{nameref}

\makeatletter
\renewcommand{\theequation}{\arabic{section}.\arabic{equation}}
\@addtoreset{equation}{section}
\makeatother

\begin{document}

\preprint{APS/123-QED}

\title{Bogoliubov-de Gennes Soliton Dynamics in Unconventional Fermi Superfluids}
\author{Daisuke A. Takahashi}\email{daisuke.takahashi.ss@riken.jp}
\affiliation{RIKEN Center for Emergent Matter Science (CEMS), Wako, Saitama 351-0198, Japan}
\affiliation{Research and Education Center for Natural Sciences, Keio University, Hiyoshi 4-1-1, Yokohama, Kanagawa 223-8521, Japan}

\date{\today}

\begin{abstract}
Exact self-consistent soliton dynamics based on the Bogoliubov-de Gennes (BdG) formalism in unconventional Fermi superfluids/superconductors possessing an $SU(d)$-symmetric two-body interaction is presented. 
The derivation is based on the ansatz having the similar form as the Gelfand-Levitan-Marchenko equation in the inverse scattering theory. Our solutions can be regarded as a multicomponent generalization of the solutions recently derived by Dunne and Thies [Phys. Rev. Lett. \textbf{111}, 121602 (2013)]. 
We also propose superpositions of occupation states, which make it possible to realize various filling rates even in one-flavor systems, and include Dirac and Majorana fermions. 
The soliton solutions in the $ d=2 $ systems, which describe the mixture of singlet $s$-wave and triplet $p$-wave superfluids, exhibit a variety of phenomena such as rotating polar phases by soliton spins, SU(2)-DHN breathers, Majorana triplet states, $s$-$p$ mixed dynamics, and so on. These solutions are illustrated by animations, where order parameters are visualized by spherical harmonic functions. The full formulation of the BdG theory is also supported, and the double-counting problem of BdG eigenstates and $N$-flavor generalization are discussed.
\end{abstract}

\maketitle

\section{Introduction}
The Bogoliubov-de Gennes (BdG) problem \cite{Bogoliubov1958,DeGennes:1999}, the self-consistent determination of gap functions and their eigenstates, appears in broad areas of physics, e.g., fermionic superfluids or superconductors in condensed matter and ultracold atom physics, organic semiconductors such as polyacetylene \cite{Su:1979ua,Takayama:1980zz}, and the mean-field study of Nambu-Jona Lasinio or Gross-Neveu models \cite{Nambu:1961tp,Gross:1974jv,Dashen:1975xh} as an effective description of quantum chromodynamics \cite{Hatsuda:1994pi}. Self-consistent fermion filling of bound states in the solitons often plays an essential role to explain various physical properties, for example, electromagnetic characteristics of polyacetylene \cite{RevModPhys.60.781}. Multi-soliton and soliton-lattice solutions have been found and applied to diverse phenomena in many physical contexts in both condensed-matter and high-energy physics \cite{Shei:1976mn,BrazovskiiGordyuninKirova,BrazovskiiKirova,Horovitz:1981,Mertsching:1981,Campbell:1981,Campbell:1981dc,OkunoOnodera,OkunoOnodera2,BrazovskiiKirovaMatveenko,Machida:1984zz,MachidaFujita,HaraNagai1986,PhysRevD.67.125015,FeinbergPLB,Feinberg:2003qz,Casalbuoni:2003wh,Schnetz2004425,MizushimaMachidaIchioka,Thies:2006ti,Basar:2008im,Basar:2008ki,Basar:2009fg,Correa:2009xa,Nickel:2009wj,Hofmann:2010gc,PhysRevD.83.085001,PhysRevB.84.024503,Takahashi:2012aw,PhysRevLett.110.131601,JPSJ.83.023703,Arancibia:2014lwa,PhysRevA.91.023616,PhysRevLett.113.130404,PhysRevA.91.023610,PhysRevD.92.034003}.
 In particular, very recently, Dunne and Thies \cite{PhysRevLett.111.121602,PhysRevA.88.062115,PhysRevD.89.025008} have presented a general class of time-dependent and self-consistent multi-soliton solutions. 
The dynamics of fermionic order parameters based on the time-dependent BdG formalism has been also numerically studied in various situations such as Bragg scattering, oscillation in trapping potentials, snake instability \cite{PhysRevLett.98.093002,PhysRevLett.106.185301,PhysRevA.88.043639}. For the applicability of mean-field theory in one dimension, see Ref.~\cite{PhysRevA.76.043605}.

The studies of exact self-consistent soliton dynamics to date have been mainly restricted to a one-component order parameter. However, our nature exhibits a variety of unconventional superconductors/superfluids, such as superfluid ${}^3\mathrm{He}$ described by the triplet $ p $-wave order parameter (Refs.~\cite{VollhardtWolfle,0953-8984-27-11-113203,arxiv1508.00787} and references therein), $s$-$p$ mixing in noncentrosymmetric superconductors due to spin-orbit coupling (e.g., Refs.~\cite{JPSJ.77.124711,JPSJ.83.013703} and references therein), and MgB${}_2$ and iron-based superconductors with multiband structure (e.g., Refs.~\cite{JPSJ.81.024712,arxiv1507.06039} and references therein). Ultracold atomic systems also possess candidates of superfluids with higher symmetries such as $ {}^6\mathrm{Li} $ with $SU(3)$-symmetric interaction and  ${}^{173}\mathrm{Yb}$ with $SU(6)$-symmetric interaction (Refs.~\cite{PhysRevLett.98.030401,JPSJ.78.012001,PhysRevA.82.063615,PhysRevA.83.063607,0034-4885-77-12-124401,PhysRevA.90.013632,arxiv1504.08113} and references therein). 
The aim of this paper is the generalization of soliton dynamics for such unconventional and multicomponent Fermi superfluids. 

Physics of mean-field solitons in fermionic systems is deeper and richer than that of bosons. While the time evolution of bosonic mean fields, e.g., Bose-Einstein condensates (BECs), is modeled by a partial differential equation (PDE) such as the Gross-Pitaevskii or nonlinear Schr\"odinger (NLS) equation, fermionic ones need to solve the BdG and gap equations self-consistently, and order parameters do not solely satisfy a simple PDE. The inverse scattering theory (IST)  of the Zakharov-Shabat (ZS) operator \cite{ZakharovShabat,ZakharovShabat2} can powerfully solve the initial-value problem of the NLS equation \cite{AKNS1974,AblowitzSegur,FaddeevTakhtajan}, but it is applicable only for a stationary problem for the fermionic BdG systems (e.g., \cite{OkunoOnodera,OkunoOnodera2,PhysRevLett.110.131601}). These circumstances may be phrased as ``mathematics underlying in bosonic soliton \textit{dynamics} $\simeq$ that in fermionic soliton \textit{statics}''. Establishing a general framework to track fermionic soliton \textit{dynamics} is therefore challenging.

In this paper, we solve the time-dependent multi-component BdG equation with $ SU(d) $-symmetric gap equation. First, we propose an efficient way to generate time-dependent reflectionless potentials and their eigenstates for multi-component PDEs based on the ansatz originating from the Gelfand-Levitan-Marchenko (GLM) equation in the IST. Next, we solve the gap equation for these potentials. We realize partial filling rates of bound states without relying on the $N$-flavor generalization using superpositions of occupied and unoccupied states, including Dirac and Majorana fermions as a special case. The results are illustrated by animation files for the $ d=2 $ case, modeling the mixture of singlet $s$-wave and triplet $p$-wave superfluids.

The organization of the paper is as follows. Section~\ref{sec:mainr} summarizes the main result of this paper. The construction of reflectionless potentials by the GLM ansatz, the reduction of the gap equation to spacetime-independent matrix equation, summary of occupation states realizing partial filling rates of bound states, and various animation examples of soliton solutions, are presented in this section. 
Sections \ref{sec:supprefless}-\ref{sec:parameters} provide supporting materials to complement the main result. 
In Sec.~\ref{sec:supprefless}, we provide mathematical proofs for properties of eigenfunctions constructed by the GLM ansatz in Subsec.~\ref{subsec:glmansatz}. In Sec.~\ref{sec:bdgformul}, we derive the BdG and gap equations for $SU(2)$- and $SU(d)$-symmetric interaction. The Andreev approximation and $N$-flavor generalization are also discussed. In Sec.~\ref{app:dblcnt}, we explain the double-counting problem of eigenstates in the BdG theory. In Sec.~\ref{sec:suppgapsupp}, we give a supplemental calculation related to the gap equation. In Sec.~\ref{sec:parameters}, we provide detailed classification of soliton solutions and show the parameters used in the generation of animation files.  Sec.~\ref{sec:summary} is devoted to a summary. Appendix~\ref{subsec:spplt} provides an explanation for spherical harmonic plot.

\section{Summary of Main Result}\label{sec:mainr}
The main findings of this paper are summarized in this section. 
\subsection{Gelfand-Levitan-Marchenko ansatz : Constructing time-dependent reflectionless potentials}\label{subsec:glmansatz}
	First, we show the construction of time-dependent reflectionless potentials, which makes a core of this work. 
 	We first prepare orthonormalized bound states. 
	Let $ w_1(x,t) , \dots , w_n(x,t) $ be $ d $-component column vectors assumed to be linearly independent of each other for every $ t $ and have the asymptotic behavior
	\begin{align}
		|w_i(x,t)| \rightarrow\begin{cases} 0 &(x\rightarrow-\infty) \\ \infty &(x\rightarrow+\infty). \end{cases} \label{eq:wiasympt}
	\end{align}
	We also use other $ d $-component column vectors $ h_1(x,t),\dots,h_n(x,t) $, and write $ W(x,t)=(w_1(x,t),\dots,w_n(x,t)) $ and $ H(x,t)=(h_1(x,t),\dots,h_n(x,t)) $. We define $ K(x,y,t):=H(x,t)W(y,t)^\dagger $ and $  \Omega(x,y,t):=W(x,t)W(y,t)^\dagger $, 
	and assume the ``GLM-like'' equation:
	\begin{align}
		K(x,y,t)+\Omega(x,y,t)+\int_{-\infty}^x\mathrm{d}z K(x,z,t)\Omega(z,y,t)=0.
	\end{align}
	Introducing the Gram matrix by $ G(x,t):=\int_{-\infty}^x\mathrm{d}z W(z,t)^\dagger W(z,t) $, 
	the functions $ h_i $'s obey the equation $ H+W+HG=0 $. 
	Since $ a^\dagger G a=\int\mathrm{d}z (\sum a_iw_i)^\dagger(\sum a_jw_j)>0 $ holds for any $ a=(a_1,\dots,a_n)^T(\ne0) $, $ I_n+G $ is positive-definite, hence invertible:
	\begin{align}
		H(x,t)=-W(x,t)[I_n+G(x,t)]^{-1}. \label{eq:bounddef}
	\end{align}
	$ h_i $'s constructed above are \textit{orthonormalized}: $ \int_{-\infty}^\infty\mathrm{d}x H^\dagger H=-[(I_n+G)^{-1}]_{-\infty}^\infty=I_n $.\\ 
%
	\indent Next, we consider a PDE which $ h_i $'s satisfy when $ w_i $'s satisfy a \textit{constant-coefficient} PDE. 
	Let us assume that $ w_i $'s satisfy 
	\begin{align}
		\partial_t w=A\partial_x w+Bw,  \quad A^\dagger=A,\ B^\dagger=-B.  \label{eq:genrlss1}
	\end{align}
	The plane-wave ansatz $ w\propto \mathrm{e}^{-\mathrm{i}(kx+\epsilon t)} $ for this equation soon reduces to  $ \det(\epsilon I_d-kA-\mathrm{i}B)=0 $. While $ w_i $'s with behavior Eq.~(\ref{eq:wiasympt}) are given by superpositions of such solutions with $ k\in\mathbb{H} $, where $ \mathbb{H}:=\{ k \,|\, \operatorname{Im}k>0 \} $, the bounded solutions, which we write $ \phi(x,t) $, are given by those with $ k\in\mathbb{R} $.\\ 
	\indent A short calculation shows that $ h_i $'s satisfy 
	\begin{align}
		\partial_th=A\partial_xh+(B+[K,A])h, \label{eq:boundstate}  
	\end{align}
	where $ K:=K(x,x,t) $. It can be regarded as a \textit{variable-coefficient} generalization of Eq. (\ref{eq:genrlss1}), where $ B $ is replaced by $ B+[K,A] $. 
	We further introduce ``scattering states'' below. Note that the equation $ H+W+HG=0 $ can be rewritten as $ h_i(x,t)=-w_i(x,t)-\int_{-\infty}^x\mathrm{d}y K(x,y,t)w_i(y,t) $, 
	which is an analog of the integral representation of Jost functions in the IST. Inspired by this expression, we define the scattering states by
	\begin{align}
		f(x,t)=\phi(x,t)+\int_{-\infty}^x\mathrm{d}yK(x,y,t)\phi(y,t), \label{eq:scatdef}
	\end{align}
	where $ \phi(x,t) $ satisfies Eq.~(\ref{eq:genrlss1}) and bounded for all $ (x,t) $. 
	Then, \textit{$ f $ also satisfies Eq. (\ref{eq:boundstate}).} 
	We can check that if $ f \sim \mathrm{e}^{\mathrm{i}k x} $ for $ x\rightarrow-\infty $, the same holds for $ x\rightarrow+\infty $.
	Therefore, the potential $ B+[K,A] $ is \textit{reflectionless}. 
	The proof that $ h_i $ and $ f $ satisfy Eq. (\ref{eq:boundstate}) is given in Sec.~\ref{sec:supprefless}, where we also provide important theorems that the set of eigenstates, $ h_i $'s and $ f $'s, satisfy the orthonormal and completeness relations [Eqs. (\ref{eq:orthobb})-(\ref{eq:compsb})]. Thus, all eigenstates are exhausted, no missing eigenstate remains.
\subsection{BdG and gap equation}
	\indent Henceforth we concentrate on the BdG systems and their gap equations. We replace $ d $ by $ 2d $, and use the notation $ \sigma_i=\tilde{\sigma}_i\otimes I_d $ with $ \tilde{\sigma}_i $ being Pauli matrices. We also use $ \sigma_\pm=(\sigma_1\pm\mathrm{i}\sigma_2)/2 $. We consider the case of $ A=-\sigma_3 $ and $ \{A,B\}=0 $. Writing $ \mathrm{i}(B+[K,A])=\Delta(x,t)\sigma_++\Delta(x,t)^\dagger\sigma_- $,  
	Eq. (\ref{eq:boundstate}) reduces to the BdG equation:
	\begin{align}
		\mathrm{i}\partial_t\begin{pmatrix}\boldsymbol{u} \\ \boldsymbol{v} \end{pmatrix}=\mathcal{L}\begin{pmatrix}\boldsymbol{u} \\ \boldsymbol{v} \end{pmatrix},\quad \mathcal{L}=\begin{pmatrix} -\mathrm{i}I_d\partial_x & \Delta(x,t) \\ \Delta(x,t)^\dagger & \mathrm{i}I_d \partial_x \end{pmatrix}. \label{eq:matrixbdg}
	\end{align}
	Here,  $ \boldsymbol{u},\boldsymbol{v} $ are $ d $-component column vectors and  $ \Delta(x,t) $ is a $ d\times d $ matrix. 
	We consider the antisymmetric (symmetric)  $ \Delta $ corresponding to $s$-wave ($p$-wave) order parameters. 
	The (anti)symmetry is expressed as $ \tau \mathcal{L}^*\tau=\mathcal{L} $ with $ \tau=\sigma_1 $ (for $\Delta=\Delta^T$) and $ \sigma_2 $ (for $\Delta=-\Delta^T$). 
	For each case, we assume the following gap equation:
	\begin{align}
		-\frac{\Delta}{g_\pm}=\frac{1}{2}\sum_j (\boldsymbol{u}_j\boldsymbol{v}_j^\dagger\pm \boldsymbol{v}_j^*\boldsymbol{u}_j^T)(2\nu_j-1) \quad \text{for } \Delta=\pm\Delta^T. \label{eq:gapsymantisym}
	\end{align}
	where $ g_\pm>0 $ is a coupling constant, and $ \nu_j=\braket{a_j^\dagger a_j} $ is a filling rate of the state $ j $.  
	Since we are interested in the low-energy physics not so far from the ground state, we consider the following occupation state: The negative (positive) scattering states are completely occupied (vacant), i.e.,  $ \nu_j=1\ (0) $, and bound states are filled partially. Note that such an excited state with specifying the occupation of bound states is different from the occupation state of the simple zero-temperature equilibrium. As we will see below, to achieve the self-consistency, adjustment of bound-state fillings is essential.

	\indent The gap equation (\ref{eq:gapsymantisym}) appears when the two-body interaction is $ SU(d) $-symmetric.  If $ d=1 $, it describes spinless $ p $-wave superconductors, a continuous analog of the Kitaev chain \cite{1063-7869-44-10S-S29}. The case $ d=2 $ and $ \Delta=-\Delta^T $ corresponds to singlet $s$-wave superconductors, and $ d=2 $ and $ \Delta=\Delta^T $ describes triplet $p$-wave superconductors/superfluids, whose prime example is the superfluid ${}^3\mathrm{He}$ (confined in one-dimensional geometry). $ d=3 $ and $ 6 $ correspond to ${}^{6}\mathrm{Li}$ and ${}^{173}\mathrm{Yb}$ with $SU(3)$- and $SU(6)$-symmetric interaction, respectively. In Eqs. (\ref{eq:matrixbdg}) and (\ref{eq:gapsymantisym}), in order to resolve the double counting of eigenstates, we use the equations with dispersion linearized at the right Fermi point $ k=k_F $ and discard those at $ k=-k_F $. 
	See Sec.~\ref{sec:bdgformul} for the formulation of the BdG theory, and Sec.~\ref{app:dblcnt} for the double-counting problem. \\ 
%
%
	\indent At the fine-tuned point $ g_+=g_-(=:g) $, we can exceptionally treat non-symmetric $ \Delta $, that is, $s$-$p$ mixed dynamics. In such case the gap equation also becomes simpler:
	\begin{align}
		-\frac{\Delta}{g}=\sum_j\boldsymbol{u}_j\boldsymbol{v}_j^\dagger(2\nu_j-1) \quad\text{ for non-symmetric $ \Delta $.} \label{eq:gapnonsym}
	\end{align}
	If we go back to the dimensionful variables, the condition $ g_+=g_- $ is rewritten as $ k_F^2g_+=g_- $ (Subsec.~\ref{subsec:andapp}) Therefore, such fine tuning will be realized by adjusting (i) coupling constants (e.g., by Feshbach resonance in cold atom systems, or by changing atomic species in a crystal) and/or (ii) the Fermi wavenumber $ k_F $ by changing the total number of particles. 
%
%

	\indent We consider the uniform boundary condition at infinity:
	\begin{align}
		\Delta(x\rightarrow\pm\infty)=m \Delta_\pm,\quad m>0,
	\end{align}
	and assume that $ \Delta_\pm $ is unitary. In such system, the uniformization variable $ s $ (\u{Z}ukowsky transform \cite{FaddeevTakhtajan}) 
	\begin{align}
		\epsilon(s)=\frac{m}{2}(s+s^{-1}),\ k(s)=\frac{m}{2}(s-s^{-1}),
	\end{align}
	is convenient in the labeling of eigenstates. Now, we prepare the eigenstates of the BdG equation. Let $ \mathrm{i}B=m(\Delta_-\sigma_++\Delta_-^\dagger\sigma_-) $ and $ w_1(x,t),\dots,w_n(x,t) $ be solutions of Eq. (\ref{eq:matrixbdg}) for a uniform gap $ \Delta(x,t)=m\Delta_- $ with the behavior (\ref{eq:wiasympt}).  $ W,H,K,\Omega $ are defined in the same way as Subsec.~\ref{subsec:glmansatz}. Then, the non-uniform gap function is constructed as
	\begin{align}
		\Delta(x,t)=m\Delta_-+2\mathrm{i}K_{12}(x,x,t), \label{eq:gapexpre}
	\end{align}
	where $ K_{12} $ represents the $ d\times d $ top-right block of $ K $,  and $ h_i $'s become normalizable bound states of Eq. (\ref{eq:matrixbdg}) for this $ \Delta(x,t) $. 
	Following Eq.~(\ref{eq:scatdef}), we introduce the scattering states
	\begin{align}
		F(x,t,\zeta)=\left[1+\!\!\int_{-\infty}^x\mathrm{d}yK(x,y,t) \mathrm{e}^{\mathrm{i}k(\zeta)(y-x)} \right]\begin{pmatrix} I_d \\ \zeta^{-1}\Delta_-^\dagger \end{pmatrix}\mathrm{e}^{\mathrm{i}[k(\zeta)x-\epsilon(\zeta)t]} \label{eq:scatstates}
	\end{align}
	for $ \zeta\in\mathbb{R} $. Every column of $ F $ becomes a solution of Eq.~(\ref{eq:matrixbdg}). 
	Using these eigenstates, after introducing a momentum cut-off $ k_c $ to avoid ultraviolet divergence, the gap equation [Eqs. (\ref{eq:gapsymantisym}) and (\ref{eq:gapnonsym})] can be rewritten as 
	\begin{align}
		&[\sigma_3,\Xi+\tau\Xi^*\tau]=0, \label{eq:gapint2} \\
		&\Xi:=HDH^\dagger+\left[ \int_{-\infty}^0\!\!-\!\int_0^\infty \right]\frac{\mathrm{d}\zeta}{4\pi}\left( mFF^\dagger+\frac{B+[K,A]}{\mathrm{i}\zeta} \right), \label{eq:gapint22}
	\end{align}
	where $ D_{ij}:=\delta_{ij}(2\nu_j-1) $, and $  \tau=0,\ \sigma_1 $, and  $\sigma_2$ for non-symmetric, symmetric, and antisymmetric  $ \Delta $'s, respectively. 
	This gap equation does not depend on $ g_\pm $ and $ k_c $, since they are eliminated through the relation $ m=2k_c\mathrm{e}^{-\pi/g_\pm} $.  


\subsection{Dunne-Thies class}
	Now we restrict the form of $ W $ to:
	\begin{align}
		W=W_0L,\quad W_0=\begin{pmatrix}U_0 \\ \Delta_-^\dagger U_0\mathcal{S} \end{pmatrix},\ U_0=(e_1\hat{p}_1,\dots,e_n\hat{p}_n),   \label{eq:dtrestrict2}
	\end{align} 
	where $ \mathcal{S}=\operatorname{diag}(s_1,\dots,s_n) $, $ s_j\in\mathbb{H} $, $ e_j:=\mathrm{e}^{-\mathrm{i}[k(s_j)x+\epsilon(s_j)t]} $,  $ \hat{p}_j $ is a normalized $ d $-component vector ($ \hat{p}_j^\dagger\hat{p}_j=1 $),  and $ L $ is an invertible constant $ n\times n $ matrix. 
	We also write $ H_0=H L^\dagger $ and $ G_0=\int_{-\infty}^x\mathrm{d}x W_0^\dagger W_0 $. 
	They satisfy $ H_0(LL^\dagger)^{-1}+W_0+H_0G_0=0 $.
	The gap function and the bound states are given by
	\begin{align}
		\Delta&=\left( mI_d-2\mathrm{i}U_0[(LL^\dagger)^{-1}+G_0]^{-1}\mathcal{S}^*U_0^\dagger \right) \Delta_-, \label{eq:deltainmain}  \\
		H&= -W_0[(LL^\dagger)^{-1}+G_0]^{-1}(L^\dagger)^{-1}. \label{eq:boundinmain}
	\end{align}
	The expression for scattering states is given in Sec.~\ref{sec:suppgapsupp} [Eqs.~(\ref{eq:dtscat1}) and (\ref{eq:dtscat2})]. For these solutions, the gap equation (\ref{eq:gapint2}) reduces to [See Subsec.~\ref{sec:suppgap} for a detailed calculation]
	\begin{gather}
		[\sigma_3,HXH^\dagger+\tau H^*X^*H^T\tau]=0, \label{eq:reducedgap2} \\
		X:=\mathcal{N}-\tfrac{1}{2}(L^\dagger \Theta (L^\dagger)^{-1}+L^{-1}\Theta^\dagger L), \label{eq:reducedgap}
	\end{gather}
	where we have introduced $ n\times n $ diagonal matrices $ \mathcal{N} $ and $ \Theta $ by
	\begin{align}
		\mathcal{N}_{ij}=\nu_i\delta_{ij},\quad \Theta_{ij}:=\frac{\theta_j+\mathrm{i}\log r_j}{\pi}\delta_{ij}, \label{eq:NandTheta}
	\end{align}
	and $ r_i,\ \theta_i $ are defined by  $ s_i=r_i\mathrm{e}^{\mathrm{i}\theta_i} $ with $ r_i>0,\ 0<\theta_i<\pi $. 
	Except for the antisymmetric $ \Delta $, finding the solution of Eq. (\ref{eq:reducedgap2}) is simply reduced to the $(x,t)$-independent matrix equation $ X=0 $. \\  
%
%
	\indent We call the above solutions \textit{``the Dunne-Thies (DT) class''}, since it reduces to their solutions \cite{PhysRevLett.111.121602,PhysRevA.88.062115,PhysRevD.89.025008} when $ \Delta $ is $ 1\times 1 $. Our work provides $ d\times d $ generalization. $ \hat{p}_j $'s are new parameters for multicomponent systems, which describe the ``angle'' of solitons. The velocity of the soliton with label $ j $ is $ V_j=\frac{1-r_j^2}{1+r_j^2} $. The Lorentz-boosted solution by velocity $ V=\frac{1-r^2}{1+r^2} $ can be obtained by replacing $ s_j \rightarrow r s_j $. 
	The one-soliton solution and its bound state are
	\begin{align}
		\Delta&=m\left[I_d-\tfrac{1}{2}(1-\mathrm{e}^{-2\mathrm{i}\theta_1})(1+\tanh y)\hat{p}_1\hat{p}_1^\dagger \right]\Delta_-, \label{eq:onesoldelta} \\
		h_1&=\begin{pmatrix}\boldsymbol{u}_1 \\ \boldsymbol{v}_1 \end{pmatrix}=\frac{-w_1}{1+G_{11}}=\frac{-\sqrt{\kappa_1}\mathrm{e}^{-\mathrm{i}y'}}{2\cosh y}\begin{pmatrix} \sqrt{1+V_1}\hat{p}_1 \\ \sqrt{1-V_1}\mathrm{e}^{\mathrm{i}\theta_1}\Delta_-^\dagger\hat{p}_1 \end{pmatrix} \label{eq:onesolbound}
	\end{align}
	with $ y=\kappa_1(x-V_1t),\ y'=\tilde{\kappa}_1(t-V_1x),\  \kappa_1=\frac{m\sin\theta_1}{\sqrt{1-V_1^2}},\ \tilde{\kappa}_1=\frac{m\cos\theta_1}{\sqrt{1-V_1^2}} $. The self-consistent filling is $ \nu_1=\theta_1/\pi $. 
	
	A self-consistent solution in non-DT-class potentials seems to be rare, though we do not have a proof. (See also perspectives.)
	

\subsection{Stationary class}\label{subsec:statclass}
	As a subset of the DT class, we define \textit{the stationary class} by $ |s_j|=r_j=1 $ for all $ j $ and diagonal $ L $ in Eq.~(\ref{eq:dtrestrict2}), where $ \Delta $ becomes time-independent and $ \epsilon(s_j) $ becomes an eigenenergy of bound states. 
	For this class, the reduction to the symmetric case is achieved by setting $ \hat{p}_j=\Delta_-\hat{p}_j^* $ and antisymmetric case by $ \hat{p}_{2j}=\Delta_-\hat{p}_{2j-1}^*,\ s_{2j}=s_{2j-1},\ L_{2j1-1,2j-1}=L_{2j,2j} $. Note that the bound states always emerge in pairs for stationary antisymmetric $ \Delta $ (hence $ n $ is even). \\
	\indent The reduced gap equation (\ref{eq:reducedgap2}) can be soon solved for the stationary class. 
	For the non-symmetric ($s$-$p$ mixed,  $ \tau=0 $) and symmetric ($p$-wave, $\tau=\sigma_1$) cases, the self-consistent condition is given by
	\begin{align}
		\nu_j=\frac{\theta_j}{\pi}. \quad ( \Delta \text{: symmetric or non-symmetric}.) \label{eq:statscsym}
	\end{align}
	For the antisymmetric case ($s$-wave, $ \tau=\sigma_2 $),
	\begin{align}
		\nu_{2j-1}+\nu_{2j}=\frac{2\theta_{2j}}{\pi}. \quad ( \Delta \text{: antisymmetric}. ) \label{eq:statscasym}
	\end{align}
	Subsection~\ref{subsec:statgap} provides a detailed derivation. Equation (\ref{eq:statscsym}) for $ d=1 $ has the same form as Ref.~\cite{PhysRevLett.111.121602}, and Eq. (\ref{eq:statscasym}) for $ d=2 $ reproduces Ref.~\cite{PhysRevLett.110.131601} after changing the convention of double-counting elimination (See Sec.~\ref{app:dblcnt}). The difference of the self-consistent condition between symmetric and antisymmetric cases crucially changes whether a fermion localized around a  $ \pi $-phase kink is Dirac or Majorana. (See next subsection.) \\
%
%
	\indent We emphasize that the stationary-class potentials are essentially the same as the ``snapshots at each time'' of the soliton solutions in the self-defocusing matrix NLS equation $ \mathrm{i}\Delta_t=-\Delta_{xx}+2\Delta\Delta^\dagger\Delta $ \cite{JPSJ.67.1175,jmp48110.10631.2423222}, including the integrable spin-1 BECs \cite{PhysRevLett.93.194102,JPSJ.73.2996,JPSJ.75.064002}, since Eq.~(\ref{eq:matrixbdg}) with $ \Delta(x,t)=\Delta(x) $,\ $ (u,v)=(u(x),v(x))\mathrm{e}^{-\mathrm{i}\epsilon t} $ reduces to the matrix-generalized ZS/AKNS eigenvalue problem. On the other hand, nonstationary potentials have no counterpart. 
	
\subsection{Partial filling and Majorana fermions}

	\begin{table}[tb]
		\begin{center}
		\caption{
		\label{ta:nonstatscsol} 
		Superpositions of occupation states in one-flavor systems realizing self-consistent solutions with partial filling rates. Here, $ n $ is a number of bound states, and the abbreviation $ (c_\xi,s_\xi)=(\cos\xi,\sin\xi),\ \xi \in \mathbb{R} $ is used. The adjective ``Dirac'' is used for filling $ \nu=0 $ or $ 1 $, and ``Majorana'' if $ \nu=1/2 $ \textit{and}  $ \ket{\Psi} $ is an eigenstate of some Majorana operator. (See the main text.) For $ n=1 $,  $ \xi=\pi/4 $ gives Majorana. For $ n=2 $, one of the two fermions always has a ``Dirac'' filling value, while the other one can take any filling. The case $ n=3 $ has three families (A)-(C). $ \beta_i $'s are complex numbers satisfying $\sum_{i=0}^3|\beta_i|^2=1$ for 3(A), $|\beta_0|^2+|\beta_3|^2=1$ for 3(B) and 3(C). 
		}
		{\small

		\begin{tabular}{|c|c|c|}
		$n$ &  $ \ket{\Psi} $ & Filling  \\ \hline
		1 & $(c_\xi+\mathrm{e}^{\mathrm{i}\varphi}s_\xi \hat{a}_1^\dagger)\ket{\mathrm{vac}}$ & $\nu_1=c_\xi^2$  \\ \hline
		2 & \begin{minipage}{11em} $(c_\xi+\mathrm{e}^{\mathrm{i}\varphi}s_\xi\hat{a}_1^\dagger)\ket{\mathrm{vac}}$ \ or \\ $(c_\xi+\mathrm{e}^{\mathrm{i}\varphi}s_\xi\hat{a}_1^\dagger)\hat{a}_2^\dagger\ket{\mathrm{vac}}$ \end{minipage} & \begin{minipage}{10em} $\nu_1=c_\xi^2$,  \\ $\nu_2=0 \text{ or }1 $. \end{minipage} \\ \hline
		3(A) & \begin{minipage}{13em} \begin{flushleft} $ \vphantom{\Big|} \tfrac{1}{\sqrt{2}}(\beta_0+\sum_{i=1}^3\beta_i\hat{a}_i^\dagger)\ket{\mathrm{vac}}$ \\ $+\tfrac{1}{\sqrt{2}}(\beta_0^*-\sum_{i=1}^3\beta_i^*\hat{a}_i)\hat{a}_1^\dagger\hat{a}_2^\dagger\hat{a}_3^\dagger\ket{\mathrm{vac}}$ \end{flushleft} \end{minipage} & \begin{minipage}{10em} $\nu_1=\nu_2=\nu_3 = \frac{1}{2}$ \\ (Majorana triplet) \end{minipage} \\ \hline
		3(B) & \begin{minipage}{13em} \begin{flushleft} $\vphantom{\big|}c_\xi(\beta_0+\beta_3\hat{a}_3^\dagger)\ket{\mathrm{vac}}$ \\ $\quad\vphantom{\Big|}+s_\xi(\beta_0^*-\beta_3^*\hat{a}_3)\hat{a}_1^\dagger\hat{a}_2^\dagger\hat{a}_3^\dagger\ket{\mathrm{vac}}$ \end{flushleft} \end{minipage} & \begin{minipage}{10em} $\nu_1=\nu_2=s_\xi^2$, \\  $\nu_3=|\beta_0|^2s_\xi^2+|\beta_3|^2c_\xi^2$. \end{minipage} \\ \hline
		3(C) & \begin{minipage}{13em} \begin{flushleft} $\vphantom{\big|}c_\xi\hat{a}_1^\dagger(\beta_0-\beta_3^*\hat{a}_3^\dagger)\ket{\mathrm{vac}}$ \\ $\quad\vphantom{\Big|}+s_\xi(\beta_3-\beta_0^*\hat{a}_3^\dagger)\hat{a}_2^\dagger\ket{\mathrm{vac}}$ \end{flushleft} \end{minipage} & \begin{minipage}{10em} $\nu_1=c_\xi^2,\ \nu_2=s_\xi^2$,\\ $\nu_3=|\beta_0|^2s_\xi^2+|\beta_3|^2c_\xi^2$. \end{minipage} \\ \hline
		\end{tabular}
		}
		\end{center}
	\end{table}

	\begin{figure}[tb]
		\begin{center}
		\begin{tabular}{cc}
		\includegraphics[scale=.85]{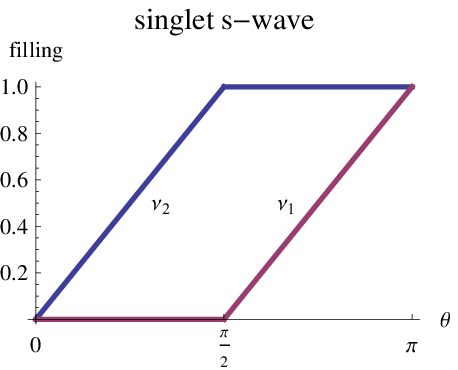} \quad&\quad \includegraphics[scale=.85]{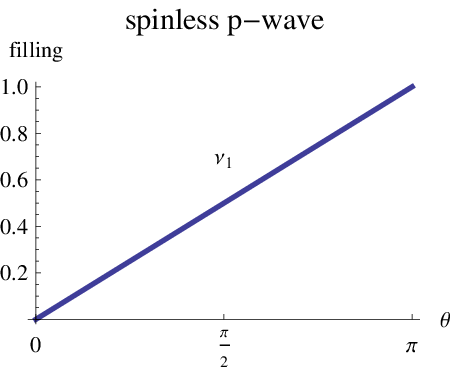} 
		\end{tabular}
		\caption{\label{fig:fillinginsp}Self-consistent filling for one-soliton state in the singlet $s$-wave ($d=2$, antisymmetric) and spinless $p$-wave ($d=1$) systems. $ 2\theta $ represents the phase shift of soliton, thus $ \theta=\frac{\pi}{2} $ corresponds to a $ \pi $-phase kink. The filling of bound states of $ \pi $-phase kink for singlet $s$-wave is ``Dirac'' ($\nu_1=0,\ \nu_2=1$), while that for spinless $p$-wave is ``Majorana''. ($\nu_1=1/2.$)}
		\end{center}
	\end{figure}

	The self-consistent condition [Eq. (\ref{eq:reducedgap2}) for the DT class, Eqs. (\ref{eq:statscsym}), (\ref{eq:statscasym}) for the stationary class] generally requires partial filling rates ($0\le \nu_j\le 1$) of bound states. 
	If we consider an $ N $-flavor model, where each eigenstate can accommodate $ N $ fermions, we can easily realize fractional filling with denominator $ N $. (This is discussed in Subsec.~\ref{app:Nflavor}.) However, the flavor number in realistic condensed-matter systems is often one. 
	One possible solution to realize partial filling rates in small-flavor systems is to construct the superposition of occupied and unoccupied states. Here we demonstrate it. 
	Let  $ \ket{\mathrm{vac}} $ be a state such that all negative scattering states are filled and others are vacant, and let us consider the linear combination  $ \ket{\Psi}=\sum_{\nu_i=0,1}C_{\nu_1\nu_2\dots\nu_n}\prod_{i=1}^n(\delta_{\nu_i,0}+\delta_{\nu_i,1}\hat{a}_i^\dagger)\ket{\mathrm{vac}} $, where $ \hat{a}_1,\dots,\hat{a}_n $ are annihilation operators of bound states. To solve the gap equation,  $ \ket{\Psi} $ must satisfy $ \bra{\Psi}\hat{a}_i\hat{a}_j\ket{\Psi}=0 $ for all $ i,j $ and $ \bra{\Psi}\hat{a}_i^\dagger\hat{a}_j\ket{\Psi}=0 $ for $ i\ne j $. Under this condition, the filling rate for the bound state $ j $ is given by $ \nu_j=\bra{\Psi}\hat{a}_j^\dagger\hat{a}_j\ket{\Psi} $. Table~\ref{ta:nonstatscsol} shows the summary for $ n\le3 $. \\
	\indent If the filling is $ \nu_i=0 $ or 1, there is no superposition between occupied and unoccupied states for the $ i $-th bound state in $ \ket{\Psi} $, and the corresponding fermion can be regarded as a conventional Dirac fermion. Therefore we call such filling \textit{``Dirac''}. We also introduce the term \textit{``Majorana''},  if the filling is $ \nu_i=\frac{1}{2} $ and $ \ket{\Psi} $ is an eigenstate of the Majorana operator. For $ n=1,2 $ and $ \xi=\frac{\pi}{4} $ in Table~\ref{ta:nonstatscsol}, if we define $ \hat{\gamma}_1=\mathrm{e}^{-\mathrm{i}\varphi}\hat{a}_1+\mathrm{e}^{\mathrm{i}\varphi}\hat{a}_1^\dagger $, then $ \hat{\gamma}_1\ket{\Psi}=\ket{\Psi} $ and $ \nu_1=\frac{1}{2} $, thus it gives an example of the ``Majorana'' filling. Such half fermion was early discussed in Ref.~\cite{Jackiw:1975fn}. For $ n=2 $, at least one of the two fermions must be ``Dirac''.  For $ n=3 $ (A) in Table~\ref{ta:nonstatscsol}, where all filling values are $ \frac{1}{2} $, we can check the relation $ \mathrm{i}\hat{\gamma}_3\hat{\gamma}_2\hat{\gamma}_1\ket{\Psi}=\ket{\Psi} $, if we define $ \hat{\gamma}_1,\hat{\gamma}_2,\hat{\gamma}_3 $ by  $ \hat{\gamma}_i=\mathrm{e}^{-\mathrm{i}\chi_i}\hat{a}_i^\dagger+\mathrm{e}^{\mathrm{i}\chi_i}\hat{a}_i $ with $ \chi_i=\arg \beta_j+\arg \beta_k+\frac{\pi}{2} $,\ $ (i,j,k)=(1,2,3),\ (2,3,1),\ (3,1,2) $ and if $ \beta_i $'s satisfy the relation $ \arg\beta_0=\arg\beta_1+\arg\beta_2+\arg\beta_3-\pi $. We refer to this $ \ket{\Psi} $ as a ``Majorana triplet'' state. $n=3$ (B) and (C) are other possible occupation states. Although there are constraints $ \nu_1=\nu_2 $ for (B) and $ \nu_1=1-\nu_2 $ for (C), two of three filling values are continuously chosen in these cases. These two reduce to (A) if we choose $ \xi=\frac{\pi}{4} $. Since the filling can be flexibly chosen in the cases $ n=3 $ (B) and (C), the soliton solutions with various fillings become meaningful even in one-flavor systems. This table suggests that, although we do not finish the classification for $ n\ge 4 $, the filling values will be also flexibly chosen for more multi-soliton solutions.  \\
	\indent Combining the self-consistent conditions (\ref{eq:statscsym}) and (\ref{eq:statscasym}) and the knowledge of Table~\ref{ta:nonstatscsol}, we can elucidate the difference of fermions localized around a kink between singlet $s$-wave and spinless $p$-wave systems. For the singlet $s$-wave case ($d=2$ and $ \Delta $ is antisymmetric), there are two degenerate bound states for one soliton. However, Table~\ref{ta:nonstatscsol} says that one of these two bound states must have a Dirac filling value. We can thus determine the possible filling values, as shown in Fig~\ref{fig:fillinginsp}. This difference is also the same for more multicomponent $s$-wave and $p$-wave superfluids.

\subsection{$2\times2$ examples}

	On the basis of the possible fillings in Table~\ref{ta:nonstatscsol}, we can carry out the concrete classification of the $n$-soliton solutions for $ n \le 3 $. This is given in Subsec~\ref{subsec:clsfysltn}.

	Here, let us see the examples in $ d=2 $ symmetric (triplet $p$-wave) or non-symmetric ($s$-$p$ mixed) cases. We use animation files \footnote{See animation files in the Supplemental Material at [URL provided by publisher]. Higher-quality gif animation files are also available at (it may need Chrome to view):  https://drive.google.com/folderview?id=0Bxd4vPIq2or8OTNySG9QVHB1V1E}. 
	Let us write $ \Delta=\left(\begin{smallmatrix} \Delta_{1,1} & \frac{1}{\sqrt{2}}(\Delta_{1,0}+\Delta_{0,0}) \\ \frac{1}{\sqrt{2}}(\Delta_{1,0}-\Delta_{0,0}) & \Delta_{1,-1} \end{smallmatrix}\right) $, where $ \Delta_{l=1,m=1,0,-1} $ and $ \Delta_{l=0,m=0} $ represent the triplet $ p $-wave and singlet $s$-wave order parameters, respectively.  Order parameters in $SU(2)$-symmetric theories are visualized by spherical harmonic functions (e.g., Ref.~\cite{Kawaguchi:2012ii}). The detail of this plot is explained in Appendix~\ref{subsec:spplt}. Drawing the spin polarization vector $ S_i=\sum_{m,n=-1,0,1}\Delta_{1,m}^*  [F_i]_{m,n} \Delta_{1,n} $ ($ F_{i=x,y,z} $: $ 3\times 3 $ spin-1 matrices) is also convenient to grasp physical picture. Parameters used in animation files are summarized in Subsec~\ref{subsec:parameters}. In all examples we set $ m=1 $, and $ \Delta_-=I_2 $. In this choice of $ \Delta_- $, the initial background condensate is purely $ p $-wave and ``polar'' phase \cite{VollhardtWolfle}. If we set $ \Delta_- $ to another unitary matrix, we can also realize the soliton dynamics with $s$-$p$ mixed background.  

\subsubsection{One-soliton solution}
	Animations 1, 2, and 3 are the examples of one-soliton solutions. Animation 1 is the most fundamental solution such that $ \theta_1=\frac{\pi}{2} $, which can be regarded as a generalization of $ \pi $-phase kink, and the localized fermion around a kink is Majorana, as discussed in the previous subsection. Animation 2 exemplifies a more general one soliton;  passing the soliton, the background superfluid is rotated by $ |\theta_1-\pi H(\theta_1-\frac{\pi}{2})| $ ($ H $: step function) about the rotation axis parallel to the spin. (In this example $ \theta_1=\frac{\pi}{3} $.) This kind of behavior, i.e., the rotation of background polar states induced by spin, is a common feature even in more complicated multi-soliton solutions. \\
	\indent If $ \hat{p}_1 $ is not proportional to a real vector ($\hat{p}_1 \not\propto \hat{p}_1^*$), the soliton induces $s$-wave-to-$p$-wave transformation, as given in Animation 3. This is a simple example of $s$-$p$ mixed dynamics.

\subsubsection{``Parallel'' and ``offset'' }
	Before going to the two- and three-soliton animations, we introduce the terms \textit{``parallel''} and \textit{``offset''} as below. If the constituent solitons of the breather or the collision phenomenon have the parallel spins, let us call such solutions ``parallel''. If not, we call them ``offset''. The parallel solutions can be realized by setting all $ \hat{p}_i $'s to the same. In particular, the parallel solution such that $ \hat{p}_i=(1,0)^T $ also represents a solution for $ d=1 $ system (spinless $p$-wave) if we only focus on $ \Delta_{1,1} $ and ignore other components. The offset solutions are, on the other hand, essentially specific to multicomponent systems.

\subsubsection{Two-soliton breather}
	Animations 4, 5, 6, and 7 show the examples of two-soliton breather solutions. In these cases, we set $ |s_1|=|s_2|=1 $. The breather period is given by  $ T=\frac{2\pi}{m|\cos\theta_1-\cos\theta_2|} $. Moving breathers can be created by the Lorentz boost, i.e., $(s_1,s_2)\rightarrow (rs_1,rs_2)$ with $ r>0 $.  Due to the constraint of fillings in the $ n=2 $ case of Table~\ref{ta:nonstatscsol}, one of the two bound states has a ``Dirac'' filling value. In particular, if $ \theta_1+\theta_2=\pi $, both fillings become ``Dirac'', i.e.,  $ (\nu_1,\nu_2)=(1,0) $ or $ (0,1) $. For this case the corresponding breather solution can be regarded as a multicomponent generalization of the Dashen-Hasslacher-Neveu (DHN) breather \cite{Dashen:1975xh,PhysRevD.89.025008}. Since we now consider $ 2\times 2 $ case,  let us call it an \textit{``SU(2)-DHN breather''}. \\
	\indent Animation 4 shows an SU(2)-DHN breather with parallel spins, and Animation 5 shows an offset one. For the parallel case, the axis of the spin breathing motion is fixed ($z$-axis in Animation 4). On the other hand, in the offset breather, the angle of the spin also oscillates, as Animation 5.  If the constituent two solitons are more separated, the amplitude of breathing motion becomes smaller, and solitons behaves like an independent stationary one-soliton solution, as shown in Animation 6. Finally, Animation 7 shows an example of the offset and non-DHN (``twisted'') breather, which is the most general two-soliton breather solution, and one of the two filling values of bound states is not necessarily ``Dirac''. (In Animation 7, we choose $ \nu_2=\frac{1}{2} $, although other values are also possible.) 

\subsubsection{Two-soliton collision}
	\indent Animations 8 and 10 show the examples of two-soliton collision phenomena. Animation 8 provides a parallel example and 10 gives an offset example. If the relative velocity between solitons is not so large, two solitons generally show breathing motions during the collision. These breathing motions are similar to the breather solutions discussed in Animations 4-7. The origin of these breathing behaviors in two-soliton collisions lies in the Dirac filling in Table~\ref{ta:nonstatscsol}; in order to achieve $ \nu=0 $ or 1, the matrix $ L $ must have an off-diagonal element, which induces the breathing. The delocalizaton of bound states also occurs for the same reason. See also the classification in Subsec.~\ref{subsec:clsfysltn}. On the other hand, several three-soliton solutions allow diagonal $ L $, hence the collision without breathing also occurs. (See the Majorana triplet solutions below.) \\
	\indent If the collision is parallel, as in Animation 8, the directions of spins remains unchanged and the background order parameter remains to be a pure $p$-wave superfluids (i.e.,  $\Delta_{0,0}=0$ holds exactly for all time.). On the other hand, in the offset collision in Animation 10, the spins are rotated during the collision and change their angles, and the contamination by the $s$-wave condensate occurs. While the former feature, i.e., the spin rotation, is similar to the soliton collision phenomena in the integrable spin-1 BEC with finite-density background \cite{JPSJ.75.064002}, the latter feature, the inevitable $s$-wave contamination, is specific to this system. This point will be more discussed in the perspective of the next subsection. 
	
	\indent The reason why complex $ \hat{p}_2 $ is chosen in Animation 10 is to realize the pure $p$-wave ($\Delta_{0,0}=0$) before collision. We can always choose $ \hat{p}_1,\hat{p}_2 $ such that the $s$-wave component vanishes \textit{either} before or after the collision. (This can be done by setting $ \Delta(x=+\infty) $ to be symmetric, where $ \Delta(x=+\infty) $ is given by replacement $ (LL^\dagger)^{-1} \rightarrow 0 $ in Eq.~(\ref{eq:deltainmain}).) However, there is no choice to realize such situation \textit{both} before and after the collision. For the $s$-wave contamination, see the perspectives in Subsec.~\ref{subsec:perspective}.
	
	\indent The detailed breathing pattern during the collision of two solitons depends on the initial relative phases of bound states. This is illustrated by Animation 9. If we only observe $ \Delta $ (both phase and amplitude), the breathing pattern seems to be difficult to predict and depend on a subtle difference of the initial condition sensitively, and the dynamics might be viewed as  ``ill-posed''. However, if we combine the information for both gap function and bound states, we can see that the breathing pattern is determined by the initial phase difference of bound states localized to each soliton. Thus the dynamics becomes predictable and  ``well-posed''.

\subsubsection{Three-soliton collision (Majorana triplet)}
	\indent Animations 11(a), 11(b), 11(c), and 12 provide the examples of ``Majorana triplet'' states in  $ n=3 $ (A) of Table~\ref{ta:nonstatscsol}, in which three-soliton collision occurs with all soliton eigenvalues satisfying $ \theta_1=\theta_2=\theta_3=\frac{\pi}{2} $. In Animation 11(a), we can observe the tunneling of the bound state from one kink to another during the collision. If the collision of three solitons occurs almost simultaneously, as Animation 11(b), the bound states temporarily delocalized to all three solitons,  but the final fate of bound states after the collision is the same as Animation 11(a). These two parallel examples also reduce to the solution of $ d=1 $, i.e.,  the spinless $p$-wave system, as with the case of parallel two-soliton solutions. Animation 11(c) shows the example of collisionless passing of solitons having antiparallel spins, realized by parameters setting $ \hat{p}_1=(1,0)^T $ and $ \hat{p}_2=(0,1)^T $. An example of offset collisions of the Majorana triplet state is shown in Animation 12. In this case, spins are rotated during the collision and the $s$-wave component inevitably mixes after collision. Solitons in Majorana triplet states show no breathing behavior in their collisions, since $ L $ is diagonal. (See Subsec.~\ref{subsec:clsfysltn} for detail.)

\subsubsection{More three-soliton examples}
	Finally, let us see the example animations of $ n=3 $ (B) and (C) in Table~\ref{ta:nonstatscsol}. Animations 13 and 14 correspond to the case (B), while 15 and 16 provide examples of the case (C). These two classes include more variety of three-soliton solutions, and  several families of analytical solutions are constructed in Subsec.~\ref{subsec:clsfysltn}. The examples shown below are all ``offset'' collisions or breathers, but we can also make ``parallel'' solutions by modifying all  $ \hat{p}_i $'s to be the same. \\
	\indent First we see the examples of the case (C), since this class can realize the Dirac-Dirac-Majorana type filling $ (\nu_1,\nu_2,\nu_3)=(1,0,\frac{1}{2}) $, where there exist two Dirac and one Majorana bound states. Animations 15 and 16 concentrate on such filling types, although more general filling values are possible.  Animation 15 shows the offset collision between one soliton and the parallel DHN breather. This animation suggests that, after the collision, the spin configuration of the breather generally becomes non-parallel. An $s$-wave component also emerges after collision, which is similar to the offset two-soliton collision given by Animation 10.  Animation 16 shows an example of the three-soliton breather, where we can observe the composite motion of an amplification similar to two-soliton breathers and unification and separation of constituent solitons. The time-dependence of three-soliton breathers is generally quasi-periodic. \\ 
	\indent Animations 13 and 14 show three-soliton solutions belonging to  $ n=3 $ (B) of Table~\ref{ta:nonstatscsol} with more general filling values.  Animation 13 shows the collision between one soliton and two-soliton non-DHN (twisted) breather. After one soliton passing through the breather, the spin angles of breathing motion changes, as similar to Animation 15. Animation 14 shows an example of three-soliton breather, where we can also observe the repeated unification and decomposition of constituent three solitons.\\
	\indent The parameter space of three-soliton solutions is too large. There might be a hidden interesting solution which cannot be covered in the present examples.

\subsection{Perspectives}\label{subsec:perspective}
\subsubsection{Finite reflection coefficient}
The GLM ansatz shown in Subsec.~\ref{subsec:glmansatz} mimics the ``genuine'' GLM equation only for the reflectionless cases. The genuine GLM equation also treats the potentials having finite reflection coefficients. These potentials generally possess small oscillations, which are called radiations (e.g., Ref.~\cite{Todahisenkeihadou}). Generalizing the GLM ansatz for the finite reflection case is left as a future task. As discussed below, such a generalization is one candidate which may solve the problem of the absence of nontrivial offset pure $p$-wave (i.e., purely symmetric $ \Delta $ ) time-dependent solutions. \\
	\indent Note that the non-zero reflection coefficient also alters the expression of the gap equation. For the stationary singlet $s$-wave problem, it is given in Ref.~\cite{takahashinittaJLTP}.

\subsubsection{How to overcome inevitable $s$-$p$ mixing}
	In the ``parallel'' solutions, i.e., Animations 4, 7, 8, 11(a), and 11(b), the gap function $ \Delta $ is exactly symmetric, i.e., it is purely $p$-wave. Animation 11(c), the antiparallel case, is also a ``cousin'' of the parallel solutions and purely $p$-wave. On the other hand,  in the ``offset'' solutions, Animations 5, 6, 10, 12, 13, 14, 15, and 16, the gap function contains an $s$-wave component ($\Delta_{0,0}\ne0$). As already mentioned before, the parallel solutions are essentially equivalent to a diagonal solution, and if we choose $ \hat{p}_i=(1,0)^T $, it reduces to the  $ 1\times 1 $ solution given by DT. Therefore, it means that the ``genuinely multicomponent'' time-dependent solutions always inevitably possess antisymmetric components, thus we cannot make an exactly symmetric solutions. On the other hand, if we confine ourselves to the time-independent problem, a family of symmetric solutions exist as shown in Subsec.~\ref{subsec:statclass}. \\
	\indent If we are interested in a system such that the $s$-wave and $p$-wave orders are energetically comparable, the above-mentioned soliton-induced $s$-wave emergence may be indeed likely and realistic, and hence our solutions provide exact examples for such phenomena, though realistic systems may not satisfy the ``fine-tuned condition'' $ g_+=g_- $. However, if we are interested in an application to pure $p$-wave systems where $s$-wave order is strongly prohibited, we must seriously consider why our solutions reported in this paper do not allow purely symmetric, time-dependent and offset soliton solutions. \\
	\indent One hint for the probable scenario is provided in the previous paragraph, i.e., the generalization of the GLM ansatz to the finite-reflection cases. We expect that, in the purely $p$-wave system, the offset collisions and the offset breathing motions are \textit{not} elastic, and hence, they are accompanied by an emission of small linear waves, e.g., phonons or magnons. If the soliton solutions are generalized with finite reflection and  thus radiations are correctly included, we expect that an exactly symmetric solution can be constructed by combination of the solitons and radiations. At this time, however, this is nothing but a conjecture. \\
	\indent The above situations are contrasted with the integrable spin-1 BECs \cite{PhysRevLett.93.194102,JPSJ.73.2996,JPSJ.75.064002}, where a large family of symmetric solutions can be realized by setting  $ \hat{p}_i=\hat{p}_i^* $ in the general soliton solutions of the matrix NLS equation. (See Subsec.~\ref{subsec:statclass} again.) However we expect that the offset collision will be possibly inelastic for \textit{non}-integrable spin-1 BECs ($ c_0\ne c_2 $), and the small radiations will be observed in these systems.

\subsubsection{More physical systems}
	In this paper, we considered the $SU(d)$-symmetric BdG systems, but it still does not cover all known examples. The one-body part of the Hamiltonian only include the kinetic term, and there is no magnetic field and spin-orbit coupling. (Note that $s$-$p$ mixing in our model is realized by an energetic degeneracy between $s$- and $p$-wave order parameters, and the mechanism is different from the known real materials, where the origin of mixing is spin-orbit interaction.)  The generalization to multiband superconductors are also worth investigating. Whether the restriction of the model can be weakened or not will be one of future problems. \\ 
%
%
	\indent The explicit soliton solutions and quasiparticle eigenstates in the stationary class (Subsec.~\ref{subsec:statclass}) will be convenient to study several static problems with external potentials, e.g., junction systems, edge states, and so on. This is because the solutions for piecewise constant external potentials (e.g., rectangular wells and barriers) can be constructed by a ``cut-and-paste'' of solutions of uniform systems. A harmonic trap, which appears in ultra cold atom systems, will be also approximated by such potentials to some extent. In this regard, we mention that a more uniform potential than harmonic ones are realized recently \cite{PhysRevLett.110.200406}. 
	Needless to say, the check of self-consistency, i.e., the gap equation, should be reconsidered for these problems. 
\subsubsection{Soliton-lattice background}
	The GLM ansatz shown in this paper will be immediately generalized to the cases of soliton-lattice background by combining the techniques constructed in this paper and the IST with uniformization variables formulated in Ref.~\cite{arxiv1304.7567}. This will be reported in future.
\subsubsection{Derivation without ansatz}
	The soliton solutions derived in this work should be reproduced without relying on the heuristic ansatz. One technical difficulty might lie in that the solution includes both $ s_j $ and $ s_j^* $, implying that the method of complex analysis, which is applicable only for meromorphic functions, might need some modification. (The stationary class [Subsec.~\ref{subsec:statclass}] has no such problem because $ s_j^*=s_j^{-1} $ and Jost functions are meromorphic.)
\subsubsection{Potentials not belonging to the DT class}
	In this paper we have focused on the DT class potentials, since it can solve the gap equation. However, the GLM ansatz include more solutions. 

	The general form of $ W $ satisfying the assumption of Subsec.~\ref{subsec:glmansatz} is
	\begin{align}
		W=\int_{s\in \mathbb{H}}\mathrm{d}s\mathrm{d}s^* \begin{pmatrix} p_1(s) & \cdots & p_n(s) \\ s\Delta_-^\dagger p_1(s) & \cdots & s\Delta_-^\dagger p_n(s) \end{pmatrix}\mathrm{e}^{-\mathrm{i}[k(s)x+\epsilon(s)t]} \label{eq:generalW11}
	\end{align} 
	where $ p_j(s) $'s are linearly independent $ d $-component weight functions with the behavior $ p_j(s)\rightarrow 0 $ for $ \operatorname{Im}s\rightarrow 0 $. If we take $ p_j(s)=\sum_i \hat{p}_i\delta(s-s_i)L_{ij} $, it reduces to the DT class (\ref{eq:dtrestrict2}). The gap function and BdG eigenstates generated from Eq.~(\ref{eq:generalW11}) satisfy the BdG equation, but they do not generally satisfy the gap equation. They could find an application in several time-dependent phenomena where self-consistency is not so important. 

\section{Eigenfunctions in the GLM ansatz} \label{sec:supprefless}
	Sections \ref{sec:supprefless}-\ref{sec:parameters} provide several technical details, which support the main result in Sec.~\ref{sec:mainr}.\\ 
	\indent In this section we prove fundamental properties of eigenfunctions of reflectionless potentials generated by the GLM ansatz introduced in Subsec.~\ref{subsec:glmansatz}.
\subsection{Proof of Eq. (\ref{eq:boundstate})}
	Here we give a proof that the bound and scattering states defined by Eqs. (\ref{eq:bounddef}) and (\ref{eq:scatdef}) satisfy Eq.~(\ref{eq:boundstate}).\\ 
	\indent Henceforth we often omit arguments of functions when it is evident. The subscripts $ x,t $ denotes the differentiation by these variables. 
	As given in Subsec.~\ref{subsec:glmansatz}, the GLM-like equation reduces to
	\begin{align}
		H(I_n+G)=-W. \label{eq:app001}
	\end{align}
	Differentiation of this equation by $ x,t $ gives
	\begin{align}
		H_x(I_n+G)&=-W_x-HG_x,\\
		H_t(I_n+G)&=-W_t-HG_t.
	\end{align}
	Since $ w_i $ satisfies Eq.~(\ref{eq:genrlss1}),  $ W $ also satisfies $ W_t=AW_x+BW $ and $ W_t^\dagger=W_x^\dagger A-W^\dagger B $. The derivatives of the Gram matrix $ G $ can be expressed as $ G_x=WW^\dagger,\ G_t=W^\dagger AW $. From these relations, we have
	\begin{align}
		(H_t-AH_x-BH)(I+G)=-[K,A]W.
	\end{align}
	However, since $ I_n+G $ is invertible and $ H=-W(I_n+G)^{-1} $, we obtain
	\begin{align}
		H_t=AH_x+(B+[K,A])H. \label{eq:appboundeq}
	\end{align}
	Thus, the equation for bound states $ h_i $'s is proved. Next we consider the scattering states. Let $ \Phi(x,t) $ be a $ d\times m $ matrix bounded for all $ (x,t) $ and satisfy the same equation with $ W $: $ \Phi_t=A\Phi_x+B\Phi $, where $ m $ need not be equal to $ n $ or $ d $. We introduce the matrix Jost function  $ F(x,t) $ by
	\begin{align}
		F(x,t)=\Phi(x,t)+\int_{-\infty}^x\!\mathrm{d}y K(x,y,t)\Phi(y,t).
	\end{align}
	Noting the relations $ K(x,y,t)=H(x,t)W(y,t)^\dagger $ and $ (W^\dagger\Phi)_t=(W^\dagger A \Phi)_x $, we find
	\begin{align}
		F_t&=\Phi_t+H_t \int_{-\infty}^x\!\mathrm{d}y W^\dagger \Phi+KA\Phi,\\
		F_x&=\Phi_x+K\Phi+H_x\int_{-\infty}^x\!\mathrm{d}y W^\dagger \Phi.
	\end{align}
	We thus obtain
	\begin{align}
		F_t-AF_x-BF&=(H_t-AH_x-BH) \int_{-\infty}^x\!\mathrm{d}y W^\dagger \Phi+[K,A]\Phi.
	\end{align}
	The RHS can be rewritten as $ [K,A]F $ using Eq.~(\ref{eq:appboundeq}). $\blacksquare$

\subsection{Orthonormal and completeness relations}
	Here we show the orthonormality and completeness of the set of eigenstates $ H $ and $ F $. \\ 
	\indent Let us assume that the scattering eigenstates of the uniform system is uniquely labeled by real variable $ \zeta $, and we write it as $ \Phi(x,t,\zeta) $. For brevity, henceforth we omit the time variable $ t $ of eigenfunctions, because it does not play a role in the proof. We assume that they satisfy the following orthogonal and completeness relations:
	\begin{align}
		\int_{-\infty}^\infty \mathrm{d}x \Phi(x,\zeta_1)^\dagger \Phi(x,\zeta_2)&=\mathcal{N}(\zeta_1,\zeta_2) \delta(k(\zeta_1)-k(\zeta_2))I_m,\\
		\int_{-\infty}^\infty \mathrm{d}\zeta \mathcal{\tilde{N}}(\zeta) \Phi(x,\zeta)\Phi(y,\zeta)^\dagger&=\delta(x-y)I_d, \label{eq:compreluni}
	\end{align}
	where $  \mathcal{N}(\zeta_1,\zeta_2) $ and $ \tilde{\mathcal{N}}(\zeta) $ are some weight functions and $ k(\zeta) $ is a wavenumber parametrized by $ \zeta $. We write the Jost function  $ F $ as follows:
	\begin{align}
		F(x,\zeta) &= \Phi(x,\zeta)+H(x)\Gamma(x,\zeta),\\
		\Gamma(x,\zeta)&:=\int_{-\infty}^x\mathrm{d}y W(y)^\dagger\Phi(y,\zeta).
	\end{align}
	Then, what we want to prove is the following: The orthonormal relations
	\begin{align}
		\int_{-\infty}^\infty \mathrm{d}x H(x)^\dagger H(x)&=I_n, \label{eq:orthobb} \\
		\int_{-\infty}^\infty\mathrm{d}x H(x)^\dagger F(x,\zeta) &=0, \label{eq:orthobs} \\
		\int_{-\infty}^\infty\mathrm{d}x  F(x,\zeta_1)^\dagger F(x,\zeta_2)&=\mathcal{N}(\zeta_1,\zeta_2)\delta(k(\zeta_1)-k(\zeta_2))I_m. \label{eq:orthoss}
	\end{align}
	and the completeness relation
	\begin{align}
		\int_{-\infty}^\infty\mathrm{d}\zeta \mathcal{\tilde{N}}(\zeta) F(x,\zeta)F(y,\zeta)^\dagger+H(x)H(y)^\dagger=\delta(x-y)I_d. \label{eq:compsb}
	\end{align}
	\indent \textit{Proof of the orthonormal relations} --- Eq. (\ref{eq:orthobb}) has been already proved in Subsec.~\ref{subsec:glmansatz} by using the relation
	\begin{align}
		H^\dagger H=-[(I_n+G)^{-1}]_x. \label{eq:hdaggerh}
	\end{align}
	Let us prove Eq.~(\ref{eq:orthobs}) and  (\ref{eq:orthoss}). We can check
	\begin{align}
		H^\dagger\Phi=-(I_n+G)^{-1}W^\dagger\Phi=-(I_n+G)^{-1}\Gamma_x. \label{eq:lemmarel}
	\end{align}
	Using Eqs. (\ref{eq:hdaggerh}) and (\ref{eq:lemmarel}), we obtain
	\begin{align}
		H(x)^\dagger F(x,\zeta)&=-\left[(I_n+G(x))^{-1}\Gamma(x,\zeta)\right]_x, \\
		F(x,\zeta_1)^\dagger F(x,\zeta_2) &= \Phi(x,\zeta_1)^\dagger\Phi(x,\zeta_2)\nonumber \\
		&\quad-\left[ \Gamma(x,\zeta_1)^\dagger(I_n+G(x))^{-1}\Gamma(x,\zeta_2) \right]_x.
	\end{align}
	Integration of these expressions provides Eqs. (\ref{eq:orthobs}) and  (\ref{eq:orthoss}). $\blacksquare$ \\
	\indent \textit{Proof of the completeness relation} --- By definition, 
	\begin{align}
		&F(x,\zeta)F(y,\zeta)^\dagger = \Phi(x,\zeta)\Phi(y,\zeta)^\dagger \nonumber \\
		&+\int_{-\infty}^x\!\mathrm{d}z H(x)W(z)^\dagger\Phi(z,\zeta)\Phi(y,\zeta)^\dagger \nonumber \\
		&+\int_{-\infty}^y\!\mathrm{d}z\Phi(x,\zeta)\Phi(z,\zeta)^\dagger W(z)H(y)^\dagger \nonumber \\
		&+\int_{-\infty}^x\!\mathrm{d}z_1\int_{-\infty}^y\!\mathrm{d}z_2 H(x)W(z_1)^\dagger\Phi(z_1,\zeta)\Phi(z_2,\zeta)^\dagger W(z_2) H(y).
	\end{align}
	After multiplying this expression by $ \mathcal{\tilde{N}}(\zeta) $, we integrate it by $ \zeta $. Using Eq.~(\ref{eq:compreluni}), we obtain
	\begin{align}
		&\int_{-\infty}^\infty\!\mathrm{d}\zeta \mathcal{\tilde{N}}(\zeta) F(x,\zeta)F(y,\zeta)^\dagger \nonumber \\
		&=\delta(x-y)I_d+H(x)W(y)^\dagger\theta(x-y)+W(x)H(y)^\dagger \theta(y-x)+\nonumber \\
		&\qquad\qquad H(x)\left[ G(x)\theta(y-x)+G(y)\theta(x-y) \right]H(y)^\dagger \nonumber \\
		&=\delta(x-y)I_d-H(x)H(y)^\dagger. 
	\end{align}
	Here we have used Eq. (\ref{eq:app001}) to obtain the last line. $\blacksquare$
%

\section{Formulation of the BdG theory}\label{sec:bdgformul}
	In this section, we formulate the BdG theory for $SU(2)$- and $SU(d)$-symmetric two-body interaction, and derive the fundamental BdG and gap equations solved in Sec.~\ref{sec:mainr}, that is, Eqs. (\ref{eq:matrixbdg}), (\ref{eq:gapsymantisym}), and (\ref{eq:gapnonsym}). \\ 
	\indent We begin with the fermionic many-body systems in the second-quantized formalism:
	\begin{align}
		\hat{H}&=\hat{H}_0+\hat{H}_{\text{int}}, \\
		\hat{H}_0&=\int\!\mathrm{d}x\,\hat{\psi}_i^\dagger(x) F_{ij}(x) \hat{\psi}_j(x) \\ 
		\hat{H}_{\text{int}}&=\frac{1}{2}\iint\mathrm{d}x\mathrm{d}y\, \hat{\psi}_j^\dagger(y)\hat{\psi}_i^\dagger(x)g_{ij,kl}(x-y)\hat{\psi}_k(x)\hat{\psi}_l(y). 
	\end{align}
	where $ x=(x_1,\dots,x_D) $ if we discuss $ D $ dimension. The subscripts $ i,j,k,l=1,\dots,S $ represent the internal degrees of freedom (spins, flavors, nuclear species, etc).  $ F_{ij}(x) $ is a one-body operator. 
\subsection{$SU(2)$-symmetric interaction}
\indent We first consider the case $ S=2 $ and they represent up/down spin: $ i,j,k,l=\uparrow, \downarrow $. The topic of this subsection can be found in many books, e.g., \cite{YamadaOhmi}. Let us assume that $ \hat{H}_{\text{int}} $ is invariant under the global $SU(2)$ transformation
	\begin{align}
		\hat{\psi}_i(x) \ \rightarrow\ U_{ij}\hat{\psi}_j(x),\quad  U\in SU(2), 
	\end{align} 
	which means that the interaction is isotropic in spin space. 
	Then $ \hat{H}_{\text{int}} $ reduces to
	\begin{align}
		\hat{H}_{\text{int}}=&\frac{1}{2}\iint\mathrm{d}x\mathrm{d}y\left(\vphantom{\sum_{m=0}^0}g_{0}(x-y)\hat{\Psi}_{0,0}^\dagger(x,y)\hat{\Psi}_{0,0}(x,y)\right. \nonumber\\
		&\qquad\left.+g_{1}(x-y)\sum_{m=-1}^1\hat{\Psi}_{1,m}^\dagger(x,y)\hat{\Psi}_{1,m}(x,y)\right), \label{eq:suppsu2hint}
	\end{align}
	where $ \hat{\Psi}_{0,0} $ and $ \hat{\Psi}_{1,m} $ are singlet and triplet components of two-body states:
	\begin{align}
		\hat{\Psi}_{0,0}(x,y)&=\tfrac{1}{\sqrt{2}}(\hat{\psi}_\downarrow(x)\hat{\psi}_\uparrow(y)-\hat{\psi}_\uparrow(x)\hat{\psi}_\downarrow(y)), \\
		\hat{\Psi}_{1,1}(x,y)&=\hat{\psi}_\uparrow(x)\hat{\psi}_\uparrow(y),\\
		\hat{\Psi}_{1,0}(x,y)&=\tfrac{1}{\sqrt{2}}(\hat{\psi}_\uparrow(x)\hat{\psi}_\downarrow(y)+\hat{\psi}_\downarrow(x)\hat{\psi}_\uparrow(y)),\\
		\hat{\Psi}_{1,-1}(x,y)&=\hat{\psi}_\downarrow(x)\hat{\psi}_\downarrow(y).
	\end{align}
	We further consider the short-range interaction, and approximate the field operator by first-order expansion: $ \hat{\Psi}_{l,m}(z+\frac{r}{2},z-\frac{r}{2})\simeq \hat{\Psi}_{l,m}(z,z)+\frac{r}{2}\cdot(\nabla_x-\nabla_y)\hat{\Psi}_{l,m}(x,y)|_{x,y\rightarrow z} $. Let the integrated coupling constants be
	\begin{align}
		\bar{g}_0:=\int g_{0}(x)\mathrm{d}x,\quad \bar{g}_1:=\frac{1}{D}\int g_{1}(x)x^2\mathrm{d}x, 
	\end{align}
	where the interaction is assumed to be isotropic in real space ($g_l(x)=g_l(|x|)$). Then, we obtain
	\begin{align}
		\hat{H}_{\text{int}}=\frac{1}{2}\int\mathrm{d}z\left[ \bar{g}_0\hat{\Phi}_{0,0}^\dagger\hat{\Phi}_{0,0}+ \bar{g}_1\sum_{m=1,0,-1}\hat{\boldsymbol{\Phi}}_{1,m}^\dagger\hat{\boldsymbol{\Phi}}_{1,m} \right]
	\end{align}
	with
	\begin{align}
		\hat{\Phi}_{0,0}&=\sqrt{2}\hat{\psi}_\downarrow\hat{\psi}_\uparrow, \label{eq:suppbdgphi00} \\
		\hat{\boldsymbol{\Phi}}_{1,1}&=\hat{\psi}_\uparrow(-\mathrm{i}\nabla)\hat{\psi}_\uparrow,\\
		\hat{\boldsymbol{\Phi}}_{1,0}&=\frac{\hat{\psi}_\uparrow(-\mathrm{i}\nabla)\hat{\psi}_\downarrow+\hat{\psi}_\downarrow(-\mathrm{i}\nabla)\hat{\psi}_\uparrow}{\sqrt{2}},\\
		\hat{\boldsymbol{\Phi}}_{1,-1}&=\hat{\psi}_\downarrow(-\mathrm{i}\nabla)\hat{\psi}_\downarrow. \label{eq:suppbdgphi1m1}
	\end{align}
	Note that $ \hat{\bm{\Phi}}_{1,m=1,0,-1} $ are $ D $-component vectors, e.g.,  $ \hat{\bm{\Phi}}_{1,1}=([\hat{\Phi}_{1,1}]_1,\dots,[\hat{\Phi}_{1,1}]_D)=(\hat{\psi}_\uparrow(-\mathrm{i}\partial_1)\hat{\psi}_\uparrow,\dots,\hat{\psi}_\uparrow(-\mathrm{i}\partial_D)\hat{\psi}_\uparrow) $. \\ 
\indent Now we go to the BdG approximation. Assuming the Cooper pair formation $ \Delta_{0,0}=\bar{g}_0\braket{\hat{\Phi}_{0,0}},\ \boldsymbol{\Delta}_{1,m}:=\bar{g}_1\braket{\hat{\boldsymbol{\Phi}}_{1,m}} $ and ignoring $ (\hat{\Phi}-\Delta/g)^2 $, we obtain the BdG Hamiltonian $ \hat{H}=\hat{H}_0+\hat{H}_{\text{int}} $, where
	\begin{align}
		\hat{H}_{\text{int}}=&\frac{1}{2}\int\mathrm{d}z\left[ \Delta_{0,0}^*\hat{\Phi}_{0,0}+\hat{\Phi}_{0,0}^\dagger\Delta_{0,0}-\frac{|\Delta_{0,0}|^2}{\bar{g}_0} \right]\nonumber \\
		&+\frac{1}{2}\int\mathrm{d}z\sum_{m}\left[ \boldsymbol{\Delta}_{1,m}^\dagger \hat{\boldsymbol{\Phi}}_{1,m}+\hat{\boldsymbol{\Phi}}_{1,m}^\dagger\boldsymbol{\Delta}_{1,m}-\frac{|\boldsymbol{\Delta}_{1,m}|^2}{\bar{g}_1} \right]. \label{eq:su2bdgint}
	\end{align}
	Here we do not include the Hartree-Fock mean fields of the type $ \braket{\hat{\psi}^\dagger\hat{\psi}} $, which will play a role in, e.g., determination of the vortex core structure in imbalanced Fermi superfluids \cite{PhysRevLett.97.180407} and charge-density-wave-superfluid transition \cite{PhysRevA.90.013632}. 
	Let us write down the BdG equation. Henceforth we use the Heisenberg picture, since it is convenient to formulate a time-dependent problem. The Heisenberg equation for the field operator is
	\begin{align}
		\mathrm{i}\frac{\partial }{\partial t}\begin{pmatrix} \hat{\psi}_\uparrow(x,t) \\ \hat{\psi}_\downarrow(x,t) \end{pmatrix}=F(x,t) \begin{pmatrix} \hat{\psi}_\uparrow(x,t) \\ \hat{\psi}_\downarrow(x,t) \end{pmatrix}+G(x,t) \begin{pmatrix} \hat{\psi}_\uparrow^\dagger(x,t) \\ \hat{\psi}_\downarrow^\dagger(x,t) \end{pmatrix} \label{eq:suppbdgbfrand}
	\end{align}
	with
	\begin{align}
		F&=\begin{pmatrix} F_{\uparrow\uparrow} & F_{\uparrow\downarrow} \\ F_{\downarrow\uparrow} & F_{\downarrow\downarrow}  \end{pmatrix}, \label{eq:suppbdgdiag} \\
		G&=\begin{pmatrix} \frac{1}{2}\left\{ -\mathrm{i}\nabla, \boldsymbol{\Delta}_{1,1} \right\} & \frac{1}{\sqrt{2}}\Delta_{0,0}+\frac{1}{2\sqrt{2}}\left\{ -\mathrm{i}\nabla, \boldsymbol{\Delta}_{1,0} \right\} \\ \frac{-1}{\sqrt{2}}\Delta_{0,0}+\frac{1}{2\sqrt{2}}\left\{ -\mathrm{i}\nabla, \boldsymbol{\Delta}_{1,0} \right\} & \frac{1}{2}\left\{ -\mathrm{i}\nabla, \boldsymbol{\Delta}_{1,-1} \right\} \end{pmatrix}, \label{eq:suppbdgoffdiag}
	\end{align}
	where we define $ \left\{\boldsymbol{A},\boldsymbol{B}\right\}f:=\boldsymbol{A}\cdot(\boldsymbol{B}f)+\boldsymbol{B}\cdot (\boldsymbol{A}f) $ for vector operators $ \boldsymbol{A} $ and $ \boldsymbol{B} $. 
	Introducing the Bogoliubov transformation
	\begin{align}
		\hat{\psi}_i(x,t)&=\sum_n u^{(n)}_i(x,t)\hat{a}_n+v^{(n)}_i(x,t)^*\hat{a}_n^\dagger, \label{eq:bogotr1}
	\end{align}
	where $ n $ represents the label of quasiparticle eigenstates (not to be confused with the internal degrees of freedom), we obtain the BdG equation 
	\begin{align}
		\mathrm{i}\frac{\partial }{\partial t}\begin{pmatrix} u_\uparrow^{(n)}(x,t) \\ u_\downarrow^{(n)}(x,t) \\ v_\uparrow^{(n)}(x,t) \\ v_\downarrow^{(n)}(x,t)  \end{pmatrix}= \begin{pmatrix} F(x,t) & G(x,t) \\ -G(x,t)^* & -F(x,t)^* \end{pmatrix}\begin{pmatrix} u_\uparrow^{(n)}(x,t) \\ u_\downarrow^{(n)}(x,t) \\ v_\uparrow^{(n)}(x,t) \\ v_\downarrow^{(n)}(x,t)  \end{pmatrix}. \label{eq:bdg10}
	\end{align}
	The gap equation can be also written down by substituting the Bogoliubov transformation (\ref{eq:bogotr1}) to the definition of the gap function $ \Delta_{l,m}=\bar{g}_l \braket{\Phi_{l,m}} $. In practical problems, we often encounter the situation such that the quasiparticle occupation state satisfies $ \braket{\hat{a}_n^\dagger\hat{a}_m}=\delta_{mn}\braket{\hat{a}_n^\dagger\hat{a}_n},\ \braket{\hat{a}_n\hat{a}_m}=\braket{\hat{a}_n^\dagger\hat{a}_m^\dagger}=0 $. If we restrict the formulation to such cases, the gap equations are given by
	\begin{align}
		\frac{\Delta_{0,0}}{\bar{g}_0}&=\frac{1}{\sqrt{2}}\sum_n\left( u_\uparrow^{(n)}v_\downarrow^{(n)*}-u_\downarrow^{(n)}v_\uparrow^{(n)*} \right)\left( 2\braket{\hat{a}_n^\dagger\hat{a}_n}-1 \right), \label{eq:gapeqswv}\\
		\frac{\boldsymbol{\Delta}_{\mathrm{1,1}}}{\bar{g}_1}&=\frac{-\mathrm{i}}{2}\sum_n\left( v_\uparrow^{(n)*}\nabla u_\uparrow^{(n)}-u_\uparrow^{(n)}\nabla v_\uparrow^{(n)*} \right)\left(2\braket{\hat{a}_n^\dagger\hat{a}_n}-1\right), \\
		\frac{\boldsymbol{\Delta}_{\mathrm{1,0}}}{\bar{g}_1}&=\frac{-\mathrm{i}}{2\sqrt{2}}\sum_n\left( v_\uparrow^{(n)*}\nabla u_\downarrow^{(n)}+v_\downarrow^{(n)*}\nabla u_\uparrow^{(n)}\right. \nonumber \\ 
		&\qquad\qquad\quad \left. -u_\uparrow^{(n)}\nabla v_\downarrow^{(n)*}-u_\downarrow^{(n)}\nabla v_\uparrow^{(n)*} \right)\left(2\braket{\hat{a}_n^\dagger\hat{a}_n}-1\right), \\
		\frac{\boldsymbol{\Delta}_{\mathrm{1,-1}}}{\bar{g}_1}&=\frac{-\mathrm{i}}{2}\sum_n\left( v_\downarrow^{(n)*}\nabla u_\downarrow^{(n)}-u_\downarrow^{(n)}\nabla v_\downarrow^{(n)*} \right)\left(2\braket{\hat{a}_n^\dagger\hat{a}_n}-1\right). \label{eq:gapeqpwvm1}
	\end{align}
	$ \Delta_{0,0} $ represents the singlet $s$-wave order parameter, while $ \boldsymbol{\Delta}_{1,m} $'s are the triplet $p$-wave ones, whose number of components is $3D$ if we consider $D$ dimension.

\subsection{Linearization at the Fermi point (Andreev approximation)}\label{subsec:andapp}
	Henceforth, we discuss one dimension and write the triplet order parameter by normal font: $ \boldsymbol{\Delta}_{1,m}=\Delta_{1,m} $.\\
	\indent In Eq. (\ref{eq:suppbdgdiag}), we set $ F_{ij}=(-\frac{\partial_x^2}{2}-\mu_{\mathrm{F}})\delta_{ij},\ \mu_F=\frac{k_{\mathrm{F}}^2}{2} $ with $ k_F $ a Fermi wavenumber. Following the Andreev approximation \cite{Andreev1964}, we linearize the BdG equation around the right and left Fermi points $ k=\pm k_{\mathrm{F}} $ [See, e.g.,  Fig.~1 of Ref.~\cite{takahashinittaJLTP}]. For the right Fermi point, substituting $ (u,v)=(u_R,v_R)\mathrm{e}^{\mathrm{i}k_{\mathrm{F}}x} $ and keeping the leading order about $ k_{\mathrm{F}} $, the BdG equation reduces to
	\begin{gather}
		\mathrm{i}\frac{\partial }{\partial t}\begin{pmatrix} u_{R\uparrow}^{(n)}(x,t) \\ u_{R\downarrow}^{(n)}(x,t) \\ v_{R\uparrow}^{(n)}(x,t) \\ v_{R\downarrow}^{(n)}(x,t)  \end{pmatrix}= \begin{pmatrix} -\mathrm{i}k_{\mathrm{F}}I_2 & G_R(x,t) \\ G_R(x,t)^\dagger & \mathrm{i}k_{\mathrm{F}}I_2 \end{pmatrix}\begin{pmatrix} u_{R\uparrow}^{(n)}(x,t) \\ u_{R\downarrow}^{(n)}(x,t) \\ v_{R\uparrow}^{(n)}(x,t) \\ v_{R\downarrow}^{(n)}(x,t)  \end{pmatrix},  \label{eq:bdg1} \\
		G_R= \begin{pmatrix} k_{\mathrm{F}} \Delta_{1,1}  & \frac{1}{\sqrt{2}}(\Delta_{0,0}+k_{\mathrm{F}}\Delta_{1,0}) \\ \frac{1}{\sqrt{2}}(-\Delta_{0,0}+k_{\mathrm{F}}\Delta_{1,0}) & k_{\mathrm{F}}\Delta_{1,-1} \end{pmatrix}. \label{eq:bdg112}
	\end{gather} 
	The differential operators for $p$-wave parts in  $ G $ are replaced by $ k_F $ by this approximation. Note that $ G_R $ is \textit{not} antisymmetric, while $ G $ in the original BdG equation [Eq.~(\ref{eq:suppbdgoffdiag})] is antisymmetric if we define the transpose of the momentum $ \boldsymbol{p}=-\mathrm{i}\nabla $ by $ \boldsymbol{p}^T=-\boldsymbol{p} $. \\
	\indent In the same way, we obtain the equations for the left Fermi point:
	\begin{align}
		\mathrm{i}\frac{\partial }{\partial t}\begin{pmatrix} u_{L\uparrow}^{(n)}(x,t) \\ u_{L\downarrow}^{(n)}(x,t) \\ v_{L\uparrow}^{(n)}(x,t) \\ v_{L\downarrow}^{(n)}(x,t)  \end{pmatrix}= \begin{pmatrix} \mathrm{i}k_{\mathrm{F}}I_2 & G_L(x,t) \\ G_L(x,t)^\dagger & -\mathrm{i}k_{\mathrm{F}}I_2 \end{pmatrix}\begin{pmatrix} u_{L\uparrow}^{(n)}(x,t) \\ u_{L\downarrow}^{(n)}(x,t) \\ v_{L\uparrow}^{(n)}(x,t) \\ v_{L\downarrow}^{(n)}(x,t)  \end{pmatrix} \label{eq:bdg1l}, 
	\end{align} 
	where $ G_L=-G_R^T $. \\ 
	\indent As discussed in Sec.~\ref{app:dblcnt} in detail, when we consider the one-dimensional system with linearized dispersion relations, one way to take into account all physically independent BdG eigenstates is to consider the eigenstates of the right-Fermi-point equation (\ref{eq:bdg1}) and ignore those of the left equation (\ref{eq:bdg1l}). If we use this convention, we can omit the subscript ``$ R $'' without confusion. In this convention, the gap equations [Eqs. (\ref{eq:gapeqswv})-(\ref{eq:gapeqpwvm1})] are approximately given as
	\begin{align}
		\frac{\Delta_{0,0}}{\bar{g}_0}&=\frac{1}{\sqrt{2}}\sum_n\left( u_\uparrow^{(n)}v_\downarrow^{(n)*}-u_\downarrow^{(n)}v_\uparrow^{(n)*} \right)\left( 2\braket{\hat{a}_n^\dagger\hat{a}_n}-1 \right) \label{eq:gapeqswvaf}\\
		\frac{\Delta_{1,1}}{\bar{g}_1}&=k_F\sum_nu_\uparrow^{(n)}v_\uparrow^{(n)*}\left(2\braket{\hat{a}_n^\dagger\hat{a}_n}-1\right) \\
		\frac{\Delta_{1,0}}{\bar{g}_1}&=\frac{k_F}{\sqrt{2}}\sum_n\left( u_\downarrow^{(n)}v_\uparrow^{(n)*} +u_\uparrow^{(n)}v_\downarrow^{(n)*} \right)\left(2\braket{\hat{a}_n^\dagger\hat{a}_n}-1\right) \\
		\frac{\Delta_{1,-1}}{\bar{g}_1}&=k_F\sum_n u_\downarrow^{(n)}v_\downarrow^{(n)*} \left(2\braket{\hat{a}_n^\dagger\hat{a}_n}-1\right), \label{eq:gapeqpwvm1af}
	\end{align}
	where the derivatives in  $ \Delta_{1,m} $'s are also replaced by $ k_\mathrm{F} $. If we use a new dimensionless unit such that $ k_{\mathrm{F}}=1 $, the BdG equations (\ref{eq:bdg1}) and (\ref{eq:bdg112}) and the gap equations (\ref{eq:gapeqswvaf})-(\ref{eq:gapeqpwvm1af}) reduce to the equation solved in Sec.~\ref{sec:mainr}. 
	In many cases, either $s$-wave or $p$-wave order only occurs, but they are degenerate when $ \bar{g}_0=k_\mathrm{F}^2\bar{g}_1 $ holds. This point realizes the ``fine-tuned $s$-$p$ mixed point'', which is introduced in Sec.~\ref{sec:mainr}, Eq.~(\ref{eq:gapnonsym}). 

\subsection{$SU(d)$ case}
	Let us consider $ d $-component fermion $ (\hat{\psi}_1,\dots,\hat{\psi}_d) $ and assume that the interaction $ \hat{H}_{\text{int}} $ is invariant under
	\begin{align}
		\hat{\psi}_i \rightarrow U_{ij}\hat{\psi}_j,\ U\in SU(d).
	\end{align}
	Then  $ \hat{H}_{\text{int}} $ reduces to symmetric and antisymmetric interaction:
	\begin{align}
		\hat{H}_{\text{int}}=\frac{1}{2}\sum_{i,j=1}^d\sum_{s=+,-}\int\mathrm{d}x\mathrm{d}y \left[g_s(x-y) \hat{\Psi}_{s,ij}(x,y)^\dagger\hat{\Psi}_{s,ij}(x,y) \right],
	\end{align}
	where $ \hat{\Psi}_{\pm,ij}(x,y)=\frac{1}{2}(\hat{\psi}_i(x)\hat{\psi}_j(y)\pm \hat{\psi}_j(x)\hat{\psi}_i(y)) $. \\
	\indent The remaining formulations (e.g., the expansion of order parameters up to first order, BdG approximation, Andreev approximation, etc.) are the same as the $SU(2)$ model, and the symmetric and antisymmetric parts describe the $p$-wave and $s$-wave order parameters, respectively. 
%

\subsection{$N$-flavor generalization}\label{app:Nflavor}
	In Sec.~\ref{sec:mainr}, we have considered the superposition of occupied and unoccupied states to realize partial filling rates of bound states.  Another way to realize this is the multi-flavor generalization, as is often done in high-energy physics. Here, we provide an $N$-flavor generalization of the $SU(2)$-symmetric interaction model. Extension to the $SU(d)$ case is straightforward. 
	Note that the $SU(d)$-symmetric interaction model is \textit{not} the $ d $-flavor generalization; we can even introduce $N$-flavor generalization for the $SU(d)$ model, where the total number of internal degrees of freedom becomes $Nd$. \\
%
	\indent Let us consider $ N $ species of spin-$1/2$ fermions, hence the total number of internal degrees of freedom is given by $ S=2N $. Let us write the field operator of this system as $ \hat{\psi}_{fi}(x) $, where $ i=\uparrow, \downarrow $ and $ f=1,\dots,N $. We call $ f $ a ``flavor''. Then, we impose the following conditions for one- and two-body operators $ \hat{H}_0 $ and $ \hat{H}_{\text{int}} $. First,  $ \hat{H}_0 $ is block-diagonal and there is no matrix element between different flavors:
	\begin{align}
		F_{(fi),(gj)}(x)=\delta_{fg}F_{ij}(x).
	\end{align}
	Second, the interaction term $ \hat{H}_{\text{int}} $ is invariant under the following global transformation:
	\begin{align}
		\hat{\psi}_{fi}(x) &\rightarrow U_{ij}\hat{\psi}_{fj}(x),\quad U\in SU(2), \\
		\hat{\psi}_{fi}(x) &\rightarrow R_{fg}\hat{\psi}_{gi}(x),\quad R\in O(N). \label{eq:suppbdgontrans}
	\end{align}
	An interaction satisfying the above condition is given by Eq.~(\ref{eq:suppsu2hint}), where  $ \hat{\Psi} $'s are modified as follows: 
	\begin{align}
		\hat{\Psi}_{0,0}(x,y)&=\sum_{f=1}^N\tfrac{1}{\sqrt{2}}(\hat{\psi}_{f\downarrow}(x)\hat{\psi}_{f\uparrow}(y)-\hat{\psi}_{f\uparrow}(x)\hat{\psi}_{f\downarrow}(y)), \\
		\hat{\Psi}_{1,1}(x,y)&=\sum_{f=1}^N\hat{\psi}_{f\uparrow}(x)\hat{\psi}_{f\uparrow}(y),\\
		\hat{\Psi}_{1,0}(x,y)&=\sum_{f=1}^N\tfrac{1}{\sqrt{2}}(\hat{\psi}_{f\uparrow}(x)\hat{\psi}_{f\downarrow}(y)+\hat{\psi}_{f\downarrow}(x)\hat{\psi}_{f\uparrow}(y)),\\
		\hat{\Psi}_{1,-1}(x,y)&=\sum_{f=1}^N\hat{\psi}_{f\downarrow}(x)\hat{\psi}_{f\downarrow}(y).
	\end{align}
	$ \hat{\Psi} $'s are invariant under the transformation (\ref{eq:suppbdgontrans}), since $ \sum_{f=1}^N \hat{\psi}_{fi}(y)\hat{\psi}_{fj}(x) $  forms an $ O(N) $ singlet. Performing the same approximation as the one-flavor system, we obtain the same  $ \hat{H}_{\text{int}} $ as Eq.~(\ref{eq:su2bdgint}), but now the physical quantities are generalized to $ N $-flavor ones, namely, Eqs. (\ref{eq:suppbdgphi00})-(\ref{eq:suppbdgphi1m1}) are replaced by 
	\begin{align}
		\hat{\Phi}_{0,0}&=\sum_{f=1}^N\sqrt{2}\hat{\psi}_{f\downarrow}\hat{\psi}_{f\uparrow},\\
		\hat{\boldsymbol{\Phi}}_{1,1}&=\sum_{f=1}^N\hat{\psi}_{f\uparrow}(-\mathrm{i}\nabla)\hat{\psi}_{f\uparrow},\\
		\hat{\boldsymbol{\Phi}}_{1,0}&=\sum_{f=1}^N\frac{\hat{\psi}_{f\uparrow}(-\mathrm{i}\nabla)\hat{\psi}_{f\downarrow}+\hat{\psi}_{f\downarrow}(-\mathrm{i}\nabla)\hat{\psi}_{f\uparrow}}{\sqrt{2}},\\
		\hat{\boldsymbol{\Phi}}_{1,-1}&=\sum_{f=1}^N\hat{\psi}_{f\downarrow}(-\mathrm{i}\nabla)\hat{\psi}_{f\downarrow},\\
		\Delta_{0,0}&=\bar{g}_0\braket{\hat{\Phi}_{0,0}},\quad \boldsymbol{\Delta}_{1,m}=\bar{g}_1\braket{\hat{\boldsymbol{\Phi}}_{1,m}}. 
	\end{align}
	The BdG equation becomes block-diagonal for each flavor and it obeys the same equation (\ref{eq:bdg10}). The dispersion linearization around the Fermi point can be done in the same way as Subsec.~\ref{subsec:andapp}. Let $ \hat{a}_{f,n} $ be an annihilation operator of the eigenstate $ n $ of flavor $ f $, 
	and the gap equation can be obtained by the following modification in Eqs. (\ref{eq:gapeqswvaf})-(\ref{eq:gapeqpwvm1af}):
	\begin{align}
		\left(2\braket{\hat{a}_n^\dagger\hat{a}_n}-1\right) \quad \rightarrow \quad \sum_{f=1}^N\left(2\braket{\hat{a}_{f,n}^\dagger\hat{a}_{f,n}}-1\right).
	\end{align} 

\section{Double-counting problem of BdG eigenstates}\label{app:dblcnt}
In this section we summarize general properties of the BdG equation, and discuss the double-counting problem.
\subsection{BdG equation and Bogoliubov transformation}
	In order to discuss double-counting problem in the next subsection, we first summarize a few general aspects of the BdG theory. For simplicity, we consider a discrete and finite-dimensional problem. The discussion below is easily generalized to continuous systems, if we interpret the h.c., transpose, and c.c. of the differential (momentum) operator $ p=-\mathrm{i}\partial $ as $ p=p^\dagger=-p^T=-p^* $. \\
	\indent The diagonalization of the BdG Hamiltonian by the Bogoliubov transformation is reduced to the following matrix eigenvalue problem:  The diagonalization of $ H $ by $ W $, where 
	\begin{align}
		H&=H^\dagger,\quad \tau H \tau=-H^*, \label{eq:bdghermite} \\
		W^{-1}&=W^\dagger,\quad \tau W \tau =W^*, \\
		\tau&:=\begin{pmatrix} & I_N \\ I_N &  \end{pmatrix}.
	\end{align}
	Due to these conditions, $ H $ and $ W $ have the form
	\begin{align}
		&H=\begin{pmatrix} F & G \\ -G^* & -F^* \end{pmatrix},\quad W=\begin{pmatrix} U & V^* \\ V & U^* \end{pmatrix}, \\
		&F^\dagger=F,\quad G^T=-G, \\
		&U^\dagger U+V^\dagger V=I_N,\ U^TV+V^TU=0, \label{eq:bdgunitary1} \\
		&UU^\dagger+V^*V^T=I_N,\ UV^\dagger+V^*U^T=0. \label{eq:bdgunitary2}
	\end{align}
	Since $ W $ and  $ H $ reduce to $O(2N,\mathbb{R})$ and its Lie algebra \cite{PhysRevB.55.1142,PhysRevB.78.195125}, the diagonalizability is always ensured, though each problem has each difficulty in practice. (This simplicity contrasts with the bosonic problem, reducing to the symplectic group $Sp(2N,\mathbb{R})$ and its Lie algebra, where classification of standard forms is complicated \cite{Colpa1978,Colpa1986,PhysRevD.91.025018,Takahashi2015101}. See also \cite{Arnold}.) \\
	\indent If $ \boldsymbol{w}=(\boldsymbol{u},\boldsymbol{v})^T $ is an eigenstate of $ H $ with eigenvalue $ \epsilon $,  $ \tau\boldsymbol{w}^*=(\boldsymbol{v}^*,\boldsymbol{u}^*)^T $ is also an eigenstate with $ -\epsilon $. Let $ W $ be a diagonalizing matrix such that
	\begin{align}
		W^{-1}HW=\begin{pmatrix} \mathcal{E} & \\ & -\mathcal{E} \end{pmatrix},\quad \mathcal{E}=\operatorname{diag}(\epsilon_1,\dots,\epsilon_N).
	\end{align}
	Then, the corresponding quantum many-body Hamiltonian
	\begin{align}
		\hat{H}&=\sum_{ij}F_{ij}\hat{c}_i^\dagger\hat{c}_j+\frac{1}{2}\sum_{ij}G_{ij}\hat{c}_i^\dagger\hat{c}_j^\dagger+\frac{1}{2}\sum_{ij}G_{ij}^*\hat{c}_j\hat{c}_i \nonumber \\
		&=\frac{1}{2}\begin{pmatrix} (\hat{\boldsymbol{c}}^\dagger)^T & \hat{\boldsymbol{c}}^T \end{pmatrix}\begin{pmatrix} F & G \\ -G^* & -F^* \end{pmatrix}\begin{pmatrix} \hat{\boldsymbol{c}} \\ \hat{\boldsymbol{c}}^\dagger \end{pmatrix}+\frac{1}{2}\operatorname{tr}F,
	\end{align}
	where $ \hat{\boldsymbol{c}}:=(\hat{c}_1,\dots,\hat{c}_N)^T $ and  the dagger symbol $ \dagger $ for annihilation operators only changes them to creation operators without changing $ N\times 1 $ matrix to $ 1\times N $, is diagonalized as
	\begin{align}
		\hat{H}&=\frac{1}{2}\begin{pmatrix} (\hat{\boldsymbol{a}}^\dagger)^T & \hat{\boldsymbol{a}}^T \end{pmatrix}\begin{pmatrix} \mathcal{E} &  \\  & -\mathcal{E} \end{pmatrix}\begin{pmatrix} \hat{\boldsymbol{a}} \\ \hat{\boldsymbol{a}}^\dagger \end{pmatrix}+\frac{1}{2}\operatorname{tr}F \nonumber \\
		&=\sum_i \epsilon_i \hat{a}_i^\dagger\hat{a}_i-\frac{1}{2}\operatorname{tr}\mathcal{E}+\frac{1}{2}\operatorname{tr}F
	\end{align}
	by the Bogoliubov transformation
	\begin{align}
		\begin{pmatrix} \hat{\boldsymbol{c}} \\ \hat{\boldsymbol{c}}^\dagger \end{pmatrix}=W\begin{pmatrix} \hat{\boldsymbol{a}} \\ \hat{\boldsymbol{a}}^\dagger \end{pmatrix} 
		\ \leftrightarrow \  \hat{c}_i = \sum_j U_{ij}\hat{a}_j+V_{ij}^*\hat{a}_j^\dagger, \label{eq:bdgtransdis1}
	\end{align}
	The inverse transformation is $ \hat{a}_i=\sum_j U_{ji}^*\hat{c}_j+V_{ji}^*\hat{c}_j^\dagger $ since $ W^{-1}=W^\dagger $. \\
	\indent For convenience, we also write down the continuous version of Eqs. (\ref{eq:bdgunitary1}), (\ref{eq:bdgunitary2}), and (\ref{eq:bdgtransdis1}). The Bogoliubov transformation:
	\begin{align}
		\hat{\psi}_i(x)&=\sum_n \left(u_i^{(n)}(x)\hat{a}_n+v_i^{(n)}(x)^*\hat{a}_n^\dagger\right),  \label{eq:Bogopsi} \\
		\hat{a}_n&=\sum_i\int\!\mathrm{d}x \left(u_i^{(n)}(x)^*\hat{\psi}_i(x)+v_i^{(n)}(x)^*\hat{\psi}_i^\dagger(x)\right), 
	\end{align}
	where the index $ i $ represents internal degrees of freedom (e.g., spin), not the label of eigenstates. 
	Orthonormality [counterpart of Eq.~(\ref{eq:bdgunitary1})]:
	\begin{align}
		\sum_i\int\!\mathrm{d}x\left( u_i^{(n)}(x)^*u_i^{(m)}(x)+v_i^{(n)}(x)^*v_i^{(m)}(x) \right)&=\delta_{nm}, \label{eq:Bogoortho1} \\
		\sum_i\int\!\mathrm{d}x\left( u_i^{(n)}(x)v_i^{(m)}(x)+v_i^{(n)}(x)u_i^{(m)}(x) \right)&=0.\label{eq:Bogoortho2}
	\end{align}
	Completeness [counterpart of Eq.~(\ref{eq:bdgunitary2})]:
	\begin{align}
		\sum_n\left( u_i^{(n)}(x)u_j^{(n)}(y)^*+v_i^{(n)}(x)^*v_j^{(n)}(y) \right)&=\delta_{ij}\delta(x-y),\label{eq:Bogoortho3} \\
		\sum_n\left( u_i^{(n)}(x)v_j^{(n)}(y)^*+v_i^{(n)}(x)^*u_j^{(n)}(y) \right)&=0. \label{eq:Bogoortho4}
	\end{align}
%

\subsection{How to eliminate double counting}
	As mentioned above, the BdG equation always has a pair of eigenstates $ (u,v) $ and $ (v^*,u^*) $ with opposite sign of eigenvalues. These two states must be identified as a different representation of the same physical state, i.e., ``creating $ (u,v) $'' $ = $ ``annihilating $(v^*,u^*)$''.  Therefore, when we define the Bogoliubov transformation, Eq (\ref{eq:bdgtransdis1}) or (\ref{eq:Bogopsi}), we should use only one eigenstate of these two. Actually, if both eigenstates were incorrectly included as independent ones, we would soon find the violation of the anticommutation relations for $ \hat{a}_i $'s. \\
	\indent For the same reason, the summation appearing in expressions of expectation values of various physical quantities, e.g., the gap equations, also should be taken only for the half of eigenstates of $ H $, not for all of them. If we consider simple occupation states, e.g., ground states or finite-temperature equilibrium, we do not need a special care for this problem, because it merely alters the effective coupling constants by twice and important physical conclusions do not change. However, if we consider some particular excited states with specific fillings, a correct identification of independent states will become important. \\
	\indent We then face the following problem: what kind of choice is convenient to resolve the above-mentioned double counting of the BdG eigenstates. The following (i) is the most general:
	\begin{enumerate}[(i)]
		\item Applicable for all systems: Ignore all positive-energy states and only use negative-energy ones. For zero-energy states, however, we need to introduce some specific rule, and what kind of rule is convenient may depend on systems and states.
	\end{enumerate}
	On the other hand, if the BdG equation has the block-diagonal form
	\begin{align}
		H=\begin{pmatrix} * &&& * \\ & * & * & \\ & * & * & \\ * &&& *  \end{pmatrix},
	\end{align}
	a more convenient rule is available: Use all eigenstates of one block for all energies, and ignore all eigenstates of the other block. In particular, we need no subtle treatment for zero-energy states. The following (ii) and (iii) correspond to this case:
	\begin{enumerate}[(i)] \setcounter{enumi}{1}
		\item Applicable for one-dimensional systems with dispersion linearization (the Andreev approximation):  Ignore left-Fermi-point eigenstates (linearized at $k=-k_F$), and only use the right ones ($k=k_F$).
		\item Applicable for $s$-wave spin-1/2 systems without spin-orbit coupling: Ignore all eigenstates $ (u_\downarrow,v_\uparrow) $, and only use $ (u_\uparrow,v_\downarrow) $. 
	\end{enumerate}
	
%
	\indent First, we explain (ii). The dispersion linearization by the Andreev approximation is already discussed in Subsec.~\ref{subsec:andapp}. We substitute $ (u,v)=\mathrm{e}^{\mathrm{i}k_Fx}(u_R,v_R) $ and $ (u,v)=\mathrm{e}^{-\mathrm{i}k_Fx}(u_L,v_L) $, and only keep the leading order about $ k_F $.  Generally, after this approximation, we obtain the BdG equations for the right and left Fermi points
	\begin{align}
		H_R\begin{pmatrix}u_R \\ v_R \end{pmatrix}=\epsilon\begin{pmatrix}u_R \\ v_R \end{pmatrix},\quad H_L\begin{pmatrix}u_L \\ v_L \end{pmatrix}=\epsilon\begin{pmatrix}u_L \\ v_L \end{pmatrix}
	\end{align}
	with $ H_R,\ H_L $ having the forms 
	\begin{align}
		H_R=\begin{pmatrix} A & B \\ B^\dagger & D \end{pmatrix},\quad H_L=\begin{pmatrix} -D^* & -B^T \\ -B^* & -A^* \end{pmatrix}
	\end{align}
	with $ A^\dagger=A,\ D^\dagger=D $. Note that $ B $ is not necessarily antisymmetric. Equations (\ref{eq:bdg1}) and (\ref{eq:bdg1l}) indeed have these forms. 
	Though $ H_R $ and $ H_L $ are hermitian, they are not the BdG-type matrix given in Eq.~(\ref{eq:bdghermite}). However, if these two are combined into the form 
	\begin{align}
		\begin{pmatrix} A & & & B \\ & -D^* & -B^T & \\ & -B^* & -A^* & \\ B^\dagger & & & D  \end{pmatrix}\begin{pmatrix} u_R \\ u_L \\ v_L \\ v_R \end{pmatrix}=\epsilon\begin{pmatrix} u_R \\ u_L \\ v_L \\ v_R \end{pmatrix},
	\end{align}
	then it recovers the condition (\ref{eq:bdghermite}). If we have one right solution $ (u_R,v_R)=(\tilde{u},\tilde{v}) $ with eigenvalue $ \epsilon $, the corresponding left solution is given by $ (u_L,v_L)=(\tilde{v}^*,\tilde{u}^*) $ with eigenvalue $ -\epsilon $. Using this correspondence, we can apply the convention (ii) to eliminate the double counting. Throughout the present work, we follow this convention and only use the eigenstates of right-Fermi-point equations [Eqs. (\ref{eq:matrixbdg})-(\ref{eq:gapnonsym})]. \\
	\indent Note that the Andreev approximation is essentially the same with the WKB approximation. (The expansion parameter is $ k_{\mathrm{F}}^{-1} $.) So, if there exists a junction or a barrier whose potential value changes rapidly, we should make a linear combination of the right-point and left-point eigenstates such as $ (u,v)=a \mathrm{e}^{\mathrm{i}k_{\mathrm{F}}x}(u_R,v_R)+b \mathrm{e}^{-\mathrm{i}k_{\mathrm{F}}x}(u_L,v_L) $ to discuss the scattering phenomena. In such case the right and left equations are not decoupled completely. \\
	\indent Next, we consider (iii). This choice becomes possible, because the BdG equation for spin-1/2 and $s$-wave system without a spin-orbit coupling term is block-diagonalized:
	\begin{align}
		H=\begin{pmatrix} -\frac{\nabla^2}{2m}-\mu-h & & & \Delta_{\uparrow\downarrow} \\ & -\frac{\nabla^2}{2m}-\mu+h & \Delta_{\downarrow\uparrow} \\ & -\Delta_{\uparrow\downarrow}^* & \frac{\nabla^2}{2m}+\mu+h & \\ -\Delta_{\downarrow\uparrow}^* &&& \frac{\nabla^2}{2m}+\mu-h \end{pmatrix}
	\end{align}
	with $ \Delta_{\downarrow\uparrow}=-\Delta_{\uparrow\downarrow} $. Different from the choice (ii), this convention is applicable even in higher dimensions. The merit of this convention is that the gap equation becomes simple; the gap equation is given by $ -\Delta_{\uparrow\downarrow}/g=\frac{1}{2}\sum_j(2\nu_j-1)(u_{j\uparrow}v_{j\downarrow}^*-u_{j\downarrow}v_{j\uparrow}^*) $, but if we set $ u_{j\downarrow}=v_{j\uparrow}=0 $ for all $ j $, we have
	\begin{align}
		-\Delta_{\uparrow\downarrow}/g=\sum_j\nu_ju_{j\uparrow}v_{j\downarrow}^*,
	\end{align}
	where the completeness relation (\ref{eq:Bogoortho4}) is used. The gap equation of this form can be found in many references. When this convention is used, the subscripts $ \uparrow,\downarrow $ are no longer necessary to label the eigenstates, so we can simply write $ (u_{j\uparrow},v_{j\downarrow})=(u_j,v_j) $ without confusion. \\
	\indent Finally, we give a remark on the difference of convention between the present work and our previous work \cite{PhysRevLett.110.131601}. For the one-dimensional spin-1/2 $s$-wave system, both (ii) and (iii) are applicable, so we have two choices. Let us fix to use $ (u_{R\uparrow},v_{R\downarrow}) $. Then, the remaining choice is $ (u_{L\uparrow},v_{L\downarrow}) $ or $ (u_{R\downarrow},v_{R\uparrow}) $. If we choose (iii) and discard $ (u_\downarrow,v_\uparrow) $, both right- and left-Fermi-point eigenstates must be included even after the use of the Andreev approximation. In order to switch from one convention to another, we should interpret in the following way:
	\begin{align}
		&\text{fill/unfill the state  $ (u_{L\uparrow},v_{L\downarrow})=(u_j,v_j) $ with $ \epsilon=\epsilon_j $. } \nonumber \\
		\leftrightarrow\quad& \text{unfill/fill the state  $ (u_{R\downarrow},v_{R\uparrow})=(v_j^*,u_j^*) $ with $ \epsilon=-\epsilon_j $}.
	\end{align}
	While the former convention has been used in Ref.~\cite{PhysRevLett.110.131601}, we use the latter convention in the present work. By this correspondence, the self-consistent condition (\ref{eq:statscasym}) becomes equivalent to Ref.~\cite{PhysRevLett.110.131601}. [Rewrite Eq.~(\ref{eq:statscasym}) by $ (\nu_{2j-1},\,\nu_{2j},\,\theta_{2j}) \rightarrow (\nu_{jR},\,1-\nu_{jL},\,\theta_j)  $.]

\section{Supplemental calculation for the gap equation}\label{sec:suppgapsupp}
In this section we show a few detailed calculations related to the derivation of the gap equation and the self-consistent condition.
\subsection{Gap equation for the DT class}\label{sec:suppgap}
	\indent In this subsection, starting from Eqs. (\ref{eq:gapint2}) and (\ref{eq:gapint22}), we derive the reduced gap equation Eqs. (\ref{eq:reducedgap2}) and (\ref{eq:reducedgap}) for the DT class. \\
	\indent The scattering states [Eq. (\ref{eq:scatstates})] $ F(x,t,\zeta),\ \zeta\in\mathbb{R} $ for the DT class  can be written as
	\begin{align}
		&F(x,t,\zeta)=[F_0+H_0Z^\dagger]\mathrm{e}^{\mathrm{i}[k(\zeta)x-\epsilon(\zeta)t]}, \label{eq:dtscat1} \\
		&F_0=\begin{pmatrix}I_d \\ \zeta^{-1} \Delta_-^\dagger \end{pmatrix},\ Z=(z_1,\dots,z_n),\ z_j=\frac{2\mathrm{i}}{m}\frac{s_je_j\hat{p}_j}{\zeta s_j-1} \label{eq:dtscat2}
	\end{align}
	We first calculate $ FF^\dagger=F_0F_0^\dagger+H_0Z^\dagger F_0^\dagger+F_0Z H_0^\dagger+H_0Z^\dagger Z H_0^\dagger $. 
	Using the relations
	\begin{align}
		&[G_0(x,t)]_{ij}=\frac{[W_0(x,t)^\dagger W_0(x,t)]_{ij}}{\mathrm{i}[k(s_i^*)-k(s_j)]}=-\frac{2\mathrm{i}}{m}\frac{s_i^*s_je_i^*e_j\hat{p}_i^\dagger \hat{p}_j }{s_i^*-s_j}, \\
		&[Z^\dagger Z]_{ij}=\frac{2\mathrm{i}}{m}[G_0]_{ij}\left( \frac{s_i^*}{1-\zeta s_i^*}-\frac{s_j}{1-\zeta s_j} \right), \label{eq:ZZtoG0} \\
		&H_0C+W_0+H_0G_0=0,\quad C:=(LL^\dagger)^{-1}, \label{eq:Hzeroeq}
	\end{align}
	we obtain
	\begin{align}
		H_0Z^\dagger Z H_0^\dagger&=-\frac{2\mathrm{i}}{m}\sum_i\left[ h_{0i}\frac{s_i^*}{1-\zeta s_i^*}\left( C_{ij}h_{0j}^\dagger+w_{0i}^\dagger \right) \right]\nonumber \\
		&\qquad+\frac{2\mathrm{i}}{m}\sum_j\left[ \left( h_{0i}C_{ij}+w_{0j} \right)\frac{s_j}{1-\zeta s_j}h_{0j}^\dagger \right],
	\end{align}
	where $ h_{0i} $ is the $ i $-th vector of  $ H_0 $. Using $ K=H_0W_0^\dagger=W_0H_0^\dagger=\sum_j w_{0j}h_{0j}^\dagger $, we also find
	\begin{align}
		F_0Z H_0^\dagger&=-\frac{2\mathrm{i}}{m}\sum_j\frac{s_j}{1-\zeta s_j}w_{0j}h_{0j}^\dagger+\frac{\mathrm{i}}{m\zeta}(\sigma_3-1)K.
	\end{align}
	Using the above relations, and recalling $ A=-\sigma_3 $ and $ \mathrm{i}B=m\left( \begin{smallmatrix} 0 & \Delta_- \\ \Delta_-^\dagger & 0 \end{smallmatrix} \right) $, the integrand of Eq.~(\ref{eq:gapint22}) is given by
	\begin{align}
		&mFF^\dagger+\frac{B+[K,A]}{\mathrm{i}\zeta} \nonumber \\
		&=2\mathrm{i}\sum_{ij}\left[ \frac{s_j}{1-\zeta s_j}-\frac{s_i^*}{1-\zeta s_i^*} \right]h_{0i}C_{ij}h_{0j}^\dagger+m\begin{pmatrix}I_d & \\ & \zeta^{-2}I_d \end{pmatrix}.
	\end{align}
	If we write $ \Xi=\tilde{\Xi}+\Xi_0 $ with $ \Xi_0 $ being a commutative part s.t. $ [\sigma_3,\Xi_0]=0 $, we obtain
	\begin{align}
		\tilde{\Xi}=HDH^\dagger+\left[\int_{-\infty}^0\!\!-\!\int_0^\infty\right]\!\frac{\mathrm{i}\mathrm{d}\zeta}{2\pi}\left( \frac{s_j}{1-\zeta s_j}-\frac{s_i^*}{1-\zeta s_i^*} \right)h_{0i}C_{ij}h_{0j}^\dagger.
	\end{align}
	Writing $ s_i=r_i\mathrm{e}^{\mathrm{i}\theta_i},\ 0<\theta_i<\pi $, we obtain 
	\begin{align}
		&\left[\int_{-\infty}^0\!\!-\!\int_0^\infty\right]\!\frac{\mathrm{i}\mathrm{d}\zeta}{2\pi}\!\left( \frac{s_j}{1-\zeta s_j}-\frac{s_i^*}{1-\zeta s_i^*}\right) \nonumber \\
		&=\frac{\pi-\theta_i-\theta_j}{\pi}+\frac{\mathrm{i}}{\pi}\log\frac{r_j}{r_i}=1-\Theta_{ii}-\Theta_{jj}^*, 
	\end{align}
	where $ \Theta_{ij} $ is defined by Eq.~(\ref{eq:NandTheta}).
	Thus,  $ \tilde{\Xi} $ is written as
	\begin{align}
		\tilde{\Xi}&=HDH^\dagger+H_0(C-\Theta C-C\Theta^\dagger)H_0^\dagger \nonumber \\
		&=H(2\mathcal{N}-L^\dagger\Theta(L^\dagger)^{-1}-L^{-1}\Theta^\dagger L)H^\dagger=2HXH^\dagger,
	\end{align}
	where $ X $ is defined by Eq.~(\ref{eq:reducedgap}). 
	Thus we obtain the reduced gap equation $[\sigma_3, \tilde{\Xi}+\tau\tilde{\Xi}\tau]=0$. 
	In many cases, the self-consistent solution can be obtained by only solving the $(x,t)$-independent equation $ X=0 \ \leftrightarrow $ 
	\begin{align}
		\mathcal{N}=\frac{1}{2}\left( L^\dagger\Theta(L^\dagger)^{-1}+L^{-1}\Theta^\dagger L \right). \label{eq:suppredgp1}
	\end{align}
	The full gap equation $[\sigma_3, \tilde{\Xi}+\tau\tilde{\Xi}\tau]=0$ becomes necessary when we derive the filling condition in the antisymmetric systems ($\Delta=-\Delta^T,\ \tau=\sigma_2$). (See next subsection.)
%
\subsection{Self-consistent condition for stationary class}\label{subsec:statgap}
	Here we solve the gap equation for the stationary class and derive the self-consistent conditions, Eqs. (\ref{eq:statscsym}) and (\ref{eq:statscasym}).\\ 
	\indent First, we consider the non-symmetric case ($\tau=0$). The equation reduces to $ HXH^\dagger=0 $, equivalent to $ X=0 $ since $ H $ has rank $ n $. We thus obtain 
	\begin{align}
		\nu_j=\frac{\theta_j}{\pi}. \label{eq:appfintndcase}
	\end{align}
	\indent Next, we consider (anti)symmetric $ \Delta=\pm\Delta^T $. We write $ \tau=\sigma_1 $ for symmetric and $ \sigma_2 $ for antisymmetric cases, respectively. 
	As mentioned in Sec.~\ref{subsec:statclass}, if the asymptotic form of $ \Delta $ is $ \Delta(x\rightarrow-\infty)=m\Delta_- $, where $ \Delta_- $ is unitary, we can realize symmetric and antisymmetric $ \Delta $ by the following conditions: 
	\begin{align}
		&\hat{p}_j=\Delta_-\hat{p}_j^* \quad(\Delta=\Delta^T), \\
		&\hat{p}_{2j}=\Delta_-\hat{p}_{2j-1}^*,\ s_{2j}=s_{2j-1},\ L_{2j1-1,2j-1}=L_{2j,2j} \quad(\Delta=-\Delta^T). \label{eq:suppstatant1}
	\end{align}
	These relations are rigorously derived by the IST. Thus, $ W $ satisfies
	\begin{align}
		\tau W^*=W\mathcal{E}(t),
	\end{align}
	where  $ \mathcal{E}(t) $ is unitary and  $ x $-independent, but possibly depends on $ t $, whose explicit form is given by
	\begin{align}
		\mathcal{E}(t)=\begin{cases} \operatorname{diag}(s_1^{-1}\mathrm{e}^{-2\mathrm{i}\epsilon(s_1)t},\dots,s_n^{-1}\mathrm{e}^{-2\mathrm{i}\epsilon(s_n)t}) &(\Delta=\Delta^T), \\ (s_1^{-1}\mathrm{e}^{-2\mathrm{i}\epsilon(s_1)t}\sigma_y)\oplus\dotsb\oplus (s_n^{-1}\mathrm{e}^{-2\mathrm{i}\epsilon(s_n)t}\sigma_y) & (\Delta=-\Delta^T). \end{cases}
	\end{align}
	The same relation for $ H $
	\begin{align}
		\tau H^*=H\mathcal{E}(t), \label{eq:statprove}
	\end{align}
	follows from its definition $ H=-W(I+\int W^\dagger W)^{-1} $. 
	Thus, $ K^*=\tau K \tau $ holds, which implies $ \Delta=\pm\Delta^T $. Note that these relations do not hold for time-dependent solutions even for a one-soliton case. \\
	\indent Let us derive the self-consistent condition using (\ref{eq:statprove}). Since $ X=X^* $ is diagonal, we find
	\begin{align}
		HXH^\dagger+\tau H^* X^* H^T\tau=H(X+\mathcal{E}X\mathcal{E}^\dagger)H^\dagger.
	\end{align}
	Thus the gap equation reduces to $ X+\mathcal{E}X\mathcal{E}^\dagger=0 $. For the symmetric case,  $ \mathcal{E} $ is diagonal and therefore the equation reduces to $ X=0 $, the same as the fine-tuned non-symmetric case (\ref{eq:appfintndcase}). For the antisymmetric case, on the other hand, $ \mathcal{E} $ is $ 2\times 2 $-block-diagonal, hence the equation reduces to $ X+(I_{n/2}\otimes \sigma_y)X(I_{n/2}\otimes \sigma_y)=0 $, yielding the condition
	\begin{align}
		\nu_{2j-1}+\nu_{2j}=\frac{\theta_{2j-1}+\theta_{2j}}{\pi}=\frac{2\theta_{2j}}{\pi}.
	\end{align}
	where $ \theta_{2j-1}=\theta_{2j} $ due to  $ s_{2j-1}=s_{2j} $ [Eq.~(\ref{eq:suppstatant1})]. Thus we obtain Eq. (\ref{eq:statscasym}).


\section{Soliton classifications and parameters in animations}\label{sec:parameters}
	In this section, we summarize the  classification of soliton solutions based on possible filling values in Table~\ref{ta:nonstatscsol}, and provide a list of parameters used in animations.

\subsection{Classification of $n$-soliton solutions for $ n \le 3 $}\label{subsec:clsfysltn}
	Here we classify $n$-soliton solutions for $ n \le 3 $. We focus on the filling conditions realizable by superpositions of occupations states in  one-flavor systems, which are summarized in Table~\ref{ta:nonstatscsol}. (If we consider an $N$-flavor system with sufficiently large $N$, the constraint for filling rates becomes weaker and the solution space becomes larger.) \\
	\indent First, we slightly change the definition of $ e_j $. Instead of $ e_j=\mathrm{e}^{-\mathrm{i}[k(s_j)x+\epsilon(s_j)t]} $, we use 
	\begin{align}
		e_j=\sqrt{\frac{m\sin\theta_j}{r_j}}\mathrm{e}^{-\mathrm{i}[k(s_j)(x-x_j)+\epsilon(s_j)t-\varphi_j]}, \label{eq:renewej}
	\end{align}
	where $ r_j,\theta_j $ is defined by $ s_j=r_j\mathrm{e}^{\mathrm{i}\theta_j} $, and $ x_j,\varphi_j $ are real parameters. Such trivial change is always possible by replacement $ L \rightarrow L'=\operatorname{diag}(\lambda_1,\dots,\lambda_n) L $ with $ \lambda_j^{-1}:=\sqrt{\frac{m\sin\theta_j}{r_j}}\mathrm{e}^{\mathrm{i}[k(s_j)x_j+\varphi_j]} $. The reduced gap equation $ X=0 $ remains unchanged because $ L^{-1}\Theta L=(L')^{-1}\Theta L' $. Roughly speaking, $ x_j $ represents the position of the soliton at $ t=0 $ up to an additive constant arising from soliton interaction, and $ \varphi_j $ shifts the ``phase'' of breathing motion of the breather.\\ 
	\indent $ e_j $ can be rewritten as
	\begin{align}
		e_j=\sqrt{\frac{m\sin\theta_j}{r_j}}\mathrm{e}^{\kappa_j[x-V_jt-x_j]-\mathrm{i}\tilde{\kappa}_j[t-V_j(x-x_j)]+\mathrm{i}\varphi_j},\\
		V_j:=\frac{1-r_j^2}{1+r_j^2},\ \kappa_j=\frac{m\sin\theta_j}{\sqrt{1-V_j^2}},\ \tilde{\kappa}_j=\frac{m\cos\theta_j}{\sqrt{1-V_j^2}}.
	\end{align}
	We also re-define $ U_0,\ W_0 $, and $ G_0 $ using new $ e_j $ [Eq. (\ref{eq:renewej})]:
	\begin{align}
		U_0&=(e_1\hat{p}_1,\dots,e_n\hat{p}_n),\\
		W_0&=\begin{pmatrix} U_0 \\ \Delta_-^\dagger U_0\mathcal{S} \end{pmatrix},\\ 
		G_0&=\int_{-\infty}^x\mathrm{d}x W_0^\dagger W_0.
	\end{align}
	By definition of $ W $, column vectors of $ U_0 $ must be linearly independent of each other as a function of $ x $. (Hence, maximum degeneracy of $ s_j $'s is $ d $.)  The matrix element of $ G_0 $ is calculated as
	\begin{align}
		[G_0]_{ij}=-\frac{2\mathrm{i}}{m}\frac{s_i^*s_j\hat{p}_i^\dagger\hat{p}_je_i^*e_j}{s_i^*-s_j}.
	\end{align}
	\indent If we re-define $ e_j $ as above, using the adjustment of $ x_j $'s and $ \varphi_j $'s, the definition of $ L $ also gets an arbitrariness such that $ L \rightarrow \Lambda L $, where $ \Lambda $ is a diagonal invertible matrix. It is desirable to construct a ``useful standard form'' of $ L $ for the classification of soliton solutions. Let the polar decomposition of $ L $ be
	\begin{align}
		L=RQ,\quad \text{$R$: positive-definite hermitian, $Q$: unitary}. \label{eq:Ldecomp}
	\end{align}
	Then, $ R $ changes as $ R^2 \rightarrow \Lambda R^2 \Lambda^\dagger $ under the transformation $ L\rightarrow \Lambda L $. One example of the convenient standard form of $ R $ for $ n=2,3 $ is as follows.  \\ 
	\indent For the two-soliton solution ($ n=2 $), by the transformation $ R^2 \rightarrow \Lambda R^2 \Lambda^\dagger $ with an appropriate choice of $ \Lambda $, we can always set 
	\begin{align}
		R=\begin{pmatrix} \cosh \alpha & \sinh\alpha \\ \sinh\alpha & \cosh\alpha \end{pmatrix},\quad \alpha\ge0. \label{eq:Rforn2}
	\end{align}
	This is just an element of $ SO(1,1) $. For the three-soliton solution ($ n=3 $), we can always choose 
	\begin{align}
		R=\mathcal{U}\begin{pmatrix} \mathrm{e}^\alpha && \\ &\mathrm{e}^{-\alpha}& \\ && 1 \end{pmatrix}\mathcal{U}^\dagger,\quad \mathcal{U} \in U(3), \label{eq:Rforn3}
	\end{align}
	where the form of $ \mathcal{U} $ is given as follows. The Euler-angle-like parametrization for $ SU(3) $ is given by Bronzan \cite{PhysRevD.38.1994}. In the current problem, the four of five phases in Bronzan's parametrization can be eliminated by using $ \Lambda $ and by multiplying an overall factor to eigenvectors of $ R $. Therefore, without loss of generality, $ \mathcal{U} $ reduces to
\begin{widetext}
	\begin{align}
		\mathcal{U}=\begin{pmatrix} -\sin\eta_2\sin\eta_3\mathrm{e}^{-\mathrm{i}\phi_5}+\cos\eta_1\cos\eta_2\cos\eta_3 & -\sin\eta_2\cos\eta_3\mathrm{e}^{-\mathrm{i}\phi_5}-\cos\eta_1\cos\eta_2\sin\eta_3 & \sin\eta_1\cos\eta_2 \\ \cos\eta_2\sin\eta_3\mathrm{e}^{-\mathrm{i}\phi_5}+\cos\eta_1\sin\eta_2\cos\eta_3 & \cos\eta_2\cos\eta_3\mathrm{e}^{-\mathrm{i}\phi_5}-\cos\eta_1\sin\eta_2\sin\eta_3 & \sin\eta_1\sin\eta_2 \\ -\sin\eta_1\cos\eta_3 & \sin\eta_1\sin\eta_3 & \cos\eta_1 \end{pmatrix},\ 0\le\phi_5,\eta_1,\eta_2,\eta_3\le\frac{\pi}{2}. \label{eq:Rforn32}
	\end{align}
\end{widetext}
Using the polar decomposition (\ref{eq:Ldecomp}) and the standard forms of  $ R $'s for $ n=2 $ [Eq. (\ref{eq:Rforn2})] and $ n=3 $ [Eqs. (\ref{eq:Rforn3}) and (\ref{eq:Rforn32})], the reduced gap equation [Eq. (\ref{eq:suppredgp1})] can be rewritten as
	\begin{align}
		\mathcal{N}=Q^\dagger \left[ \tfrac{1}{2}(R\Theta R^{-1}+R^{-1}\Theta^\dagger R) \right] Q \label{eq:reducedgapap2}
	\end{align}
	Taking the trace of both sides, we immediately obtain the necessary condition
	\begin{align}
		\sum_j \nu_j = \sum_j \frac{\theta_j}{\pi}. \label{eq:tracegapeq}
	\end{align}
	The gap function $ \Delta $ and the array of the bound states $ H=(h_1,\dots,h_n) $ are given by
	\begin{align}
		\Delta&=(mI_d-2\mathrm{i}U_0[R^{-2}+G_0]^{-1}\mathcal{S}^*U_0^\dagger)\Delta_-, \label{eq:suppdelta} \\
		H&=-W_0[R^{-2}+G_0]^{-1}R^{-1}Q. \label{eq:suppbdst}
	\end{align}
	If one only wants to plot $ \Delta $, the determination of $ Q $ is unnecessary. \\
\indent Henceforth we solve the diagonalization problem of the reduced gap equation (\ref{eq:reducedgapap2}) and provide explicit forms of solutions for $ n\le 3$.
\subsubsection{ $ n=1 $ }
\indent  For $ n=1 $, there is no diagonalization problem. The gap function $ \Delta $ and the bound state $ H=h_1 $ is given by
	\begin{align}
		\Delta&=m\left[I_d-\hat{p}_1\hat{p}_1^\dagger+\hat{p}_1\hat{p}_1^\dagger \mathrm{e}^{-\mathrm{i}\theta_1}(\cos\theta_1-\mathrm{i}\sin\theta_1 \tanh y) \right]\Delta_-, \label{eq:supp1soldelta}\\
		h_1&=\frac{-w_1}{1+G_{11}}=\frac{-\sqrt{\kappa_1}\mathrm{e}^{-\mathrm{i}y'}}{2\cosh y}\begin{pmatrix} \sqrt{1+V_1}\hat{p}_1 \\ \sqrt{1-V_1}\mathrm{e}^{\mathrm{i}\theta_1}\Delta_-^\dagger\hat{p}_1 \end{pmatrix} \label{eq:supp1solbdst}
	\end{align}
	with $ y=\kappa_1(x-V_1t-x_1),\ y'=\tilde{\kappa}_1[t-V_1(x-x_1)]-\varphi_1 $. If we set $ x_1=\varphi_1=0 $, it reduces to Eqs. (\ref{eq:onesoldelta}) and (\ref{eq:onesolbound}). 
\subsubsection{ $ n=2 $ }
\indent Next we consider $ n=2 $. As given in Table~\ref{ta:nonstatscsol}, if we restrict our consideration to one-flavor systems, one of the filling values must be ``Dirac'', i.e., $ \nu_1=0 $ or 1. Since $ \nu_i \in [0,1] $ and $ \nu_1+\nu_2=\frac{\theta_1+\theta_2}{\pi} $ [Eq. (\ref{eq:tracegapeq})], the possible filling eigenvalues of $ \mathcal{N} $ in Eq. (\ref{eq:reducedgapap2}) are
\begin{align}
	&\nu_1=0,\ \nu_2=\frac{\theta_1+\theta_2}{\pi} && \text{if \ $ 0<\theta_1+\theta_2\le\pi $,} \\
	&\nu_1=1,\ \nu_2=\frac{\theta_1+\theta_2-\pi}{\pi} && \text{if \ $ \pi<\theta_1+\theta_2<2\pi $.}
\end{align}
From this filling constraint, the parameter $ \alpha $ in $ R $ [Eq. (\ref{eq:Rforn2})] is determined to be
	\begin{align}
		\sinh 2\alpha=\begin{cases} \sqrt{\frac{4\theta_1\theta_2}{(\theta_1-\theta_2)^2+(\log r_2/r_1)^2}} & \text{if \ $ 0<\theta_1+\theta_2\le\pi $,} \\[1ex] \sqrt{\frac{4(\pi-\theta_1)(\pi-\theta_2)}{(\theta_1-\theta_2)^2+(\log r_2/r_1)^2}} &\text{if \ $ \pi<\theta_1+\theta_2<2\pi $.} \end{cases} \label{eq:supp2solalpha}
	\end{align}
	The diagonalizing unitary matrix $ Q $ is given by
	\begin{align}
		&Q=\begin{pmatrix} q_- & q_+ \\ q_+ & q_- \end{pmatrix}, \label{eq:supp2solQ}\\
		&q_\pm=\nonumber \\
		&\begin{cases} \frac{\theta_1+\theta_2\pm(\theta_1-\theta_2)\cosh2\alpha\pm\mathrm{i}\log(r_2/r_1)\sinh2\alpha}{2(\theta_1+\theta_2)} & \text{if \ $ 0<\theta_1+\theta_2\le\pi $,} \\[1ex] \frac{\theta_1+\theta_2-2\pi\pm(\theta_1-\theta_2)\cosh2\alpha\pm\mathrm{i}\log(r_2/r_1)\sinh2\alpha}{2(\theta_1+\theta_2-2\pi)} & \text{if \ $ \pi<\theta_1+\theta_2<2\pi $.} \end{cases} \label{eq:supp2solQ2}
	\end{align}
	Using these matrices,  the filling matrix in the reduced gap equation [Eq.~(\ref{eq:reducedgapap2})] is diagonalized as
	\begin{align}
		\mathcal{N}=\begin{cases} \operatorname{diag}(0,\frac{\theta_1+\theta_2}{\pi}) & \text{if \ $ 0<\theta_1+\theta_2\le\pi $.} \\ \operatorname{diag}(1,\frac{\theta_1+\theta_2-\pi}{\pi}) & \text{if \ $ \pi<\theta_1+\theta_2<2\pi $.} \end{cases} \label{eq:supp2solN}
	\end{align}
\subsubsection{$ n=3 $, case (A):  ``Majorana triplet''}
Table~\ref{ta:nonstatscsol} shows that the three-soliton solution has three families of filling conditions. First we consider (A), which we call the Majorana triplet states. Since $ \mathcal{N}=\frac{1}{2}I_3 $, the reduced gap equation [Eq. (\ref{eq:reducedgapap2})] is rewritten as
	\begin{align}
		R^2\Theta R^{-2}=I-\Theta^\dagger,
	\end{align}
	which means that $ \Theta $ and $ I-\Theta^\dagger $ are similar, and the transformation matrix between them is the positive-definite hermitian matrix $ R^2 $. Under such condition, we soon conclude $ \Theta = I-\Theta^\dagger $, hence $ \theta_1=\theta_2=\theta_3=\frac{\pi}{2} $, and $ L=R=Q=I_3 $. Since $ L $ is diagonal, solitons in the Majorana-triplet states shows no breathing behavior during their collision.

\subsubsection{$n=3$, cases (B) and (C): preliminary remark}
	Next we consider $ n=3 $ (B) and (C) in Table~\ref{ta:nonstatscsol}, where filling values are more flexibly chosen than (A) but still there is a constraint such that $ \nu_1=\nu_2 $ for (B) and $ \nu_1=1-\nu_2 $ for (C). 
	We must find $ R $ and $ \Theta $ such that the eigenvalues of $ \frac{1}{2}(R\Theta R^{-1}+R^{-1}\Theta^\dagger R) $ satisfy these constraints. \\
	\indent  We concentrate on the following three analytically tractable cases:
	\begin{enumerate}[(i)]\setlength{\itemsep}{0ex}
		\item  $ (\phi_5,\eta_3,\eta_1,\eta_2)=(0,\frac{\pi}{4},0,0) $ 
		\item $ (\phi_5,\eta_3,\eta_2)=(0,\frac{\pi}{4},\frac{\pi}{4}) $ and $ r_1=r_2=r_3 $ 
		\item  $ (\phi_5,\eta_3,\eta_1)=(0,\frac{\pi}{4},\frac{\pi}{2}) $ and $ r_1=r_2=r_3 $ 
	\end{enumerate}
	Though these (i)-(iii) may not exhaust all possibilities, the resulting solutions sufficiently tell us the diversity of three-soliton solutions. \\
	\indent While the case (i) can treat three solitons having mutually different velocities, the cases (ii) and (iii) can only support breather-type solutions due to the constraint $ r_1=r_2=r_3 $.  \\
	\indent As given in Table~\ref{ta:nonstatscsol}, the third filling eigenvalue is given by $ \nu_3=\nu_1|\beta_0|^2+(1-\nu_1)|\beta_3|^2 $ with $ |\beta_0|^2+|\beta_3|^2=1 $, and hence it must be bound by the following inequality:
	\begin{align}
		\min(\nu_1,1-\nu_1)\le \nu_3 \le \max(\nu_1,1-\nu_1). \label{eq:suppnu3ineq}
	\end{align}
	In particular, $ \nu_3=1/2 $ is always included. (However, it may not be Majorana unless $ \nu_1=0 \text{ or } 1$.) \\
	\indent Henceforth, we derive three solutions (B)-(i), (C)-(i), and (C)-(i)' from (i). We also derive (B)-(ii), (C)-(ii), and (C)-(ii)' from (ii). For (iii), we only treat the case (B).

\subsubsection{$n=3$, case (i)}
	We first consider the case (i) $ (\phi_5,\eta_3,\eta_1,\eta_2)=(0,\frac{\pi}{4},0,0) $ without imposing constraint between filling eigenvalues. The cases (B)-(i), (C)-(i), and (C)-(i)' are later given as a special reduction. The matrix $ R $ [Eq.~(\ref{eq:Rforn3})] reduces to 
	\begin{align}
		R=\begin{pmatrix} \cosh\alpha & \sinh\alpha & \\ \sinh\alpha & \cosh\alpha & \\ && 1 \end{pmatrix}. \label{eq:supp3solb1Rc} 
	\end{align}
	Let us parametrize $ \alpha $ by an auxiliary parameter  $ \tilde{\theta} $ such that $ \tilde{\theta}\ne(\theta_1+\theta_2)/2 $ by
	\begin{align}
		\sinh 2\alpha=\sqrt{\frac{4(\theta_1-\tilde{\theta})(\theta_2-\tilde{\theta})}{(\theta_1-\theta_2)^2+(\log r_2/r_1)^2}}. \label{eq:supp3solb1Qalphac}
	\end{align}
	Then, the diagonalizing unitary matrix $ Q $ is given by
	\begin{align}
		Q=\begin{pmatrix} q_- & q_+ & \\ q_+ & q_- & \\ && 1 \end{pmatrix}
	\end{align}
	with
	\begin{align}
		q_\pm=\frac{\theta_1+\theta_2-2\tilde{\theta}\pm(\theta_1-\theta_2)\cosh2\alpha\pm\mathrm{i}\log(r_2/r_1)\sinh2\alpha}{2(\theta_1+\theta_2-2\tilde{\theta})}. \label{eq:supp3solb1Q2c}
	\end{align}
	Note that parameters of the two-soliton solution are revisited by setting $ \tilde{\theta}=0 $ and $ \pi $. 
	Using these $ R $ and $ Q $, the filling matrix $ \mathcal{N} $ is diagonalized as
	\begin{align}
		&\mathcal{N}=\operatorname{diag}(\nu_1,\nu_3,\nu_2), \\
		&\nu_1=\frac{\tilde{\theta}}{\pi},\ \nu_2=\frac{\theta_3}{\pi},\ \nu_3=\frac{\theta_1+\theta_2-\tilde{\theta}}{\pi}. \label{eq:supp3solb1N2c}
	\end{align}
	Below, we provide a reduction to (B) and (C). 

\subsubsection{$n=3$, case (B)-(i)}
	This case is obtained by the following reduction in Eqs. (\ref{eq:supp3solb1Rc})-(\ref{eq:supp3solb1N2c}):
	\begin{align}
		\tilde{\theta}=\theta_3. \label{eq:supp3solb1redce}
	\end{align}
	Then filling eigenvalues of $ \mathcal{N} $ are given by
	\begin{align}
		\nu_1=\nu_2=\frac{\theta_3}{\pi},\ \nu_3=\frac{\theta_1+\theta_2-\theta_3}{\pi}.
	\end{align} 
	Since $ \nu_i $'s satisfy (\ref{eq:suppnu3ineq}) and $ \alpha $ must be real,  $ \theta_i $'s must be chosen to satisfy 
	\begin{gather}
		\min(2\theta_3,\pi)\le\theta_1+\theta_2\le \max(2\theta_3,\pi), \label{eq:supp3solb1ineq1} \\
		(\theta_1-\theta_3)(\theta_2-\theta_3)\ge0. \label{eq:supp3solb1ineq2}
	\end{gather}
	\indent If we focus on the case $ \nu_3=1/2 $ and if $ \theta_3<\frac{\pi}{2} $, the candidate of $ \theta_1,\theta_2 $ based on the above inequalities is
	\begin{align}
		\theta_1=\theta_3+\delta,\quad \theta_2=\frac{\pi}{2}-\delta,\quad 0 \le \delta \le \frac{\pi}{2}-\theta_3.
	\end{align}
	The parameter of Animation 13 corresponds to $ \theta_3=\frac{\pi}{4},\ \delta=\frac{\pi}{6} $. 

\subsubsection{$n=3$, case (C)-(i)}
	 This case is obtained by the following reduction in Eqs. (\ref{eq:supp3solb1Rc})-(\ref{eq:supp3solb1N2c}): 
	\begin{align}
		\tilde{\theta}=\pi-\theta_3. \label{eq:supp3solc1redce}
	\end{align}
	Filling values are reduced to
	\begin{align}
		\nu_2=1-\nu_1=\frac{\theta_3}{\pi},\ \nu_3=\frac{\theta_1+\theta_2+\theta_3-\pi}{\pi}.
	\end{align}
	Since $ \nu_i $'s satisfy (\ref{eq:suppnu3ineq}) and $ \alpha $ is real,  $ \theta_i $'s  satisfy the constraints
	\begin{gather}
		\min(2(\pi-\theta_3),\pi)\le\theta_1+\theta_2\le \max(2(\pi-\theta_3),\pi), \label{eq:supp3solc1ineq1} \\
		(\theta_1+\theta_3-\pi)(\theta_2+\theta_3-\pi)\ge0. \label{eq:supp3solc1ineq2}
	\end{gather}
%
\subsubsection{$n=3$, case (C)-(i)' }
	The case $ n=3 $ (C) allows another choice; while we have set $ \nu_1=1-\nu_2 $ in (C)-(i), $ \nu_1=1-\nu_3 $ is also possible. In this case, we impose
	\begin{align}
		\theta_2=\pi-\theta_1, \label{eq:supp3solcd1redce}
	\end{align}
	in Eqs. (\ref{eq:supp3solb1Rc})-(\ref{eq:supp3solb1N2c}). Instead, $ \tilde{\theta} $ remains an adjustable parameter such that $ 0\le \tilde{\theta}\le\pi $ and $ \tilde{\theta}\ne\frac{\pi}{2} $. 
	The filling values become
	 \begin{align}
	 	\nu_1=\frac{\tilde{\theta}}{\pi},\ \nu_2=\frac{\theta_3}{\pi},\ \nu_3=\frac{\pi-\tilde{\theta}}{\pi}.
	 \end{align} 
	For this choice,  $ \nu_3 $ in Eq.~(\ref{eq:suppnu3ineq}) must be replaced by $ \nu_2 $. Therefore, the inequalities to be satisfied are
	\begin{gather}
		\min(\tilde{\theta},\pi-\tilde{\theta})\le\theta_3\le \max(\tilde{\theta},\pi-\tilde{\theta}), \label{eq:supp3solcd1ineq1} \\
		(\theta_1-\tilde{\theta})(\pi-\theta_1-\tilde{\theta})\ge0. \label{eq:supp3solcd1ineq2}
	\end{gather}
	The case (C)-(i)' includes several basic solutions such as independent three solitons realized by setting $ \tilde{\theta}=\theta_1 $ and ``one DHN breather + one soliton'' realized by setting $ \tilde{\theta}=0, r_1=r_2 $.

\begin{widetext}
\subsubsection{$n=3$, case (ii)}
	Next we consider $ (\phi_5,\eta_3,\eta_2)=(0,\frac{\pi}{4},\frac{\pi}{4}) $ and $ r_1=r_2=r_3 $. As with the case (i), we first consider the general diagonalization of $ \mathcal{N} $  and later consider the reduction to (B) and (C). For this case the matrix $ R $ is
\begin{align}
	R=\begin{pmatrix} \left( \cosh\frac{\alpha}{2}-\cos\eta_1\sinh\frac{\alpha}{2} \right)^2 & -\sin^2\eta_1\sinh^2\frac{\alpha}{2} & \frac{1}{\sqrt{2}}\left( \sin\eta_1\sinh\alpha-\sin2\eta_1\sinh^2\frac{\alpha}{2} \right) \\ -\sin^2\eta_1\sinh^2\frac{\alpha}{2} & \left( \cosh\frac{\alpha}{2}+\cos\eta_1\sinh\frac{\alpha}{2} \right)^2 & \frac{-1}{\sqrt{2}}\left( \sin\eta_1\sinh\alpha+\sin2\eta_1\sinh^2\frac{\alpha}{2} \right) \\ \frac{1}{\sqrt{2}}\left( \sin\eta_1\sinh\alpha-\sin2\eta_1\sinh^2\frac{\alpha}{2} \right) & \frac{-1}{\sqrt{2}}\left( \sin\eta_1\sinh\alpha+\sin2\eta_1\sinh^2\frac{\alpha}{2} \right) & \cos^2\eta_1+\cosh\alpha\sin^2\eta_1 \end{pmatrix} \label{eq:suppn3b2R}
\end{align}
Note that $ R^n $ can be easily obtained by replacement $ \alpha\rightarrow n\alpha $ in Eq. (\ref{eq:suppn3b2R}), since it is originally defined by Eq. (\ref{eq:Rforn3}). This rule is convenient when we need $ R^{-1} $ and $ R^{-2} $ to plot $ \Delta $ and $ H $ [Eqs. (\ref{eq:suppdelta}) and (\ref{eq:suppbdst})]. \\
\indent For brevity, we write $ M:=\frac{1}{2}(R\Theta R^{-1}+R^{-1}\Theta^\dagger R) $. We also introduce notations
\begin{align}
	\theta_0=\frac{\theta_1+\theta_2+\theta_3}{3},\ \theta_8=\frac{\theta_1+\theta_2-2\theta_3}{6},\ \xi=\frac{\theta_1-\theta_2}{\theta_1+\theta_2-2\theta_3},\ S=\sin\eta_1\sinh\alpha 
\end{align}
Then, we can check that eigenvalues of $ M $ are
\begin{align}
	\nu_1=\frac{\theta_0-\theta_8(2+3S^2)}{\pi},\quad \nu_\pm=\frac{2\theta_0+\theta_8(2+3S^2)\pm3\theta_8\sqrt{S^4+4\xi^2(1+S^2)}}{2\pi}
\end{align}
and corresponding (unnormalized) eigenvectors are
\begin{align}
	q_1&=\frac{1}{\sqrt{2(1+\sin^2\eta_1\sinh^2\alpha)}}\begin{pmatrix} -\sin 2\eta_1\sinh^2\frac{\alpha}{2} \\ -\sin 2\eta_1\sinh^2\frac{\alpha}{2} \\ \sqrt{2}(\cos^2\eta_1+\sin^2\eta_1\cosh\alpha) \end{pmatrix}, \label{eq:solclan3iigenQq1} \\
	 \tilde{q}_\pm&=\begin{pmatrix} M_{12}M_{23}-M_{13}(M_{22}-\nu_\pm) \\ M_{13}M_{21}-(M_{11}-\nu_\pm)M_{23} \\ (M_{11}-\nu_\pm)(M_{22}-\nu_\pm)-M_{12}M_{21} \end{pmatrix}, \label{eq:solclan3iigenQ}
\end{align}
where $ q_1 $ is normalized but $ \tilde{q}_\pm $ are not normalized. If we define $ Q=(q_1,q_-,q_+) $ with $ q_\pm=\tilde{q}_\pm/|\tilde{q}_\pm| $, we have $ \mathcal{N}=\operatorname{diag}(\nu_1,\nu_-,\nu_+) $. \\
\indent Below, we consider reduction to (B) and (C).
\subsubsection{$n=3$, case (B)-(ii)}
The case (B) has a degenerate filling eigenvalue. Let us impose $ \nu_1=\nu_- $. This is realized by $ \xi^2=1+2S^2 $, yielding
\begin{align}
	\sinh\alpha=\sqrt{\frac{\xi^2-1}{2\sin^2\eta_1}},\quad \xi=\frac{\theta_1-\theta_2}{\theta_1+\theta_2-2\theta_3}. \label{eq:suppn3b2R2} 
\end{align}
Then, the filling eigenvalues are reduced to
\begin{align}
	\nu_1=\nu_-=\frac{\theta_1\theta_2-\theta_3^2}{\pi(\theta_1+\theta_2-2\theta_3)},\quad \nu_3:=\nu_+=\frac{\theta_1^2+\theta_2^2-\theta_1\theta_3-\theta_2\theta_3}{\pi(\theta_1+\theta_2-2\theta_3)}. \label{eq:suppn3b2N2}
\end{align}
	The matrix $ Q $ is determined as  $ Q=(q_1,q_2,q_3) $, where normalized eigenvectors are
	\begin{align}
		q_1&=\frac{1}{\sqrt{1+\xi^2}}\begin{pmatrix} -\sin 2\eta_1\sinh^2\frac{\alpha}{2} \\ -\sin 2\eta_1\sinh^2\frac{\alpha}{2} \\ \sqrt{2}(\cos^2\eta_1+\sin^2\eta_1\cosh\alpha) \end{pmatrix}, \label{eq:suppn3b2Q1}\\
		q_2&=\frac{1}{\sqrt{2(1+\xi^2)(1+3\xi^2)}}\begin{pmatrix} -1-\xi^2+2\xi\left[ \cos^2\eta_1+\sin^2\eta_1\cosh\alpha \right] \\ 1+\xi^2+2\xi\left[ \cos^2\eta_1+\sin^2\eta_1\cosh\alpha \right] \\ 2\sqrt{2}\xi\sin2\eta_1\sinh^2\frac{\alpha}{2} \end{pmatrix},\\
		q_3&=\frac{1}{\sqrt{1+3\xi^2}}\begin{pmatrix} \xi+\cos^2\eta_1+\sin^2\eta_1\cosh\alpha \\ -\xi+\cos^2\eta_1+\sin^2\eta_1\cosh\alpha \\ \sqrt{2}\sin2\eta_1\sinh^2\frac{\alpha}{2} \end{pmatrix}. \label{eq:suppn3b2Q3}
	\end{align}
	Using these $ R $ and $ Q $, the filling matrix is diagonalized as $ \mathcal{N}=\operatorname{diag}(\nu_1,\nu_1,\nu_3) $. The inequality $ \xi^2\ge 1 $ must hold in order for $ \alpha $ to be real. Also, $ 0\le \nu_1\le 1 $ and Eq.~(\ref{eq:suppnu3ineq})  hold.  $ \theta_i $'s should be chosen under these constraints. \\
	\indent The other choice,  $ \nu_-=\nu_+ $, reduces to the case (i) since $ R $ becomes diagonal. 
%
\subsubsection{$n=3$, case (C)-(ii)}
	Either $ \nu_1=1-\nu_+ $ or $ \nu_1=1-\nu_- $ is realized by setting
	\begin{align}
		\sinh\alpha=\sqrt{\frac{\gamma^2-\xi^2}{(\xi^2-\gamma)\sin^2\eta_1}},\quad \xi=\frac{\theta_1-\theta_2}{\theta_1+\theta_2-2\theta_3},\quad \gamma=\frac{2\pi-\theta_1-\theta_2-2\theta_3}{\theta_1+\theta_2-2\theta_3} \label{eq:suppn3c2R2}
	\end{align}
	and filling values are reduced to
	\begin{align}
		&\nu_1=\frac{1}{2}+\frac{3\theta_8(\gamma-1)(\gamma+\xi^2)}{2\pi(\gamma-\xi^2)},\ \nu_2=1-\nu_1,\ \nu_3=\frac{3\theta_0}{\pi}-1. \\
		\leftrightarrow\quad& \nu_1=\frac{1}{2}+\frac{(\pi-\theta_1-\theta_2)[\pi(\theta_1+\theta_2-2\theta_3)-2\theta_1\theta_2+2\theta_3^2]}{2\pi[\pi(\theta_1+\theta_2-2\theta_3)-\theta_1^2-\theta_2^2+2\theta_3^2]},\ \nu_2=1-\nu_1,\ \nu_3=\frac{\theta_1+\theta_2+\theta_3}{\pi}-1. \label{eq:suppn3c2N5}
	\end{align}
	Corresponding normalized eigenvectors are given by $ q_1,q_2 $, and $q_3$, where $ q_1 $ is given by Eq.~(\ref{eq:solclan3iigenQ}), and 
	\begin{align}
		q_3&=\frac{1}{\sqrt{N}}\begin{pmatrix} \left[ \xi(\gamma^2-\xi^2)+2(\gamma-\xi^2)(\gamma-\xi\cos2\eta_1)\sinh^2\frac{\alpha}{2} \right]\sin\eta_1 \\[1ex] \left[ \xi(\gamma^2-\xi^2)-2(\gamma-\xi^2)(\gamma+\xi\cos2\eta_1)\sinh^2\frac{\alpha}{2} \right]\sin\eta_1 \\ \sqrt{2}\xi\left[ \gamma^2-\xi^2+4(\gamma-\xi^2)\sin^2\eta_1\sinh^2\frac{\alpha}{2} \right]\cos\eta_1 \end{pmatrix},  \label{eq:suppn3c2Qq3} \\
		N&=2\xi^2(\gamma^2-\xi^2)^2+8(\gamma-\xi^2)\left[\xi^2(\gamma^2-\xi^2)+(\gamma^2+\xi^2)(\gamma-\xi^2)\sinh^2\frac{\alpha}{2}\right]\sin^2\eta_1\sinh^2\frac{\alpha}{2}, \label{eq:suppn3c2Qq32}
	\end{align}
	and $ q_2=q_1\times q_3 $. Using $ Q=(q_1,q_2,q_3) $, the filling matrix is diagonalized as $ \mathcal{N}=\operatorname{diag}(\nu_1,\nu_2,\nu_3) $.  
	In order for $ \alpha $ to be real, $ (\gamma^2-\xi^2)(\xi^2-\gamma)>0 $.  $ 0\le \nu_1 \le 1 $ and Eq.~(\ref{eq:suppnu3ineq}) are also necessary. \\
	\indent If we are interested in the Dirac-Dirac-Majorana filling $ (\nu_1,\nu_2,\nu_3)=(1,0,\frac{1}{2}) $,  $ \theta_1 $ and $ \theta_2 $ are parametrized by  $ \theta_3 $ as
	\begin{align}
		\theta_1,\ \theta_2=\frac{3\pi-2\theta_3}{4}\pm\frac{\pi-2\theta_3}{4}\sqrt{\frac{3(3\pi-2\theta_3)}{\pi+2\theta_3}}. \label{eq:suppn3c2ddm}  
	\end{align}
\subsubsection{$n=3$, case (C)-(ii)'}
	Another choice to realize (C) is $ \nu_++\nu_-=1 $, yielding 
	\begin{align}
		\sinh\alpha=\sqrt{\frac{\theta_{12}}{3\theta_8\sin^2\eta_1}},\quad \theta_{12}:=\pi-2(\theta_0+\theta_8) \label{eq;suppn3cd2alph}
	\end{align}
	then filling eigenvalues become
	\begin{align}
		\nu_1=\frac{3\theta_0}{\pi}-1,\quad \nu_\pm=\frac{1}{2}\pm\frac{\sqrt{\xi^2(36\theta_8^2+12\theta_8\theta_{12})+\theta_{12}^2}}{2\pi}. \label{eq;suppn3cd2N3}
	\end{align}
	There seems to be no simpler expression for eigenvectors $ q_i $'s than the general expression Eq.~(\ref{eq:solclan3iigenQ}). 
	The inequalities to be satisfied are $ \theta_8\theta_{12}>0 $ ($ \leftrightarrow $ realness of $ \alpha $),\  $  \xi^2(36\theta_8^2+12\theta_8\theta_{12})+\theta_{12}^2<\pi^2  $ ( $\leftrightarrow \nu_\pm\in[0,1] $ ),\ and $ \nu_-<\nu_1<\nu_+ $ (the counterpart of Eq.~(\ref{eq:suppnu3ineq}); in the current notation, $ \nu_3 $ in Eq. (\ref{eq:suppnu3ineq}) corresponds to the current $ \nu_1 $). We can check that these inequalities exclude the possibility of $ \nu_1=1/2 $. So, there is no Dirac-Dirac-Majorana filling in (C)-(ii)'.
\subsubsection{$n=3$, case (B)-(iii)}
Finally we consider the case (iii) $ (\phi_5,\eta_3,\eta_1)=(0,\frac{\pi}{4},\frac{\pi}{2}) $ and $ r_1=r_2=r_3 $. Here we only discuss (B). The reduction for (C)-(iii) and (C)-(iii)' can be done in the similar way as before. The matrix $ R $ becomes
	\begin{align}
		R=\begin{pmatrix} \cos^2\eta_2+\sin^2\eta_2\cosh\alpha & \cos\eta_2\sin\eta_2(1-\cosh\alpha) & \sin\eta_2\sinh\alpha \\ \cos\eta_2\sin\eta_2(1-\cosh\alpha) & \cos^2\eta_2\cosh\alpha+\sin^2\eta_2 & -\cos\eta_2\sinh\alpha \\ \sin\eta_2\sinh\alpha & -\cos\eta_2\sinh\alpha & \cosh\alpha \end{pmatrix} \label{eq:suppn3b3R}
	\end{align}
 $ R^n $ can be easily obtained by replacement $ \alpha\rightarrow n\alpha $ in Eq. (\ref{eq:suppn3b3R}), since the original definition of $R$ is Eq. (\ref{eq:Rforn3}). The condition of degenerated eigenvalue for $ \mathcal{N} $ fixes $ \alpha $ as:  
	\begin{align}
		\cosh 2\alpha = \frac{(2c^2+1)\xi^2-2c\xi-1+\sqrt{(1+\xi^2)^2-4c\xi(-1+3\xi^2)+4c^2\xi^2(-1+2\xi^2)}}{2(-1+c\xi)^2},\quad c=\cos2\eta_2,\ \xi=\frac{\theta_1-\theta_2}{\theta_1+\theta_2-2\theta_3}. \label{eq:suppn3b3R2}
	\end{align}
	In order for $ \alpha $ to be real,  $ \cosh 2\alpha \ge 1 \leftrightarrow \xi^2\ge 1 $. The matrix $ Q $ is given by
		\begin{align}
		Q=\begin{pmatrix} \cos(\beta+\eta_2) & -\sin(\beta+\eta_2) & \\ \sin(\beta+\eta_2) & \cos(\beta+\eta_2) & \\ && 1 \end{pmatrix},\quad \tan\beta=\frac{(\xi\cos2\eta_2-1)\cosh2\alpha}{\xi\cosh\alpha\sin2\eta_2}. \label{eq:suppn3b3Q}
	\end{align}
	Using them, the filling matrix is diagonalized as $ \mathcal{N}=\operatorname{diag}(\nu_3,\nu_1,\nu_1) $, where 
	\begin{align}
		\nu_1&=\frac{(\theta_1+\theta_2-(\theta_1-\theta_2)\cos2\eta_2)(1-\cosh2\alpha)+2\theta_3(1+\cosh2\alpha)}{4\pi}, \label{eq:suppn3b3N2} \\
		\nu_3&=\frac{(\theta_1+\theta_2)(1+\cosh2\alpha)+(\theta_1-\theta_2)\cos2\eta_2(1-\cosh2\alpha)-2\theta_3\cosh2\alpha}{2\pi}. \label{eq:suppn3b3N3}
	\end{align}
	The inequalities to be satisfied are $ \xi^2\ge1,\ 0\le \nu_1 \le 1 $, and Eq. (\ref{eq:suppnu3ineq}).
\end{widetext} 
\subsection{Parameters used in animation files}\label{subsec:parameters}
	Here we show soliton parameters used to generate animation files. Several parameters ($ s_j $'s and $ \hat{p}_j $'s) are already described in animations, but full information about other parameters such as $ x_j $'s,  $ \varphi_j $'s, matrices $R$ and $ Q $, are necessary to reproduce the animation. \\
	\indent In all examples shown below, the matrix size of $ \Delta $ is $ d=2 $, the gap is set $ m=1 $ and $ \Delta_-=I_2 $. Therefore the asymptotic form at $ x=-\infty $ is given by $ \Delta(x=-\infty)=I_2 $. 
\subsubsection{One-soliton solution}
Animation 1, 2, and 3 correspond to this category. \\
\indent One-soliton solution has independent parameters $ s_1=r_1\mathrm{e}^{\mathrm{i}\theta_1},\ \hat{p}_1,\ x_1,\, \varphi_1 $. The gap function and bound states are given by Eqs. (\ref{eq:supp1soldelta}) and (\ref{eq:supp1solbdst}).
In all examples, the filling rate is given by $ \nu_1=\theta_1/\pi $. 
\begin{enumerate}[$\hspace{1ex}\bullet$]
	\item Animation 1: a multicomponent analog of $ \pi $-phase kink with Majorana. \\
	$ s_1=1.1\,\mathrm{i},\ \hat{p}_1=\begin{pmatrix} 0 \\ 1 \end{pmatrix},\ x_1=0,\ \varphi_1=0 $. 
	\item Animation 2: an example of more general one soliton \\ 
	$ s_1=0.9\,\mathrm{e}^{\mathrm{i}\pi/3},\ \hat{p}_1=\begin{pmatrix} \cos\frac{3\pi}{5} \\ \sin\frac{3\pi}{5} \end{pmatrix},\ x_1=0,\ \varphi_1=0 $. 
	\item Animation 3: $s$-$p$ mixed case, $ \hat{p}_1 $ is not real \\
	$ s_1=1.15\,\mathrm{i},\ \hat{p}_1=\frac{1}{\sqrt{2}}\begin{pmatrix} 1 \\ \mathrm{e}^{-5\mathrm{i}\pi/8} \end{pmatrix},\ x_1=0,\ \varphi_1=0 $. 
\end{enumerate}
\subsubsection{Two-soliton solution}
Animation 4, 5, 6, 7, 8, 9, and 10 correspond to this category. \\
\indent Two-soliton solution has independent parameters $ s_1=r_1\mathrm{e}^{\mathrm{i}\theta_1},\ s_2=r_2\mathrm{e}^{\mathrm{i}\theta_2},\ \hat{p}_1,\ \hat{p}_2,\ x_1,\ x_2,\ \varphi_1,\ \varphi_2 $. \\
\indent The gap function and bound states are given by Eqs.~(\ref{eq:suppdelta}) and (\ref{eq:suppbdst}). The matrix $ R $ and its parameter $ \alpha $ are given by Eqs.~(\ref{eq:Rforn2}) and (\ref{eq:supp2solalpha}). The matrix $ Q $ is given by Eq.~(\ref{eq:supp2solQ}) with (\ref{eq:supp2solQ2}). The filling rates of bound states are given by Eq.~(\ref{eq:supp2solN}). \\
\indent The period of the breather solution such that $ r_1=r_2=1 $ is calculated by the following formula
\begin{align}
	T=\frac{2\pi}{|\tilde{\kappa}_1-\tilde{\kappa}_2|}=\frac{2\pi}{m|\cos\theta_1-\cos\theta_2|}, \label{eq:breatherperiod}
\end{align}
since the breathing motion is originated from the off-diagonal element of the matrix $ L $.
\begin{enumerate}[$\hspace{1ex}\bullet$]
	\item Animation 4: Parallel SU(2)-DHN breather \\
	$ s_1=\mathrm{e}^{\mathrm{i}\pi/3},\ s_2=\mathrm{e}^{2\mathrm{i}\pi/3},\ \hat{p}_1=\hat{p}_2=\begin{pmatrix} 1 \\ 0 \end{pmatrix}$, \\ $x_1=\frac{1}{3}-0.4002,\ x_2=-\frac{1}{3}-0.4002,\ \varphi_1=\varphi_2=0 $. \\[.5ex] ($-0.4002$ is added to locate the breather at $ x=0 $.)
	\item Animation 5: Offset SU(2)-DHN breather \\
	$ s_1=\mathrm{e}^{\mathrm{i}\pi/4},\ s_2=\mathrm{e}^{3\mathrm{i}\pi/4},\ \hat{p}_1=\begin{pmatrix} 1 \\ 0 \end{pmatrix},\ \hat{p}_2=\begin{pmatrix} -\frac{\sqrt{3}}{2} \\ \frac{1}{2} \end{pmatrix}, $ \\  $ x_1=-\frac{1}{2}-0.1662,\ x_2=\frac{1}{2}-0.1662,\ \varphi_1=\varphi_2=0 $.  \\[.5ex] ($-0.1662$ is added to locate the breather at $ x=0 $.)
	\item Animation 6: Solitons more separated in Animation 5 \\
	$x_1=-2-0.1898,\ x_2=2-0.1898$. \\ Other parameters are the same as Animation 5. 
	\item Animation 7: Non-DHN offset breather  \\
	$ s_1=\mathrm{e}^{\mathrm{i}\pi/3},\ s_2=\mathrm{e}^{\mathrm{i}\pi/6},\ \hat{p}_1=\begin{pmatrix}1 \\ 0\end{pmatrix},\ \hat{p}_2=\begin{pmatrix} \cos\frac{\pi}{8} \\ -\sin\frac{\pi}{8} \end{pmatrix} $, \\  $ x_1=-\frac{1}{4},\ x_2=\frac{1}{4},\ \varphi_1=\varphi_2=0 $. 
	\item Animation 8: Parallel two-soliton collision \\
	$ s_1=1.2 \,\mathrm{e}^{\mathrm{i}\pi/4},\ s_2=0.8\,\mathrm{e}^{3\mathrm{i}\pi/4},\ \hat{p}_1=\hat{p}_2=\begin{pmatrix}1 \\ 0\end{pmatrix} $, \\  $ x_1=x_2=\varphi_1=\varphi_2=0 $. 
	\item Animation 9: Breathing patterns and relative phases of bound states \\
	9(a): the same as Animation 8. \\
	9(b):  $ \varphi_2=\frac{\pi}{2} $ and others are the same as 9(a). \\
	(Note: $ u_{i,\downarrow}=v_{i,\downarrow}=0 $ because $ \hat{p}_i=(1,0)^T $.)
	\item Animation 10: Offset two-soliton collision \\
	$ s_1=0.9 \,\mathrm{e}^{2\mathrm{i}\pi/3},\ s_2=1.1\,\mathrm{e}^{5\mathrm{i}\pi/6},\ \hat{p}_1=\begin{pmatrix}1 \\ 0\end{pmatrix},\ $ \\ $ \hat{p}_2=\begin{pmatrix} \cos\frac{\pi}{5} \\ \mathrm{e}^{0.855\mathrm{i}\pi}\sin\frac{\pi}{5} \end{pmatrix},\  x_1=x_2=\varphi_1=\varphi_2=0 $. \\[.5ex]
	The phase factor in $ \hat{p}_2 $ is attached in order to realize the situation such that the initial state becomes a pure $p$-wave.
	
\end{enumerate}

\subsubsection{Three-soliton solution (A): ``Majorana triplet''}
Animation 11(a), 11(b), 11(c), 12 correspond to this category. \\
\indent The gap function and bound states are given by Eqs.~(\ref{eq:suppdelta}) and (\ref{eq:suppbdst}) with $ R=Q=I_3 $. In this class, $ \theta_1=\theta_2=\theta_3=\frac{\pi}{2} $. $ \varphi_i $'s do not appear in $ \Delta $ and they only change the overall phase of bound states.  Thus, the essential independent parameters are $ r_1,\ r_2,\ r_3,\ \hat{p}_1,\ \hat{p}_2,\ \hat{p}_3,\ x_1,\ x_2,\ x_3 $. 
\begin{enumerate}[$\hspace{1ex}\bullet$]
	\item Animation 11(a): Parallel three-soliton collision \\
	$s_1=1.05\,\mathrm{i},\ s_2=0.9\,\mathrm{i},\ s_3=0.95\,\mathrm{i},\ \hat{p}_1=\hat{p}_2=\hat{p}_3=\begin{pmatrix}1 \\ 0\end{pmatrix}$, \\
	$x_1=x_2=0,\ x_3=-3,\ \varphi_1=\varphi_2=\varphi_3=0$.
	\item Animation 11(b): Parallel and almost simultaneous three-soliton collision \\
	$s_1=1.05\,\mathrm{i},\ s_2=0.9\,\mathrm{i},\ s_3=0.95\,\mathrm{i},\ \hat{p}_1=\hat{p}_2=\hat{p}_3=\begin{pmatrix}1 \\ 0\end{pmatrix}$, \\
	$x_1=-1,\ x_2=-\frac{1}{2},\ x_3=-1,\ \varphi_1=\varphi_2=\varphi_3=0$.
	\item Animation 11(c): Antiparallel case (collisionless) \\
	$s_1=1.05\,\mathrm{i},\ s_2=0.9\,\mathrm{i},\ s_3=0.95\,\mathrm{i},\ \hat{p}_1=\hat{p}_3=\begin{pmatrix}1 \\ 0\end{pmatrix},$ \\
	$ \hat{p}_2=\begin{pmatrix}0 \\ 1\end{pmatrix},\ x_1=-1,\ x_2=0,\ x_3=-1,\ \varphi_1=\varphi_2=\varphi_3=0$.
	\item Animation 12: Offset three-soliton collision \\
	$s_1=1.05\,\mathrm{i},\ s_2=0.9\,\mathrm{i},\ s_3=0.95\,\mathrm{i},\ \hat{p}_1=\begin{pmatrix}\frac{1}{2} \\ \frac{\sqrt{3}}{2}\end{pmatrix},\ \hat{p}_2=\begin{pmatrix}1 \\ 0\end{pmatrix},$ \\
	$ \hat{p}_3=\begin{pmatrix}\frac{1}{\sqrt{2}} \\ \frac{1}{\sqrt{2}}\end{pmatrix},\ x_1=2,\ x_2=0,\ x_3=-3,\ \varphi_1=\varphi_2=\varphi_3=0$.
	
\end{enumerate}
\subsubsection{Three-soliton solution (B)-(i)}
Animation 13 corresponds to this category. \\
\indent This class has independent parameters $ s_1=r_1\mathrm{e}^{\mathrm{i}\theta_1},\ s_2=r_2\mathrm{e}^{\mathrm{i}\theta_2},\ s_3=r_3\mathrm{e}^{\mathrm{i}\theta_3},\ \hat{p}_1,\ \hat{p}_2,\ \hat{p}_3,\ x_1,\ x_2,\ x_3,\ \varphi_1,\ \varphi_2,\ \varphi_3 $.  The gap function and bound states are given by Eqs.~(\ref{eq:suppdelta}) and (\ref{eq:suppbdst}), where the matrix  $ R $ and $ Q $ are given by Eqs.~(\ref{eq:supp3solb1Rc})-(\ref{eq:supp3solb1Q2c}) with the reduction $ \tilde{\theta}=\theta_3 $ [Eq.~(\ref{eq:supp3solb1redce})].  $ \theta_i $'s must satisfy the inequalities (\ref{eq:supp3solb1ineq1}) and (\ref{eq:supp3solb1ineq2}). \\
\indent In Animation 13, we focus on the case $ r_1=r_2\ne r_3 $ to realize the collision between the soliton and the breather, though this class can generally set all $ r_i $'s different values. 
\begin{enumerate}[$\hspace{1ex}\bullet$]
	\item Animation 13: Offset collision of one soliton and non-DHN breather (general filling values) \\
	$s_1=\mathrm{e}^{5\mathrm{i}\pi/12},\ s_2=\mathrm{e}^{\mathrm{i}\pi/3},\ s_3=1.2\,\mathrm{e}^{\mathrm{i}\pi/4},\ \hat{p}_1=\begin{pmatrix} 1 \\ 0 \end{pmatrix}, $ \\
	$\hat{p}_2=\begin{pmatrix} \cos\frac{\pi}{12} \\ \sin\frac{\pi}{12} \end{pmatrix},\ \hat{p}_3=\begin{pmatrix} \cos\frac{\pi}{12} \\ -\sin\frac{\pi}{12} \end{pmatrix} $, \\
	$x_1=x_2=x_3=\varphi_1=\varphi_2=\varphi_3=0$.
\end{enumerate}

\subsubsection{Three-soliton solution (B)-(ii)}
Animation 14 corresponds to this category. This class has independent parameters $ s_1=r_1\mathrm{e}^{\mathrm{i}\theta_1},\ s_2=r_1\mathrm{e}^{\mathrm{i}\theta_2},\ s_3=r_1\mathrm{e}^{\mathrm{i}\theta_3},\ \hat{p}_1,\ \hat{p}_2,\ \hat{p}_3,\ x_1,\ x_2,\ x_3,\ \varphi_1,\ \varphi_2,\ \varphi_3 $, and $ \eta_1 $. The gap function and bound states are given by Eqs.~(\ref{eq:suppdelta}) and (\ref{eq:suppbdst}), where the matrix $ R $ is given by Eq.~(\ref{eq:suppn3b2R}) and the parameter $ \alpha $ by Eq.~(\ref{eq:suppn3b2R2}), and the matrix $ Q $ is given by $ Q=(q_1,q_2,q_3) $ with Eqs. (\ref{eq:suppn3b2Q1})-(\ref{eq:suppn3b2Q3}). The filling rates of bound states are given by  $ (\nu_1,\nu_1,\nu_3) $ with  Eq.~(\ref{eq:suppn3b2N2}).  $ \theta_i $'s should be chosen to satisfy the inequalities $ \xi^2\ge1,\ 0\le \nu_1 \le 1 $, and Eq.~(\ref{eq:suppnu3ineq}).  
\begin{enumerate}[$\hspace{1ex}\bullet$]
	\item Animation 14: Offset three-soliton breather \\
	$s_1=\mathrm{e}^{\mathrm{i}\pi/3},\ s_2=\mathrm{e}^{\mathrm{i}\pi/6},\ s_3=\mathrm{e}^{2\mathrm{i}\pi/9},\ \hat{p}_1=\begin{pmatrix} 1 \\ 0 \end{pmatrix}, $ \\
	$\hat{p}_2=\begin{pmatrix} \cos\frac{\pi}{10} \\ \sin\frac{\pi}{10} \end{pmatrix},\ \hat{p}_3=\begin{pmatrix} \cos\frac{\pi}{15} \\ \sin\frac{\pi}{15} \end{pmatrix} $, \\
	$x_1=x_2=-2,\ x_3=-\frac{5}{2},\ \varphi_1=\varphi_2=\varphi_3=0$,\ $\eta_1=\frac{\pi}{4}$. 
\end{enumerate}
\subsubsection{Three-soliton solution (B)-(iii)}
No animation has been prepared for this category. The independent parameters of this class are $ s_1=r_1\mathrm{e}^{\mathrm{i}\theta_1},\ s_2=r_1\mathrm{e}^{\mathrm{i}\theta_2},\ s_3=r_1\mathrm{e}^{\mathrm{i}\theta_3},\ \hat{p}_1,\ \hat{p}_2,\ \hat{p}_3,\ x_1,\ x_2,\ x_3,\ \varphi_1,\ \varphi_2,\ \varphi_3 $, and $ \eta_2 $. The gap function and bound states are given by Eqs.~(\ref{eq:suppdelta}) and (\ref{eq:suppbdst}). The matrix $ R $ is given by Eq. (\ref{eq:suppn3b3R}) with (\ref{eq:suppn3b3R2}). $ Q $ is given by Eq.~(\ref{eq:suppn3b3Q}). The filling rates are given by Eqs.~(\ref{eq:suppn3b3N2}) and (\ref{eq:suppn3b3N3}). $ \theta_i $'s should satisfy inequalities $ \xi^2\ge1,\ 0\le \nu_1 \le 1 $, and Eq.~(\ref{eq:suppnu3ineq}).  

\subsubsection{Three-soliton solution (C)-(i)}
\indent This class has independent parameters $ s_1=r_1\mathrm{e}^{\mathrm{i}\theta_1},\ s_2=r_2\mathrm{e}^{\mathrm{i}\theta_2},\ s_3=r_3\mathrm{e}^{\mathrm{i}\theta_3},\ \hat{p}_1,\ \hat{p}_2,\ \hat{p}_3,\ x_1,\ x_2,\ x_3,\ \varphi_1,\ \varphi_2,\ \varphi_3 $. The gap function and bound states are given by Eqs.~(\ref{eq:suppdelta}) and (\ref{eq:suppbdst}), where the matrices  $ R $ and $ Q $ are given by Eqs.~(\ref{eq:supp3solb1Rc})-(\ref{eq:supp3solb1Q2c}) with the reduction $ \tilde{\theta}=\pi-\theta_3 $ [Eq.~(\ref{eq:supp3solc1redce})].  $ \theta_i $'s  must satisfy the inequalities (\ref{eq:supp3solc1ineq1}) and (\ref{eq:supp3solc1ineq2}). \\
	\indent No animation has been prepared for this class.

\subsubsection{Three-soliton solution (C)-(i)'}
Animation 15 corresponds to this category. \indent This class has independent parameters $ s_1=r_1\mathrm{e}^{\mathrm{i}\theta_1},\ s_2=r_2\mathrm{e}^{\mathrm{i}(\pi-\theta_1)},\ s_3=r_3\mathrm{e}^{\mathrm{i}\theta_3},\ \hat{p}_1,\ \hat{p}_2,\ \hat{p}_3,\ x_1,\ x_2,\ x_3,\ \varphi_1,\ \varphi_2,\ \varphi_3 $, and $ \tilde{\theta} $. Note that the arguments of $ s_1 $ and $ s_2 $ are not independent. The gap function and bound states are given by Eqs.~(\ref{eq:suppdelta}) and (\ref{eq:suppbdst}), where the matrices  $ R $ and $ Q $ are given by Eqs.~(\ref{eq:supp3solb1Rc})-(\ref{eq:supp3solb1Q2c}) with the reduction $ \theta_2=\pi-\theta_1 $ [Eq.~(\ref{eq:supp3solcd1redce})].  $ \theta_1,\ \theta_3 $, and $ \tilde{\theta} $  must satisfy the inequalities (\ref{eq:supp3solcd1ineq1}) and (\ref{eq:supp3solcd1ineq2}), and $ 0\le \tilde{\theta}<\frac{\pi}{2},\ \frac{\pi}{2}<\tilde{\theta}\le\pi $.
\begin{enumerate}[$\hspace{1ex}\bullet$]
	\item Animation 15: Offset collision of one soliton and parallel SU(2)-DHN breather \\
	$s_1=\mathrm{e}^{\mathrm{i}\pi/3},\ s_2=\mathrm{e}^{2\mathrm{i}\pi/3},\ s_3=1.2\,\mathrm{i},\ \hat{p}_1=\hat{p}_2=\begin{pmatrix} 1 \\ 0 \end{pmatrix}, $ \\
	$\hat{p}_3=\begin{pmatrix} \frac{\sqrt{3}}{2} \\ \frac{1}{2} \end{pmatrix}$, $x_1=x_2=x_3=\varphi_1=\varphi_2=\varphi_3=0$, and $ \tilde{\theta}=0 $. 
\end{enumerate}
\subsubsection{Three-soliton solution (C)-(ii)}
Animation 16 corresponds to this category. This class has independent parameters $ s_1=r_1\mathrm{e}^{\mathrm{i}\theta_1},\ s_2=r_1\mathrm{e}^{\mathrm{i}\theta_2},\ s_3=r_1\mathrm{e}^{\mathrm{i}\theta_3},\ \hat{p}_1,\ \hat{p}_2,\ \hat{p}_3,\ x_1,\ x_2,\ x_3,\ \varphi_1,\ \varphi_2,\ \varphi_3 $, and $ \eta_1 $. The gap function and bound states are given by Eqs.~(\ref{eq:suppdelta}) and (\ref{eq:suppbdst}), where the matrix $ R $ is given by Eq.~(\ref{eq:suppn3b2R}) and the parameter $ \alpha $ by Eq.~(\ref{eq:suppn3c2R2}), and the matrix $ Q $ is given by $ Q=(q_1,q_2,q_3) $ with Eqs. (\ref{eq:solclan3iigenQq1}), (\ref{eq:suppn3c2Qq3}), (\ref{eq:suppn3c2Qq32}), and $ q_2=q_1\times q_3 $. The filling rates of bound states are given by Eq.~(\ref{eq:suppn3c2N5}).  $ \theta_i $'s should be chosen to satisfy the inequalities $ (\gamma^2-\xi^2)(\xi^2-\gamma)>0 $,\  $ 0\le \nu_1 \le 1 $, and Eq.~(\ref{eq:suppnu3ineq}).  If one wants to realize  $ (\nu_1,\nu_2,\nu_3)=(1,0,\frac{1}{2}) $ or $ (0,1,\frac{1}{2}) $,  $ \theta_1 $ and $ \theta_2 $ should be given by Eq. (\ref{eq:suppn3c2ddm}).
\begin{enumerate}[$\hspace{1ex}\bullet$]
	\item Animation 16: Offset three-soliton breather (Dirac-Dirac-Majorana filling) \\
	$s_1=\mathrm{e}^{\mathrm{i}\theta_1},\ \theta_1=\frac{35-\sqrt{105}}{84}\pi=0.295\pi, $ \\ $  s_2=\mathrm{e}^{\mathrm{i}\theta_2},\ \theta_2=\frac{35+\sqrt{105}}{84}\pi=0.539\pi,\ s_3=\mathrm{e}^{2\mathrm{i}\pi/3}, $ \\ $  \hat{p}_1=\begin{pmatrix} \frac{1}{\sqrt{2}} \\ \frac{1}{\sqrt{2}} \end{pmatrix},\ \hat{p}_2=\begin{pmatrix} \cos\frac{\pi}{12} \\ \sin\frac{\pi}{12} \end{pmatrix},\ \hat{p}_3=\begin{pmatrix} 1 \\ 0 \end{pmatrix} $, \\
	$x_1=-\frac{1}{2},\ x_2=x_3=\varphi_1=\varphi_2=\varphi_3=0$,\ $\eta_1=\frac{\pi}{4}$. 
\end{enumerate}
\subsubsection{Three-soliton solution (C)-(ii)'}
No animation has been prepared for this class. Here we show a summary of parameters. The independent parameters are  $ s_1=r_1\mathrm{e}^{\mathrm{i}\theta_1},\ s_2=r_1\mathrm{e}^{\mathrm{i}\theta_2},\ s_3=r_1\mathrm{e}^{\mathrm{i}\theta_3},\ \hat{p}_1,\ \hat{p}_2,\ \hat{p}_3,\ x_1,\ x_2,\ x_3,\ \varphi_1,\ \varphi_2,\ \varphi_3 $, and $ \eta_1 $. The gap function and bound states are given by Eqs.~(\ref{eq:suppdelta}) and (\ref{eq:suppbdst}), where the matrix $ R $ is given by Eq.~(\ref{eq:suppn3b2R}) and the parameter $ \alpha $ by Eq.~(\ref{eq;suppn3cd2alph}), and the matrix $ Q $ is given by $ Q=(q_1,q_-,q_+) $ with  Eqs. (\ref{eq:solclan3iigenQq1}) and (\ref{eq:solclan3iigenQ}) and $ q_\pm:=\tilde{q}_\pm/|\tilde{q}_\pm| $. The filling rates of bound states are given by Eq.~(\ref{eq;suppn3cd2N3}).  $ \theta_i $'s should be chosen to satisfy the inequalities  $ \theta_8\theta_{12}>0,\ \xi^2(36\theta_8^2+12\theta_8\theta_{12})+\theta_{12}^2<\pi^2 $, and $ \nu_-<\nu_1<\nu_+ $.

\section{Summary}\label{sec:summary}
	In this paper, we have reported the exact time-dependent and self-consistent multi-soliton solutions for the BdG equations satisfying the $SU(d)$-symmetric gap equation. The main result and examples of solutions are summarized in Sec.~\ref{sec:mainr}, and Secs. \ref{sec:supprefless}-\ref{sec:parameters} provide technical details which support the main result. \\
	\indent The new soliton solutions are constructed using the ansatz originating from the GLM equation in the IST. The result can be regard as a matrix generalization of the solution recently derived by Dunne and Thies. We have also considered the superposition of occupation states which can realize partial filling rates even in one-flavor systems, including Dirac and Majorana fermions as a special case. The examples in the $ d=2 $ system, which models the mixture of singlet $s$-wave and triplet $p$-wave superfluids, are presented using animations, and various collision phenomena and breathers with nontrivial spin dynamics emerging due to multicomponent characters are elucidated. \\
	\indent Possible future works and perspectives are summarized in Subsec.~\ref{subsec:perspective}. 
%
	Investgation of correspondence of soliton solutions between the BdG systems and higher-dimensional integrable systems \cite{Tsuchidapc,Anker529} will be also an interesting and important future task.

\begin{acknowledgments}
The author is grateful to M. Nitta for many suggestions at the starting stage of this research. \\

\end{acknowledgments}

\appendix

\makeatletter
\renewcommand{\theequation}{\Alph{section}\arabic{equation}}
\@addtoreset{equation}{section}
\makeatother

\section{Visualization by spherical harmonic functions}\label{subsec:spplt}
	\indent Here, we explain the spherical harmonic plot \cite{Kawaguchi:2012ii} for a given order parameter $ (\Delta_{0,0},\Delta_{1,1},\Delta_{1,0},\Delta_{1,-1}) $. Let the polar coordinate be $ (r,\theta,\varphi) $. Let us define
	\begin{align}
		&Y(\theta,\varphi)=\Delta_{0,0}Y_{0,0}(\theta,\varphi)+\sum_{m=-1}^1\Delta_{1,m}Y_{1,m}(\theta,\varphi), \\
		&Y_{0,0}=\frac{1}{\sqrt{4\pi}},\ Y_{1,\pm1}=\mp\mathrm{e}^{\pm\mathrm{i}\varphi}\sqrt{\frac{3}{8\pi}}\sin\theta,\ Y_{1,0}=\sqrt{\frac{3}{4\pi}}\cos\theta.
	\end{align}
	Then, the radius and the color at $ (\theta,\varphi) $ is given by
	\begin{align}
		r(\theta,\varphi)&=|Y(\theta,\varphi)|^2,\\
		\text{color}(\theta,\varphi)&=\arg Y(\theta,\varphi).
	\end{align}
	What color is specified for each arg is arbitrary. In this work I used the ``Hue'' function in Mathematica. \\
	\indent The singlet order parameter $(\Delta_{0,0},\Delta_{1,1},\Delta_{1,0},\Delta_{1,-1})=(1,0,0,0) $ is isotropic hence represented by a sphere. $ (\Delta_{0,0},\Delta_{1,1},\Delta_{1,0},\Delta_{1,-1})=(0,1,0,0) $ gives the doughnut picture representing the ``$\beta$ phase'' \cite{Barton:1974ae,PhysRevB.34.131,VollhardtWolfle}, which is an analog of the ferromagnetic phase of the spin-1 BEC \cite{Ho:1998zz,JPSJ.67.1822}. $ (\Delta_{0,0},\Delta_{1,1},\Delta_{1,0},\Delta_{1,-1})=(0,0,1,0) $ gives the figure-eight picture corresponds to the ``polar phase''. (This name is shared for both helium 3 and spin-1 BEC.) $(\Delta_{0,0},\Delta_{1,1},\Delta_{1,0},\Delta_{1,-1})=(0,\frac{1}{\sqrt{2}},0,\frac{1}{\sqrt{2}}) $ also represents the polar phase at a different angle.  \\
	\indent In addition to the spherical harmonic plots, we also plot the spin $ (S_x,S_y,S_z) $, which is defined by
	\begin{align}
		S_i=\sum_{m,n=-1,0,1}\Delta_{1,m}^*  [F_i]_{mn} \Delta_{1,n},
	\end{align}
	where $ F_x,\ F_y, $ and $ F_z $ are $ 3\times 3 $ spin-1 matrices. 

\begin{thebibliography}{101}%
\makeatletter
\providecommand \@ifxundefined [1]{%
 \@ifx{#1\undefined}
}%
\providecommand \@ifnum [1]{%
 \ifnum #1\expandafter \@firstoftwo
 \else \expandafter \@secondoftwo
 \fi
}%
\providecommand \@ifx [1]{%
 \ifx #1\expandafter \@firstoftwo
 \else \expandafter \@secondoftwo
 \fi
}%
\providecommand \natexlab [1]{#1}%
\providecommand \enquote  [1]{``#1''}%
\providecommand \bibnamefont  [1]{#1}%
\providecommand \bibfnamefont [1]{#1}%
\providecommand \citenamefont [1]{#1}%
\providecommand \href@noop [0]{\@secondoftwo}%
\providecommand \href [0]{\begingroup \@sanitize@url \@href}%
\providecommand \@href[1]{\@@startlink{#1}\@@href}%
\providecommand \@@href[1]{\endgroup#1\@@endlink}%
\providecommand \@sanitize@url [0]{\catcode `\\12\catcode `\$12\catcode
  `\&12\catcode `\#12\catcode `\^12\catcode `\_12\catcode `\%12\relax}%
\providecommand \@@startlink[1]{}%
\providecommand \@@endlink[0]{}%
\providecommand \url  [0]{\begingroup\@sanitize@url \@url }%
\providecommand \@url [1]{\endgroup\@href {#1}{\urlprefix }}%
\providecommand \urlprefix  [0]{URL }%
\providecommand \Eprint [0]{\href }%
\providecommand \doibase [0]{http://dx.doi.org/}%
\providecommand \selectlanguage [0]{\@gobble}%
\providecommand \bibinfo  [0]{\@secondoftwo}%
\providecommand \bibfield  [0]{\@secondoftwo}%
\providecommand \translation [1]{[#1]}%
\providecommand \BibitemOpen [0]{}%
\providecommand \bibitemStop [0]{}%
\providecommand \bibitemNoStop [0]{.\EOS\space}%
\providecommand \EOS [0]{\spacefactor3000\relax}%
\providecommand \BibitemShut  [1]{\csname bibitem#1\endcsname}%
\let\auto@bib@innerbib\@empty
\bibitem [{\citenamefont {Bogoliubov}(1958)}]{Bogoliubov1958}%
  \BibitemOpen
  \bibfield  {author} {\bibinfo {author} {\bibfnamefont {N.~N.}\ \bibnamefont
  {Bogoliubov}},\ }\href@noop {} {\bibfield  {journal} {\bibinfo  {journal}
  {Sov. Phys. JETP}\ }\textbf {\bibinfo {volume} {7}},\ \bibinfo {pages} {41}
  (\bibinfo {year} {1958})}\BibitemShut {NoStop}%
\bibitem [{\citenamefont {de~Gennes}(1999)}]{DeGennes:1999}%
  \BibitemOpen
  \bibfield  {author} {\bibinfo {author} {\bibfnamefont {P.~G.}\ \bibnamefont
  {de~Gennes}},\ }\href@noop {} {\emph {\bibinfo {title} {Superconductivity of
  metals and alloys}}}\ (\bibinfo  {publisher} {Westview Press},\ \bibinfo
  {address} {Boulder},\ \bibinfo {year} {1999})\BibitemShut {NoStop}%
\bibitem [{\citenamefont {Su}\ \emph {et~al.}(1979)\citenamefont {Su},
  \citenamefont {Schrieffer},\ and\ \citenamefont {Heeger}}]{Su:1979ua}%
  \BibitemOpen
  \bibfield  {author} {\bibinfo {author} {\bibfnamefont {W.}~\bibnamefont
  {Su}}, \bibinfo {author} {\bibfnamefont {J.}~\bibnamefont {Schrieffer}}, \
  and\ \bibinfo {author} {\bibfnamefont {A.}~\bibnamefont {Heeger}},\ }\href
  {\doibase 10.1103/PhysRevLett.42.1698} {\bibfield  {journal} {\bibinfo
  {journal} {Phys.Rev.Lett.}\ }\textbf {\bibinfo {volume} {42}},\ \bibinfo
  {pages} {1698} (\bibinfo {year} {1979})}\BibitemShut {NoStop}%
\bibitem [{\citenamefont {Takayama}\ \emph {et~al.}(1980)\citenamefont
  {Takayama}, \citenamefont {Lin-Liu},\ and\ \citenamefont
  {Maki}}]{Takayama:1980zz}%
  \BibitemOpen
  \bibfield  {author} {\bibinfo {author} {\bibfnamefont {H.}~\bibnamefont
  {Takayama}}, \bibinfo {author} {\bibfnamefont {Y.}~\bibnamefont {Lin-Liu}}, \
  and\ \bibinfo {author} {\bibfnamefont {K.}~\bibnamefont {Maki}},\ }\href
  {\doibase 10.1103/PhysRevB.21.2388} {\bibfield  {journal} {\bibinfo
  {journal} {Phys. Rev. B}\ }\textbf {\bibinfo {volume} {21}},\ \bibinfo
  {pages} {2388} (\bibinfo {year} {1980})}\BibitemShut {NoStop}%
\bibitem [{\citenamefont {Nambu}\ and\ \citenamefont
  {Jona-Lasinio}(1961)}]{Nambu:1961tp}%
  \BibitemOpen
  \bibfield  {author} {\bibinfo {author} {\bibfnamefont {Y.}~\bibnamefont
  {Nambu}}\ and\ \bibinfo {author} {\bibfnamefont {G.}~\bibnamefont
  {Jona-Lasinio}},\ }\href {\doibase 10.1103/PhysRev.122.345} {\bibfield
  {journal} {\bibinfo  {journal} {Phys.Rev.}\ }\textbf {\bibinfo {volume}
  {122}},\ \bibinfo {pages} {345} (\bibinfo {year} {1961})}\BibitemShut
  {NoStop}%
\bibitem [{\citenamefont {Gross}\ and\ \citenamefont
  {Neveu}(1974)}]{Gross:1974jv}%
  \BibitemOpen
  \bibfield  {author} {\bibinfo {author} {\bibfnamefont {D.~J.}\ \bibnamefont
  {Gross}}\ and\ \bibinfo {author} {\bibfnamefont {A.}~\bibnamefont {Neveu}},\
  }\href {\doibase 10.1103/PhysRevD.10.3235} {\bibfield  {journal} {\bibinfo
  {journal} {Phys. Rev D.}\ }\textbf {\bibinfo {volume} {10}},\ \bibinfo
  {pages} {3235} (\bibinfo {year} {1974})}\BibitemShut {NoStop}%
\bibitem [{\citenamefont {Dashen}\ \emph {et~al.}(1975)\citenamefont {Dashen},
  \citenamefont {Hasslacher},\ and\ \citenamefont {Neveu}}]{Dashen:1975xh}%
  \BibitemOpen
  \bibfield  {author} {\bibinfo {author} {\bibfnamefont {R.~F.}\ \bibnamefont
  {Dashen}}, \bibinfo {author} {\bibfnamefont {B.}~\bibnamefont {Hasslacher}},
  \ and\ \bibinfo {author} {\bibfnamefont {A.}~\bibnamefont {Neveu}},\ }\href
  {\doibase 10.1103/PhysRevD.12.2443} {\bibfield  {journal} {\bibinfo
  {journal} {Phys. Rev. D}\ }\textbf {\bibinfo {volume} {12}},\ \bibinfo
  {pages} {2443} (\bibinfo {year} {1975})}\BibitemShut {NoStop}%
\bibitem [{\citenamefont {Hatsuda}\ and\ \citenamefont
  {Kunihiro}(1994)}]{Hatsuda:1994pi}%
  \BibitemOpen
  \bibfield  {author} {\bibinfo {author} {\bibfnamefont {T.}~\bibnamefont
  {Hatsuda}}\ and\ \bibinfo {author} {\bibfnamefont {T.}~\bibnamefont
  {Kunihiro}},\ }\href {\doibase 10.1016/0370-1573(94)90022-1} {\bibfield
  {journal} {\bibinfo  {journal} {Phys. Rept.}\ }\textbf {\bibinfo {volume}
  {247}},\ \bibinfo {pages} {221} (\bibinfo {year} {1994})}\BibitemShut
  {NoStop}%
\bibitem [{\citenamefont {Heeger}\ \emph {et~al.}(1988)\citenamefont {Heeger},
  \citenamefont {Kivelson}, \citenamefont {Schrieffer},\ and\ \citenamefont
  {Su}}]{RevModPhys.60.781}%
  \BibitemOpen
  \bibfield  {author} {\bibinfo {author} {\bibfnamefont {A.~J.}\ \bibnamefont
  {Heeger}}, \bibinfo {author} {\bibfnamefont {S.}~\bibnamefont {Kivelson}},
  \bibinfo {author} {\bibfnamefont {J.~R.}\ \bibnamefont {Schrieffer}}, \ and\
  \bibinfo {author} {\bibfnamefont {W.~P.}\ \bibnamefont {Su}},\ }\href
  {\doibase 10.1103/RevModPhys.60.781} {\bibfield  {journal} {\bibinfo
  {journal} {Rev. Mod. Phys.}\ }\textbf {\bibinfo {volume} {60}},\ \bibinfo
  {pages} {781} (\bibinfo {year} {1988})}\BibitemShut {NoStop}%
\bibitem [{\citenamefont {Shei}(1976)}]{Shei:1976mn}%
  \BibitemOpen
  \bibfield  {author} {\bibinfo {author} {\bibfnamefont {S.-S.}\ \bibnamefont
  {Shei}},\ }\href {\doibase 10.1103/PhysRevD.14.535} {\bibfield  {journal}
  {\bibinfo  {journal} {Phys. Rev. D}\ }\textbf {\bibinfo {volume} {14}},\
  \bibinfo {pages} {535} (\bibinfo {year} {1976})}\BibitemShut {NoStop}%
\bibitem [{\citenamefont {Brazovskii}\ \emph {et~al.}(1980)\citenamefont
  {Brazovskii}, \citenamefont {Gordynin},\ and\ \citenamefont
  {Kirova}}]{BrazovskiiGordyuninKirova}%
  \BibitemOpen
  \bibfield  {author} {\bibinfo {author} {\bibfnamefont {S.~A.}\ \bibnamefont
  {Brazovskii}}, \bibinfo {author} {\bibfnamefont {S.~A.}\ \bibnamefont
  {Gordynin}}, \ and\ \bibinfo {author} {\bibfnamefont {N.~N.}\ \bibnamefont
  {Kirova}},\ }\href@noop {} {\bibfield  {journal} {\bibinfo  {journal} {JETP
  Lett.}\ }\textbf {\bibinfo {volume} {31}},\ \bibinfo {pages} {456} (\bibinfo
  {year} {1980})}\BibitemShut {NoStop}%
\bibitem [{\citenamefont {Brazovskii}\ and\ \citenamefont
  {Kirova}(1981)}]{BrazovskiiKirova}%
  \BibitemOpen
  \bibfield  {author} {\bibinfo {author} {\bibfnamefont {S.~A.}\ \bibnamefont
  {Brazovskii}}\ and\ \bibinfo {author} {\bibfnamefont {N.~N.}\ \bibnamefont
  {Kirova}},\ }\href@noop {} {\bibfield  {journal} {\bibinfo  {journal} {JETP
  Lett.}\ }\textbf {\bibinfo {volume} {33}},\ \bibinfo {pages} {4} (\bibinfo
  {year} {1981})}\BibitemShut {NoStop}%
\bibitem [{\citenamefont {Horovitz}(1981)}]{Horovitz:1981}%
  \BibitemOpen
  \bibfield  {author} {\bibinfo {author} {\bibfnamefont {B.}~\bibnamefont
  {Horovitz}},\ }\href {\doibase 10.1103/PhysRevLett.46.742} {\bibfield
  {journal} {\bibinfo  {journal} {Phys. Rev. Lett.}\ }\textbf {\bibinfo
  {volume} {46}},\ \bibinfo {pages} {742} (\bibinfo {year} {1981})}\BibitemShut
  {NoStop}%
\bibitem [{\citenamefont {Mertsching}\ and\ \citenamefont
  {Fischbeck}(1981)}]{Mertsching:1981}%
  \BibitemOpen
  \bibfield  {author} {\bibinfo {author} {\bibfnamefont {J.}~\bibnamefont
  {Mertsching}}\ and\ \bibinfo {author} {\bibfnamefont {H.~J.}\ \bibnamefont
  {Fischbeck}},\ }\href {\doibase 10.1002/pssb.2221030242} {\bibfield
  {journal} {\bibinfo  {journal} {Phys. Status Solidi}\ }\textbf {\bibinfo
  {volume} {103}},\ \bibinfo {pages} {783} (\bibinfo {year}
  {1981})}\BibitemShut {NoStop}%
\bibitem [{\citenamefont {Campbell}\ and\ \citenamefont
  {Bishop}(1981)}]{Campbell:1981}%
  \BibitemOpen
  \bibfield  {author} {\bibinfo {author} {\bibfnamefont {D.}~\bibnamefont
  {Campbell}}\ and\ \bibinfo {author} {\bibfnamefont {A.}~\bibnamefont
  {Bishop}},\ }\href {\doibase 10.1103/PhysRevB.24.4859} {\bibfield  {journal}
  {\bibinfo  {journal} {Phys. Rev. B}\ }\textbf {\bibinfo {volume} {24}},\
  \bibinfo {pages} {4859} (\bibinfo {year} {1981})}\BibitemShut {NoStop}%
\bibitem [{\citenamefont {Campbell}\ and\ \citenamefont
  {Bishop}(1982)}]{Campbell:1981dc}%
  \BibitemOpen
  \bibfield  {author} {\bibinfo {author} {\bibfnamefont {D.}~\bibnamefont
  {Campbell}}\ and\ \bibinfo {author} {\bibfnamefont {A.}~\bibnamefont
  {Bishop}},\ }\href {\doibase 10.1016/0550-3213(82)90089-X} {\bibfield
  {journal} {\bibinfo  {journal} {Nucl. Phys. B}\ }\textbf {\bibinfo {volume}
  {200}},\ \bibinfo {pages} {297} (\bibinfo {year} {1982})}\BibitemShut
  {NoStop}%
\bibitem [{\citenamefont {Okuno}\ and\ \citenamefont
  {Onodera}(1983)}]{OkunoOnodera}%
  \BibitemOpen
  \bibfield  {author} {\bibinfo {author} {\bibfnamefont {S.}~\bibnamefont
  {Okuno}}\ and\ \bibinfo {author} {\bibfnamefont {Y.}~\bibnamefont
  {Onodera}},\ }\href {\doibase 10.1143/JPSJ.52.3495} {\bibfield  {journal}
  {\bibinfo  {journal} {J. Phys. Soc. Jpn.}\ }\textbf {\bibinfo {volume}
  {52}},\ \bibinfo {pages} {3495} (\bibinfo {year} {1983})}\BibitemShut
  {NoStop}%
\bibitem [{\citenamefont {Onodera}\ and\ \citenamefont
  {Okuno}(1983)}]{OkunoOnodera2}%
  \BibitemOpen
  \bibfield  {author} {\bibinfo {author} {\bibfnamefont {Y.}~\bibnamefont
  {Onodera}}\ and\ \bibinfo {author} {\bibfnamefont {S.}~\bibnamefont
  {Okuno}},\ }\href {\doibase 10.1143/JPSJ.52.2478} {\bibfield  {journal}
  {\bibinfo  {journal} {J. Phys. Soc. Jpn.}\ }\textbf {\bibinfo {volume}
  {52}},\ \bibinfo {pages} {2478} (\bibinfo {year} {1983})}\BibitemShut
  {NoStop}%
\bibitem [{\citenamefont {Brazovskii}\ \emph {et~al.}(1984)\citenamefont
  {Brazovskii}, \citenamefont {Kirova},\ and\ \citenamefont
  {Matveenko}}]{BrazovskiiKirovaMatveenko}%
  \BibitemOpen
  \bibfield  {author} {\bibinfo {author} {\bibfnamefont {S.~A.}\ \bibnamefont
  {Brazovskii}}, \bibinfo {author} {\bibfnamefont {N.~N.}\ \bibnamefont
  {Kirova}}, \ and\ \bibinfo {author} {\bibfnamefont {S.~I.}\ \bibnamefont
  {Matveenko}},\ }\href@noop {} {\bibfield  {journal} {\bibinfo  {journal}
  {Sov. Phys. JETP}\ }\textbf {\bibinfo {volume} {59}},\ \bibinfo {pages} {434}
  (\bibinfo {year} {1984})}\BibitemShut {NoStop}%
\bibitem [{\citenamefont {Machida}\ and\ \citenamefont
  {Nakanishi}(1984)}]{Machida:1984zz}%
  \BibitemOpen
  \bibfield  {author} {\bibinfo {author} {\bibfnamefont {K.}~\bibnamefont
  {Machida}}\ and\ \bibinfo {author} {\bibfnamefont {H.}~\bibnamefont
  {Nakanishi}},\ }\href {\doibase 10.1103/PhysRevB.30.122} {\bibfield
  {journal} {\bibinfo  {journal} {Phys. Rev. B}\ }\textbf {\bibinfo {volume}
  {30}},\ \bibinfo {pages} {122} (\bibinfo {year} {1984})}\BibitemShut
  {NoStop}%
\bibitem [{\citenamefont {Machida}\ and\ \citenamefont
  {Fujita}(1984)}]{MachidaFujita}%
  \BibitemOpen
  \bibfield  {author} {\bibinfo {author} {\bibfnamefont {K.}~\bibnamefont
  {Machida}}\ and\ \bibinfo {author} {\bibfnamefont {M.}~\bibnamefont
  {Fujita}},\ }\href {\doibase 10.1103/PhysRevB.30.5284} {\bibfield  {journal}
  {\bibinfo  {journal} {Phys. Rev. B}\ }\textbf {\bibinfo {volume} {30}},\
  \bibinfo {pages} {5284} (\bibinfo {year} {1984})}\BibitemShut {NoStop}%
\bibitem [{\citenamefont {Hara}\ and\ \citenamefont
  {Nagai}(1986)}]{HaraNagai1986}%
  \BibitemOpen
  \bibfield  {author} {\bibinfo {author} {\bibfnamefont {J.}~\bibnamefont
  {Hara}}\ and\ \bibinfo {author} {\bibfnamefont {K.}~\bibnamefont {Nagai}},\
  }\href@noop {} {\bibfield  {journal} {\bibinfo  {journal} {Prog. Theor.
  Phys.}\ }\textbf {\bibinfo {volume} {76}},\ \bibinfo {pages} {1237} (\bibinfo
  {year} {1986})}\BibitemShut {NoStop}%
\bibitem [{\citenamefont {Thies}\ and\ \citenamefont
  {Urlichs}(2003)}]{PhysRevD.67.125015}%
  \BibitemOpen
  \bibfield  {author} {\bibinfo {author} {\bibfnamefont {M.}~\bibnamefont
  {Thies}}\ and\ \bibinfo {author} {\bibfnamefont {K.}~\bibnamefont
  {Urlichs}},\ }\href {\doibase 10.1103/PhysRevD.67.125015} {\bibfield
  {journal} {\bibinfo  {journal} {Phys. Rev. D}\ }\textbf {\bibinfo {volume}
  {67}},\ \bibinfo {pages} {125015} (\bibinfo {year} {2003})}\BibitemShut
  {NoStop}%
\bibitem [{\citenamefont {Feinberg}(2003)}]{FeinbergPLB}%
  \BibitemOpen
  \bibfield  {author} {\bibinfo {author} {\bibfnamefont {J.}~\bibnamefont
  {Feinberg}},\ }\href@noop {} {\bibfield  {journal} {\bibinfo  {journal}
  {Phys. Lett. B}\ }\textbf {\bibinfo {volume} {569}},\ \bibinfo {pages} {204}
  (\bibinfo {year} {2003})}\BibitemShut {NoStop}%
\bibitem [{\citenamefont {Feinberg}(2004)}]{Feinberg:2003qz}%
  \BibitemOpen
  \bibfield  {author} {\bibinfo {author} {\bibfnamefont {J.}~\bibnamefont
  {Feinberg}},\ }\href {\doibase 10.1016/j.aop.2003.08.004} {\bibfield
  {journal} {\bibinfo  {journal} {Ann. Phys.}\ }\textbf {\bibinfo {volume}
  {309}},\ \bibinfo {pages} {166} (\bibinfo {year} {2004})}\BibitemShut
  {NoStop}%
\bibitem [{\citenamefont {Casalbuoni}\ and\ \citenamefont
  {Nardulli}(2004)}]{Casalbuoni:2003wh}%
  \BibitemOpen
  \bibfield  {author} {\bibinfo {author} {\bibfnamefont {R.}~\bibnamefont
  {Casalbuoni}}\ and\ \bibinfo {author} {\bibfnamefont {G.}~\bibnamefont
  {Nardulli}},\ }\href {\doibase 10.1103/RevModPhys.76.263} {\bibfield
  {journal} {\bibinfo  {journal} {Rev. Mod. Phys.}\ }\textbf {\bibinfo {volume}
  {76}},\ \bibinfo {pages} {263} (\bibinfo {year} {2004})}\BibitemShut
  {NoStop}%
\bibitem [{\citenamefont {Schnetz}\ \emph {et~al.}(2004)\citenamefont
  {Schnetz}, \citenamefont {Thies},\ and\ \citenamefont
  {Urlichs}}]{Schnetz2004425}%
  \BibitemOpen
  \bibfield  {author} {\bibinfo {author} {\bibfnamefont {O.}~\bibnamefont
  {Schnetz}}, \bibinfo {author} {\bibfnamefont {M.}~\bibnamefont {Thies}}, \
  and\ \bibinfo {author} {\bibfnamefont {K.}~\bibnamefont {Urlichs}},\ }\href
  {\doibase http://dx.doi.org/10.1016/j.aop.2004.06.009} {\bibfield  {journal}
  {\bibinfo  {journal} {Ann. Phys.}\ }\textbf {\bibinfo {volume} {314}},\
  \bibinfo {pages} {425 } (\bibinfo {year} {2004})}\BibitemShut {NoStop}%
\bibitem [{\citenamefont {Mizushima}\ \emph {et~al.}(2005)\citenamefont
  {Mizushima}, \citenamefont {Machida},\ and\ \citenamefont
  {Ichioka}}]{MizushimaMachidaIchioka}%
  \BibitemOpen
  \bibfield  {author} {\bibinfo {author} {\bibfnamefont {T.}~\bibnamefont
  {Mizushima}}, \bibinfo {author} {\bibfnamefont {K.}~\bibnamefont {Machida}},
  \ and\ \bibinfo {author} {\bibfnamefont {M.}~\bibnamefont {Ichioka}},\
  }\href@noop {} {\bibfield  {journal} {\bibinfo  {journal} {Phys.\ Rev.\
  Lett.}\ }\textbf {\bibinfo {volume} {94}},\ \bibinfo {pages} {060404}
  (\bibinfo {year} {2005})}\BibitemShut {NoStop}%
\bibitem [{\citenamefont {Thies}(2006)}]{Thies:2006ti}%
  \BibitemOpen
  \bibfield  {author} {\bibinfo {author} {\bibfnamefont {M.}~\bibnamefont
  {Thies}},\ }\href {\doibase 10.1088/0305-4470/39/41/S04} {\bibfield
  {journal} {\bibinfo  {journal} {J. Phys. A}\ }\textbf {\bibinfo {volume}
  {39}},\ \bibinfo {pages} {12707} (\bibinfo {year} {2006})}\BibitemShut
  {NoStop}%
\bibitem [{\citenamefont {Basar}\ and\ \citenamefont
  {Dunne}(2008{\natexlab{a}})}]{Basar:2008im}%
  \BibitemOpen
  \bibfield  {author} {\bibinfo {author} {\bibfnamefont {G.}~\bibnamefont
  {Basar}}\ and\ \bibinfo {author} {\bibfnamefont {G.~V.}\ \bibnamefont
  {Dunne}},\ }\href {\doibase 10.1103/PhysRevLett.100.200404} {\bibfield
  {journal} {\bibinfo  {journal} {Phys. Rev. Lett.}\ }\textbf {\bibinfo
  {volume} {100}},\ \bibinfo {pages} {200404} (\bibinfo {year}
  {2008}{\natexlab{a}})}\BibitemShut {NoStop}%
\bibitem [{\citenamefont {Basar}\ and\ \citenamefont
  {Dunne}(2008{\natexlab{b}})}]{Basar:2008ki}%
  \BibitemOpen
  \bibfield  {author} {\bibinfo {author} {\bibfnamefont {G.}~\bibnamefont
  {Basar}}\ and\ \bibinfo {author} {\bibfnamefont {G.~V.}\ \bibnamefont
  {Dunne}},\ }\href {\doibase 10.1103/PhysRevD.78.065022} {\bibfield  {journal}
  {\bibinfo  {journal} {Phys. Rev. D}\ }\textbf {\bibinfo {volume} {78}},\
  \bibinfo {pages} {065022} (\bibinfo {year} {2008}{\natexlab{b}})}\BibitemShut
  {NoStop}%
\bibitem [{\citenamefont {Basar}\ \emph {et~al.}(2009)\citenamefont {Basar},
  \citenamefont {Dunne},\ and\ \citenamefont {Thies}}]{Basar:2009fg}%
  \BibitemOpen
  \bibfield  {author} {\bibinfo {author} {\bibfnamefont {G.}~\bibnamefont
  {Basar}}, \bibinfo {author} {\bibfnamefont {G.~V.}\ \bibnamefont {Dunne}}, \
  and\ \bibinfo {author} {\bibfnamefont {M.}~\bibnamefont {Thies}},\ }\href
  {\doibase 10.1103/PhysRevD.79.105012} {\bibfield  {journal} {\bibinfo
  {journal} {Phys. Rev. D}\ }\textbf {\bibinfo {volume} {79}},\ \bibinfo
  {pages} {105012} (\bibinfo {year} {2009})}\BibitemShut {NoStop}%
\bibitem [{\citenamefont {Correa}\ \emph {et~al.}(2009)\citenamefont {Correa},
  \citenamefont {Dunne},\ and\ \citenamefont {Plyushchay}}]{Correa:2009xa}%
  \BibitemOpen
  \bibfield  {author} {\bibinfo {author} {\bibfnamefont {F.}~\bibnamefont
  {Correa}}, \bibinfo {author} {\bibfnamefont {G.~V.}\ \bibnamefont {Dunne}}, \
  and\ \bibinfo {author} {\bibfnamefont {M.~S.}\ \bibnamefont {Plyushchay}},\
  }\href {\doibase 10.1016/j.aop.2009.06.005} {\bibfield  {journal} {\bibinfo
  {journal} {Ann. Phys.}\ }\textbf {\bibinfo {volume} {324}},\ \bibinfo {pages}
  {2522} (\bibinfo {year} {2009})}\BibitemShut {NoStop}%
\bibitem [{\citenamefont {Nickel}(2009)}]{Nickel:2009wj}%
  \BibitemOpen
  \bibfield  {author} {\bibinfo {author} {\bibfnamefont {D.}~\bibnamefont
  {Nickel}},\ }\href {\doibase 10.1103/PhysRevD.80.074025} {\bibfield
  {journal} {\bibinfo  {journal} {Phys. Rev. D}\ }\textbf {\bibinfo {volume}
  {80}},\ \bibinfo {pages} {074025} (\bibinfo {year} {2009})}\BibitemShut
  {NoStop}%
\bibitem [{\citenamefont {Hofmann}(2010)}]{Hofmann:2010gc}%
  \BibitemOpen
  \bibfield  {author} {\bibinfo {author} {\bibfnamefont {J.}~\bibnamefont
  {Hofmann}},\ }\href {\doibase 10.1103/PhysRevD.82.125027} {\bibfield
  {journal} {\bibinfo  {journal} {Phys.Rev. D}\ }\textbf {\bibinfo {volume}
  {82}},\ \bibinfo {pages} {125027} (\bibinfo {year} {2010})}\BibitemShut
  {NoStop}%
\bibitem [{\citenamefont {Fitzner}\ and\ \citenamefont
  {Thies}(2011)}]{PhysRevD.83.085001}%
  \BibitemOpen
  \bibfield  {author} {\bibinfo {author} {\bibfnamefont {C.}~\bibnamefont
  {Fitzner}}\ and\ \bibinfo {author} {\bibfnamefont {M.}~\bibnamefont
  {Thies}},\ }\href {\doibase 10.1103/PhysRevD.83.085001} {\bibfield  {journal}
  {\bibinfo  {journal} {Phys. Rev. D}\ }\textbf {\bibinfo {volume} {83}},\
  \bibinfo {pages} {085001} (\bibinfo {year} {2011})}\BibitemShut {NoStop}%
\bibitem [{\citenamefont {Yoshii}\ \emph {et~al.}(2011)\citenamefont {Yoshii},
  \citenamefont {Tsuchiya}, \citenamefont {Marmorini},\ and\ \citenamefont
  {Nitta}}]{PhysRevB.84.024503}%
  \BibitemOpen
  \bibfield  {author} {\bibinfo {author} {\bibfnamefont {R.}~\bibnamefont
  {Yoshii}}, \bibinfo {author} {\bibfnamefont {S.}~\bibnamefont {Tsuchiya}},
  \bibinfo {author} {\bibfnamefont {G.}~\bibnamefont {Marmorini}}, \ and\
  \bibinfo {author} {\bibfnamefont {M.}~\bibnamefont {Nitta}},\ }\href
  {\doibase 10.1103/PhysRevB.84.024503} {\bibfield  {journal} {\bibinfo
  {journal} {Phys. Rev. B}\ }\textbf {\bibinfo {volume} {84}},\ \bibinfo
  {pages} {024503} (\bibinfo {year} {2011})}\BibitemShut {NoStop}%
\bibitem [{\citenamefont {Takahashi}\ \emph {et~al.}(2012)\citenamefont
  {Takahashi}, \citenamefont {Tsuchiya}, \citenamefont {Yoshii},\ and\
  \citenamefont {Nitta}}]{Takahashi:2012aw}%
  \BibitemOpen
  \bibfield  {author} {\bibinfo {author} {\bibfnamefont {D.~A.}\ \bibnamefont
  {Takahashi}}, \bibinfo {author} {\bibfnamefont {S.}~\bibnamefont {Tsuchiya}},
  \bibinfo {author} {\bibfnamefont {R.}~\bibnamefont {Yoshii}}, \ and\ \bibinfo
  {author} {\bibfnamefont {M.}~\bibnamefont {Nitta}},\ }\href {\doibase
  10.1016/j.physletb.2012.10.058} {\bibfield  {journal} {\bibinfo  {journal}
  {Phys. Lett. B}\ }\textbf {\bibinfo {volume} {718}},\ \bibinfo {pages} {632}
  (\bibinfo {year} {2012})}\BibitemShut {NoStop}%
\bibitem [{\citenamefont {Takahashi}\ and\ \citenamefont
  {Nitta}(2013)}]{PhysRevLett.110.131601}%
  \BibitemOpen
  \bibfield  {author} {\bibinfo {author} {\bibfnamefont {D.~A.}\ \bibnamefont
  {Takahashi}}\ and\ \bibinfo {author} {\bibfnamefont {M.}~\bibnamefont
  {Nitta}},\ }\href {\doibase 10.1103/PhysRevLett.110.131601} {\bibfield
  {journal} {\bibinfo  {journal} {Phys. Rev. Lett.}\ }\textbf {\bibinfo
  {volume} {110}},\ \bibinfo {pages} {131601} (\bibinfo {year}
  {2013})}\BibitemShut {NoStop}%
\bibitem [{\citenamefont {Mizushima}\ \emph {et~al.}(2014)\citenamefont
  {Mizushima}, \citenamefont {Takahashi},\ and\ \citenamefont
  {Machida}}]{JPSJ.83.023703}%
  \BibitemOpen
  \bibfield  {author} {\bibinfo {author} {\bibfnamefont {T.}~\bibnamefont
  {Mizushima}}, \bibinfo {author} {\bibfnamefont {M.}~\bibnamefont
  {Takahashi}}, \ and\ \bibinfo {author} {\bibfnamefont {K.}~\bibnamefont
  {Machida}},\ }\href {\doibase 10.7566/JPSJ.83.023703} {\bibfield  {journal}
  {\bibinfo  {journal} {J. Phys. Soc. Jpn.}\ }\textbf {\bibinfo {volume}
  {83}},\ \bibinfo {pages} {023703} (\bibinfo {year} {2014})}\BibitemShut
  {NoStop}%
\bibitem [{\citenamefont {Arancibia}\ and\ \citenamefont
  {Plyushchay}(2014)}]{Arancibia:2014lwa}%
  \BibitemOpen
  \bibfield  {author} {\bibinfo {author} {\bibfnamefont {A.}~\bibnamefont
  {Arancibia}}\ and\ \bibinfo {author} {\bibfnamefont {M.~S.}\ \bibnamefont
  {Plyushchay}},\ }\href {\doibase 10.1103/PhysRevD.90.025008} {\bibfield
  {journal} {\bibinfo  {journal} {Phys. Rev. D}\ }\textbf {\bibinfo {volume}
  {90}},\ \bibinfo {pages} {025008} (\bibinfo {year} {2014})}\BibitemShut
  {NoStop}%
\bibitem [{\citenamefont {Efimkin}\ and\ \citenamefont
  {Galitski}(2015)}]{PhysRevA.91.023616}%
  \BibitemOpen
  \bibfield  {author} {\bibinfo {author} {\bibfnamefont {D.~K.}\ \bibnamefont
  {Efimkin}}\ and\ \bibinfo {author} {\bibfnamefont {V.}~\bibnamefont
  {Galitski}},\ }\href {\doibase 10.1103/PhysRevA.91.023616} {\bibfield
  {journal} {\bibinfo  {journal} {Phys. Rev. A}\ }\textbf {\bibinfo {volume}
  {91}},\ \bibinfo {pages} {023616} (\bibinfo {year} {2015})}\BibitemShut
  {NoStop}%
\bibitem [{\citenamefont {Xu}\ \emph {et~al.}(2014)\citenamefont {Xu},
  \citenamefont {Mao}, \citenamefont {Wu},\ and\ \citenamefont
  {Zhang}}]{PhysRevLett.113.130404}%
  \BibitemOpen
  \bibfield  {author} {\bibinfo {author} {\bibfnamefont {Y.}~\bibnamefont
  {Xu}}, \bibinfo {author} {\bibfnamefont {L.}~\bibnamefont {Mao}}, \bibinfo
  {author} {\bibfnamefont {B.}~\bibnamefont {Wu}}, \ and\ \bibinfo {author}
  {\bibfnamefont {C.}~\bibnamefont {Zhang}},\ }\href {\doibase
  10.1103/PhysRevLett.113.130404} {\bibfield  {journal} {\bibinfo  {journal}
  {Phys. Rev. Lett.}\ }\textbf {\bibinfo {volume} {113}},\ \bibinfo {pages}
  {130404} (\bibinfo {year} {2014})}\BibitemShut {NoStop}%
\bibitem [{\citenamefont {Liu}(2015)}]{PhysRevA.91.023610}%
  \BibitemOpen
  \bibfield  {author} {\bibinfo {author} {\bibfnamefont {X.-J.}\ \bibnamefont
  {Liu}},\ }\href {\doibase 10.1103/PhysRevA.91.023610} {\bibfield  {journal}
  {\bibinfo  {journal} {Phys. Rev. A}\ }\textbf {\bibinfo {volume} {91}},\
  \bibinfo {pages} {023610} (\bibinfo {year} {2015})}\BibitemShut {NoStop}%
\bibitem [{\citenamefont {Hidaka}\ \emph {et~al.}(2015)\citenamefont {Hidaka},
  \citenamefont {Kamikado}, \citenamefont {Kanazawa},\ and\ \citenamefont
  {Noumi}}]{PhysRevD.92.034003}%
  \BibitemOpen
  \bibfield  {author} {\bibinfo {author} {\bibfnamefont {Y.}~\bibnamefont
  {Hidaka}}, \bibinfo {author} {\bibfnamefont {K.}~\bibnamefont {Kamikado}},
  \bibinfo {author} {\bibfnamefont {T.}~\bibnamefont {Kanazawa}}, \ and\
  \bibinfo {author} {\bibfnamefont {T.}~\bibnamefont {Noumi}},\ }\href
  {\doibase 10.1103/PhysRevD.92.034003} {\bibfield  {journal} {\bibinfo
  {journal} {Phys. Rev. D}\ }\textbf {\bibinfo {volume} {92}},\ \bibinfo
  {pages} {034003} (\bibinfo {year} {2015})}\BibitemShut {NoStop}%
\bibitem [{\citenamefont {Dunne}\ and\ \citenamefont
  {Thies}(2013{\natexlab{a}})}]{PhysRevLett.111.121602}%
  \BibitemOpen
  \bibfield  {author} {\bibinfo {author} {\bibfnamefont {G.~V.}\ \bibnamefont
  {Dunne}}\ and\ \bibinfo {author} {\bibfnamefont {M.}~\bibnamefont {Thies}},\
  }\href {\doibase 10.1103/PhysRevLett.111.121602} {\bibfield  {journal}
  {\bibinfo  {journal} {Phys. Rev. Lett.}\ }\textbf {\bibinfo {volume} {111}},\
  \bibinfo {pages} {121602} (\bibinfo {year} {2013}{\natexlab{a}})}\BibitemShut
  {NoStop}%
\bibitem [{\citenamefont {Dunne}\ and\ \citenamefont
  {Thies}(2013{\natexlab{b}})}]{PhysRevA.88.062115}%
  \BibitemOpen
  \bibfield  {author} {\bibinfo {author} {\bibfnamefont {G.~V.}\ \bibnamefont
  {Dunne}}\ and\ \bibinfo {author} {\bibfnamefont {M.}~\bibnamefont {Thies}},\
  }\href {\doibase 10.1103/PhysRevA.88.062115} {\bibfield  {journal} {\bibinfo
  {journal} {Phys. Rev. A}\ }\textbf {\bibinfo {volume} {88}},\ \bibinfo
  {pages} {062115} (\bibinfo {year} {2013}{\natexlab{b}})}\BibitemShut
  {NoStop}%
\bibitem [{\citenamefont {Dunne}\ and\ \citenamefont
  {Thies}(2014)}]{PhysRevD.89.025008}%
  \BibitemOpen
  \bibfield  {author} {\bibinfo {author} {\bibfnamefont {G.~V.}\ \bibnamefont
  {Dunne}}\ and\ \bibinfo {author} {\bibfnamefont {M.}~\bibnamefont {Thies}},\
  }\href {\doibase 10.1103/PhysRevD.89.025008} {\bibfield  {journal} {\bibinfo
  {journal} {Phys. Rev. D}\ }\textbf {\bibinfo {volume} {89}},\ \bibinfo
  {pages} {025008} (\bibinfo {year} {2014})}\BibitemShut {NoStop}%
\bibitem [{\citenamefont {Challis}\ \emph {et~al.}(2007)\citenamefont
  {Challis}, \citenamefont {Ballagh},\ and\ \citenamefont
  {Gardiner}}]{PhysRevLett.98.093002}%
  \BibitemOpen
  \bibfield  {author} {\bibinfo {author} {\bibfnamefont {K.~J.}\ \bibnamefont
  {Challis}}, \bibinfo {author} {\bibfnamefont {R.~J.}\ \bibnamefont
  {Ballagh}}, \ and\ \bibinfo {author} {\bibfnamefont {C.~W.}\ \bibnamefont
  {Gardiner}},\ }\href {\doibase 10.1103/PhysRevLett.98.093002} {\bibfield
  {journal} {\bibinfo  {journal} {Phys. Rev. Lett.}\ }\textbf {\bibinfo
  {volume} {98}},\ \bibinfo {pages} {093002} (\bibinfo {year}
  {2007})}\BibitemShut {NoStop}%
\bibitem [{\citenamefont {Scott}\ \emph {et~al.}(2011)\citenamefont {Scott},
  \citenamefont {Dalfovo}, \citenamefont {Pitaevskii},\ and\ \citenamefont
  {Stringari}}]{PhysRevLett.106.185301}%
  \BibitemOpen
  \bibfield  {author} {\bibinfo {author} {\bibfnamefont {R.~G.}\ \bibnamefont
  {Scott}}, \bibinfo {author} {\bibfnamefont {F.}~\bibnamefont {Dalfovo}},
  \bibinfo {author} {\bibfnamefont {L.~P.}\ \bibnamefont {Pitaevskii}}, \ and\
  \bibinfo {author} {\bibfnamefont {S.}~\bibnamefont {Stringari}},\ }\href
  {\doibase 10.1103/PhysRevLett.106.185301} {\bibfield  {journal} {\bibinfo
  {journal} {Phys. Rev. Lett.}\ }\textbf {\bibinfo {volume} {106}},\ \bibinfo
  {pages} {185301} (\bibinfo {year} {2011})}\BibitemShut {NoStop}%
\bibitem [{\citenamefont {Cetoli}\ \emph {et~al.}(2013)\citenamefont {Cetoli},
  \citenamefont {Brand}, \citenamefont {Scott}, \citenamefont {Dalfovo},\ and\
  \citenamefont {Pitaevskii}}]{PhysRevA.88.043639}%
  \BibitemOpen
  \bibfield  {author} {\bibinfo {author} {\bibfnamefont {A.}~\bibnamefont
  {Cetoli}}, \bibinfo {author} {\bibfnamefont {J.}~\bibnamefont {Brand}},
  \bibinfo {author} {\bibfnamefont {R.~G.}\ \bibnamefont {Scott}}, \bibinfo
  {author} {\bibfnamefont {F.}~\bibnamefont {Dalfovo}}, \ and\ \bibinfo
  {author} {\bibfnamefont {L.~P.}\ \bibnamefont {Pitaevskii}},\ }\href
  {\doibase 10.1103/PhysRevA.88.043639} {\bibfield  {journal} {\bibinfo
  {journal} {Phys. Rev. A}\ }\textbf {\bibinfo {volume} {88}},\ \bibinfo
  {pages} {043639} (\bibinfo {year} {2013})}\BibitemShut {NoStop}%
\bibitem [{\citenamefont {Liu}\ \emph {et~al.}(2007)\citenamefont {Liu},
  \citenamefont {Hu},\ and\ \citenamefont {Drummond}}]{PhysRevA.76.043605}%
  \BibitemOpen
  \bibfield  {author} {\bibinfo {author} {\bibfnamefont {X.-J.}\ \bibnamefont
  {Liu}}, \bibinfo {author} {\bibfnamefont {H.}~\bibnamefont {Hu}}, \ and\
  \bibinfo {author} {\bibfnamefont {P.~D.}\ \bibnamefont {Drummond}},\ }\href
  {\doibase 10.1103/PhysRevA.76.043605} {\bibfield  {journal} {\bibinfo
  {journal} {Phys. Rev. A}\ }\textbf {\bibinfo {volume} {76}},\ \bibinfo
  {pages} {043605} (\bibinfo {year} {2007})}\BibitemShut {NoStop}%
\bibitem [{\citenamefont {Vollhardt}\ and\ \citenamefont
  {W{\"o}lfle}(2013)}]{VollhardtWolfle}%
  \BibitemOpen
  \bibfield  {author} {\bibinfo {author} {\bibfnamefont {D.}~\bibnamefont
  {Vollhardt}}\ and\ \bibinfo {author} {\bibfnamefont {P.}~\bibnamefont
  {W{\"o}lfle}},\ }\href@noop {} {\emph {\bibinfo {title} {The Superfluid
  Phases of Helium 3}}}\ (\bibinfo  {publisher} {Dover Publications},\ \bibinfo
  {address} {Mineola, New York},\ \bibinfo {year} {2013})\BibitemShut {NoStop}%
\bibitem [{\citenamefont {Mizushima}\ \emph
  {et~al.}(2015{\natexlab{a}})\citenamefont {Mizushima}, \citenamefont
  {Tsutsumi}, \citenamefont {Sato},\ and\ \citenamefont
  {Machida}}]{0953-8984-27-11-113203}%
  \BibitemOpen
  \bibfield  {author} {\bibinfo {author} {\bibfnamefont {T.}~\bibnamefont
  {Mizushima}}, \bibinfo {author} {\bibfnamefont {Y.}~\bibnamefont {Tsutsumi}},
  \bibinfo {author} {\bibfnamefont {M.}~\bibnamefont {Sato}}, \ and\ \bibinfo
  {author} {\bibfnamefont {K.}~\bibnamefont {Machida}},\ }\href@noop {}
  {\bibfield  {journal} {\bibinfo  {journal} {Journal of Physics: Condensed
  Matter}\ }\textbf {\bibinfo {volume} {27}},\ \bibinfo {pages} {113203}
  (\bibinfo {year} {2015}{\natexlab{a}})}\BibitemShut {NoStop}%
\bibitem [{\citenamefont {Mizushima}\ \emph
  {et~al.}(2015{\natexlab{b}})\citenamefont {Mizushima}, \citenamefont
  {Tsutsumi}, \citenamefont {Kawakami}, \citenamefont {Sato}, \citenamefont
  {Ichioka},\ and\ \citenamefont {Machida}}]{arxiv1508.00787}%
  \BibitemOpen
  \bibfield  {author} {\bibinfo {author} {\bibfnamefont {T.}~\bibnamefont
  {Mizushima}}, \bibinfo {author} {\bibfnamefont {Y.}~\bibnamefont {Tsutsumi}},
  \bibinfo {author} {\bibfnamefont {T.}~\bibnamefont {Kawakami}}, \bibinfo
  {author} {\bibfnamefont {M.}~\bibnamefont {Sato}}, \bibinfo {author}
  {\bibfnamefont {M.}~\bibnamefont {Ichioka}}, \ and\ \bibinfo {author}
  {\bibfnamefont {K.}~\bibnamefont {Machida}},\ }\href@noop {} {\bibfield
  {journal} {\bibinfo  {journal} {arXiv:1508.00787}\ } (\bibinfo {year}
  {2015}{\natexlab{b}})}\BibitemShut {NoStop}%
\bibitem [{\citenamefont {Yanase}\ and\ \citenamefont
  {Sigrist}(2008)}]{JPSJ.77.124711}%
  \BibitemOpen
  \bibfield  {author} {\bibinfo {author} {\bibfnamefont {Y.}~\bibnamefont
  {Yanase}}\ and\ \bibinfo {author} {\bibfnamefont {M.}~\bibnamefont
  {Sigrist}},\ }\href {\doibase 10.1143/JPSJ.77.124711} {\bibfield  {journal}
  {\bibinfo  {journal} {J. Phys. Soc. Jpn.}\ }\textbf {\bibinfo {volume}
  {77}},\ \bibinfo {pages} {124711} (\bibinfo {year} {2008})}\BibitemShut
  {NoStop}%
\bibitem [{\citenamefont {Yoshida}\ \emph {et~al.}(2014)\citenamefont
  {Yoshida}, \citenamefont {Sigrist},\ and\ \citenamefont
  {Yanase}}]{JPSJ.83.013703}%
  \BibitemOpen
  \bibfield  {author} {\bibinfo {author} {\bibfnamefont {T.}~\bibnamefont
  {Yoshida}}, \bibinfo {author} {\bibfnamefont {M.}~\bibnamefont {Sigrist}}, \
  and\ \bibinfo {author} {\bibfnamefont {Y.}~\bibnamefont {Yanase}},\ }\href
  {\doibase 10.7566/JPSJ.83.013703} {\bibfield  {journal} {\bibinfo  {journal}
  {J. Phys. Soc. Jpn.}\ }\textbf {\bibinfo {volume} {83}},\ \bibinfo {pages}
  {013703} (\bibinfo {year} {2014})}\BibitemShut {NoStop}%
\bibitem [{\citenamefont {Yanagisawa}\ \emph {et~al.}(2012)\citenamefont
  {Yanagisawa}, \citenamefont {Tanaka}, \citenamefont {Hase},\ and\
  \citenamefont {Yamaji}}]{JPSJ.81.024712}%
  \BibitemOpen
  \bibfield  {author} {\bibinfo {author} {\bibfnamefont {T.}~\bibnamefont
  {Yanagisawa}}, \bibinfo {author} {\bibfnamefont {Y.}~\bibnamefont {Tanaka}},
  \bibinfo {author} {\bibfnamefont {I.}~\bibnamefont {Hase}}, \ and\ \bibinfo
  {author} {\bibfnamefont {K.}~\bibnamefont {Yamaji}},\ }\href {\doibase
  10.1143/JPSJ.81.024712} {\bibfield  {journal} {\bibinfo  {journal} {J. Phys.
  Soc. Jpn.}\ }\textbf {\bibinfo {volume} {81}},\ \bibinfo {pages} {024712}
  (\bibinfo {year} {2012})}\BibitemShut {NoStop}%
\bibitem [{\citenamefont {Nagai}\ and\ \citenamefont
  {Nakamura}(2015)}]{arxiv1507.06039}%
  \BibitemOpen
  \bibfield  {author} {\bibinfo {author} {\bibfnamefont {Y.}~\bibnamefont
  {Nagai}}\ and\ \bibinfo {author} {\bibfnamefont {H.}~\bibnamefont
  {Nakamura}},\ }\href@noop {} {\bibfield  {journal} {\bibinfo  {journal}
  {arXiv:1507.06039}\ } (\bibinfo {year} {2015})}\BibitemShut {NoStop}%
\bibitem [{\citenamefont {Fukuhara}\ \emph {et~al.}(2007)\citenamefont
  {Fukuhara}, \citenamefont {Takasu}, \citenamefont {Kumakura},\ and\
  \citenamefont {Takahashi}}]{PhysRevLett.98.030401}%
  \BibitemOpen
  \bibfield  {author} {\bibinfo {author} {\bibfnamefont {T.}~\bibnamefont
  {Fukuhara}}, \bibinfo {author} {\bibfnamefont {Y.}~\bibnamefont {Takasu}},
  \bibinfo {author} {\bibfnamefont {M.}~\bibnamefont {Kumakura}}, \ and\
  \bibinfo {author} {\bibfnamefont {Y.}~\bibnamefont {Takahashi}},\ }\href
  {\doibase 10.1103/PhysRevLett.98.030401} {\bibfield  {journal} {\bibinfo
  {journal} {Phys. Rev. Lett.}\ }\textbf {\bibinfo {volume} {98}},\ \bibinfo
  {pages} {030401} (\bibinfo {year} {2007})}\BibitemShut {NoStop}%
\bibitem [{\citenamefont {Takasu}\ and\ \citenamefont
  {Takahashi}(2009)}]{JPSJ.78.012001}%
  \BibitemOpen
  \bibfield  {author} {\bibinfo {author} {\bibfnamefont {Y.}~\bibnamefont
  {Takasu}}\ and\ \bibinfo {author} {\bibfnamefont {Y.}~\bibnamefont
  {Takahashi}},\ }\href {\doibase 10.1143/JPSJ.78.012001} {\bibfield  {journal}
  {\bibinfo  {journal} {J. Phys. Soc. Jpn.}\ }\textbf {\bibinfo {volume}
  {78}},\ \bibinfo {pages} {012001} (\bibinfo {year} {2009})}\BibitemShut
  {NoStop}%
\bibitem [{\citenamefont {Ozawa}\ and\ \citenamefont
  {Baym}(2010)}]{PhysRevA.82.063615}%
  \BibitemOpen
  \bibfield  {author} {\bibinfo {author} {\bibfnamefont {T.}~\bibnamefont
  {Ozawa}}\ and\ \bibinfo {author} {\bibfnamefont {G.}~\bibnamefont {Baym}},\
  }\href {\doibase 10.1103/PhysRevA.82.063615} {\bibfield  {journal} {\bibinfo
  {journal} {Phys. Rev. A}\ }\textbf {\bibinfo {volume} {82}},\ \bibinfo
  {pages} {063615} (\bibinfo {year} {2010})}\BibitemShut {NoStop}%
\bibitem [{\citenamefont {Yip}(2011)}]{PhysRevA.83.063607}%
  \BibitemOpen
  \bibfield  {author} {\bibinfo {author} {\bibfnamefont {S.-K.}\ \bibnamefont
  {Yip}},\ }\href {\doibase 10.1103/PhysRevA.83.063607} {\bibfield  {journal}
  {\bibinfo  {journal} {Phys. Rev. A}\ }\textbf {\bibinfo {volume} {83}},\
  \bibinfo {pages} {063607} (\bibinfo {year} {2011})}\BibitemShut {NoStop}%
\bibitem [{\citenamefont {Cazalilla}\ and\ \citenamefont
  {Rey}(2014)}]{0034-4885-77-12-124401}%
  \BibitemOpen
  \bibfield  {author} {\bibinfo {author} {\bibfnamefont {M.}~\bibnamefont
  {Cazalilla}}\ and\ \bibinfo {author} {\bibfnamefont {A.}~\bibnamefont
  {Rey}},\ }\href {http://stacks.iop.org/0034-4885/77/i=12/a=124401} {\bibfield
   {journal} {\bibinfo  {journal} {Rep. Prog. Phys.}\ }\textbf {\bibinfo
  {volume} {77}},\ \bibinfo {pages} {124401} (\bibinfo {year}
  {2014})}\BibitemShut {NoStop}%
\bibitem [{\citenamefont {Sakaida}\ and\ \citenamefont
  {Kawakami}(2014)}]{PhysRevA.90.013632}%
  \BibitemOpen
  \bibfield  {author} {\bibinfo {author} {\bibfnamefont {M.}~\bibnamefont
  {Sakaida}}\ and\ \bibinfo {author} {\bibfnamefont {N.}~\bibnamefont
  {Kawakami}},\ }\href {\doibase 10.1103/PhysRevA.90.013632} {\bibfield
  {journal} {\bibinfo  {journal} {Phys. Rev. A}\ }\textbf {\bibinfo {volume}
  {90}},\ \bibinfo {pages} {013632} (\bibinfo {year} {2014})}\BibitemShut
  {NoStop}%
\bibitem [{\citenamefont {Higashikawa}\ and\ \citenamefont
  {Ueda}(2015)}]{arxiv1504.08113}%
  \BibitemOpen
  \bibfield  {author} {\bibinfo {author} {\bibfnamefont {S.}~\bibnamefont
  {Higashikawa}}\ and\ \bibinfo {author} {\bibfnamefont {M.}~\bibnamefont
  {Ueda}},\ }\href@noop {} {\bibfield  {journal} {\bibinfo  {journal}
  {arXiv:1504.08113}\ } (\bibinfo {year} {2015})}\BibitemShut {NoStop}%
\bibitem [{\citenamefont {Zakharov}\ and\ \citenamefont
  {Shabat}(1972)}]{ZakharovShabat}%
  \BibitemOpen
  \bibfield  {author} {\bibinfo {author} {\bibfnamefont {V.~E.}\ \bibnamefont
  {Zakharov}}\ and\ \bibinfo {author} {\bibfnamefont {A.~B.}\ \bibnamefont
  {Shabat}},\ }\href@noop {} {\bibfield  {journal} {\bibinfo  {journal} {Sov.
  Phys. JETP}\ }\textbf {\bibinfo {volume} {34}},\ \bibinfo {pages} {62}
  (\bibinfo {year} {1972})}\BibitemShut {NoStop}%
\bibitem [{\citenamefont {Zakharov}\ and\ \citenamefont
  {Shabat}(1973)}]{ZakharovShabat2}%
  \BibitemOpen
  \bibfield  {author} {\bibinfo {author} {\bibfnamefont {V.~E.}\ \bibnamefont
  {Zakharov}}\ and\ \bibinfo {author} {\bibfnamefont {A.~B.}\ \bibnamefont
  {Shabat}},\ }\href@noop {} {\bibfield  {journal} {\bibinfo  {journal} {Sov.
  Phys. JETP}\ }\textbf {\bibinfo {volume} {37}},\ \bibinfo {pages} {823}
  (\bibinfo {year} {1973})}\BibitemShut {NoStop}%
\bibitem [{\citenamefont {Ablowitz}\ \emph {et~al.}(1974)\citenamefont
  {Ablowitz}, \citenamefont {Kaup}, \citenamefont {Newell},\ and\ \citenamefont
  {Segur}}]{AKNS1974}%
  \BibitemOpen
  \bibfield  {author} {\bibinfo {author} {\bibfnamefont {M.~J.}\ \bibnamefont
  {Ablowitz}}, \bibinfo {author} {\bibfnamefont {D.~J.}\ \bibnamefont {Kaup}},
  \bibinfo {author} {\bibfnamefont {A.~C.}\ \bibnamefont {Newell}}, \ and\
  \bibinfo {author} {\bibfnamefont {H.}~\bibnamefont {Segur}},\ }\href@noop {}
  {\bibfield  {journal} {\bibinfo  {journal} {Stud. Appl. Math.}\ }\textbf
  {\bibinfo {volume} {53}},\ \bibinfo {pages} {249} (\bibinfo {year}
  {1974})}\BibitemShut {NoStop}%
\bibitem [{\citenamefont {Ablowitz}\ and\ \citenamefont
  {Segur}(1981)}]{AblowitzSegur}%
  \BibitemOpen
  \bibfield  {author} {\bibinfo {author} {\bibfnamefont {M.~J.}\ \bibnamefont
  {Ablowitz}}\ and\ \bibinfo {author} {\bibfnamefont {H.}~\bibnamefont
  {Segur}},\ }\href@noop {} {\emph {\bibinfo {title} {Solitons and the Inverse
  Scattering Transform}}}\ (\bibinfo  {publisher} {SIAM},\ \bibinfo {address}
  {Philadelphia},\ \bibinfo {year} {1981})\BibitemShut {NoStop}%
\bibitem [{\citenamefont {Faddeev}\ and\ \citenamefont
  {Takhtajan}(1987)}]{FaddeevTakhtajan}%
  \BibitemOpen
  \bibfield  {author} {\bibinfo {author} {\bibfnamefont {L.~D.}\ \bibnamefont
  {Faddeev}}\ and\ \bibinfo {author} {\bibfnamefont {L.~A.}\ \bibnamefont
  {Takhtajan}},\ }\href@noop {} {\emph {\bibinfo {title} {Hamiltonian Methods
  in the Theory of Solitons}}}\ (\bibinfo  {publisher} {Springer},\ \bibinfo
  {address} {Berlin, Heidelberg},\ \bibinfo {year} {1987})\BibitemShut
  {NoStop}%
\bibitem [{\citenamefont {Kitaev}(2001)}]{1063-7869-44-10S-S29}%
  \BibitemOpen
  \bibfield  {author} {\bibinfo {author} {\bibfnamefont {A.~Y.}\ \bibnamefont
  {Kitaev}},\ }\href {http://stacks.iop.org/1063-7869/44/i=10S/a=S29}
  {\bibfield  {journal} {\bibinfo  {journal} {Physics-Uspekhi}\ }\textbf
  {\bibinfo {volume} {44}},\ \bibinfo {pages} {131} (\bibinfo {year}
  {2001})}\BibitemShut {NoStop}%
\bibitem [{\citenamefont {Tsuchida}\ and\ \citenamefont
  {Wadati}(1998)}]{JPSJ.67.1175}%
  \BibitemOpen
  \bibfield  {author} {\bibinfo {author} {\bibfnamefont {T.}~\bibnamefont
  {Tsuchida}}\ and\ \bibinfo {author} {\bibfnamefont {M.}~\bibnamefont
  {Wadati}},\ }\href {\doibase 10.1143/JPSJ.67.1175} {\bibfield  {journal}
  {\bibinfo  {journal} {J. Phys. Soc. Jpn.}\ }\textbf {\bibinfo {volume}
  {67}},\ \bibinfo {pages} {1175} (\bibinfo {year} {1998})}\BibitemShut
  {NoStop}%
\bibitem [{\citenamefont {Ieda}\ \emph {et~al.}(2007)\citenamefont {Ieda},
  \citenamefont {Uchiyama},\ and\ \citenamefont
  {Wadati}}]{jmp48110.10631.2423222}%
  \BibitemOpen
  \bibfield  {author} {\bibinfo {author} {\bibfnamefont {J.}~\bibnamefont
  {Ieda}}, \bibinfo {author} {\bibfnamefont {M.}~\bibnamefont {Uchiyama}}, \
  and\ \bibinfo {author} {\bibfnamefont {M.}~\bibnamefont {Wadati}},\ }\href
  {\doibase http://dx.doi.org/10.1063/1.2423222} {\bibfield  {journal}
  {\bibinfo  {journal} {J. Math. Phys.}\ }\textbf {\bibinfo {volume} {48}},\
  \bibinfo {eid} {013507} (\bibinfo {year} {2007})}\BibitemShut {NoStop}%
\bibitem [{\citenamefont {Ieda}\ \emph
  {et~al.}(2004{\natexlab{a}})\citenamefont {Ieda}, \citenamefont {Miyakawa},\
  and\ \citenamefont {Wadati}}]{PhysRevLett.93.194102}%
  \BibitemOpen
  \bibfield  {author} {\bibinfo {author} {\bibfnamefont {J.}~\bibnamefont
  {Ieda}}, \bibinfo {author} {\bibfnamefont {T.}~\bibnamefont {Miyakawa}}, \
  and\ \bibinfo {author} {\bibfnamefont {M.}~\bibnamefont {Wadati}},\ }\href
  {\doibase 10.1103/PhysRevLett.93.194102} {\bibfield  {journal} {\bibinfo
  {journal} {Phys. Rev. Lett.}\ }\textbf {\bibinfo {volume} {93}},\ \bibinfo
  {pages} {194102} (\bibinfo {year} {2004}{\natexlab{a}})}\BibitemShut
  {NoStop}%
\bibitem [{\citenamefont {Ieda}\ \emph
  {et~al.}(2004{\natexlab{b}})\citenamefont {Ieda}, \citenamefont {Miyakawa},\
  and\ \citenamefont {Wadati}}]{JPSJ.73.2996}%
  \BibitemOpen
  \bibfield  {author} {\bibinfo {author} {\bibfnamefont {J.}~\bibnamefont
  {Ieda}}, \bibinfo {author} {\bibfnamefont {T.}~\bibnamefont {Miyakawa}}, \
  and\ \bibinfo {author} {\bibfnamefont {M.}~\bibnamefont {Wadati}},\ }\href
  {\doibase 10.1143/JPSJ.73.2996} {\bibfield  {journal} {\bibinfo  {journal}
  {J. Phys. Soc. Jpn.}\ }\textbf {\bibinfo {volume} {73}},\ \bibinfo {pages}
  {2996} (\bibinfo {year} {2004}{\natexlab{b}})}\BibitemShut {NoStop}%
\bibitem [{\citenamefont {Uchiyama}\ \emph {et~al.}(2006)\citenamefont
  {Uchiyama}, \citenamefont {Ieda},\ and\ \citenamefont
  {Wadati}}]{JPSJ.75.064002}%
  \BibitemOpen
  \bibfield  {author} {\bibinfo {author} {\bibfnamefont {M.}~\bibnamefont
  {Uchiyama}}, \bibinfo {author} {\bibfnamefont {J.}~\bibnamefont {Ieda}}, \
  and\ \bibinfo {author} {\bibfnamefont {M.}~\bibnamefont {Wadati}},\ }\href
  {\doibase 10.1143/JPSJ.75.064002} {\bibfield  {journal} {\bibinfo  {journal}
  {J. Phys. Soc. Jpn.}\ }\textbf {\bibinfo {volume} {75}},\ \bibinfo {pages}
  {064002} (\bibinfo {year} {2006})}\BibitemShut {NoStop}%
\bibitem [{\citenamefont {Jackiw}\ and\ \citenamefont
  {Rebbi}(1976)}]{Jackiw:1975fn}%
  \BibitemOpen
  \bibfield  {author} {\bibinfo {author} {\bibfnamefont {R.}~\bibnamefont
  {Jackiw}}\ and\ \bibinfo {author} {\bibfnamefont {C.}~\bibnamefont {Rebbi}},\
  }\href {\doibase 10.1103/PhysRevD.13.3398} {\bibfield  {journal} {\bibinfo
  {journal} {Phys. Rev. D}\ }\textbf {\bibinfo {volume} {13}},\ \bibinfo
  {pages} {3398} (\bibinfo {year} {1976})}\BibitemShut {NoStop}%
\bibitem [{Note1()}]{Note1}%
  \BibitemOpen
  \bibinfo {note} {See animations in the Ancillary files
  }\BibitemShut {NoStop}%
\bibitem [{\citenamefont {Kawaguchi}\ and\ \citenamefont
  {Ueda}(2012)}]{Kawaguchi:2012ii}%
  \BibitemOpen
  \bibfield  {author} {\bibinfo {author} {\bibfnamefont {Y.}~\bibnamefont
  {Kawaguchi}}\ and\ \bibinfo {author} {\bibfnamefont {M.}~\bibnamefont
  {Ueda}},\ }\href@noop {} {\bibfield  {journal} {\bibinfo  {journal} {Phys.
  Rept.}\ }\textbf {\bibinfo {volume} {520}},\ \bibinfo {pages} {253} (\bibinfo
  {year} {2012})}\BibitemShut {NoStop}%
\bibitem [{\citenamefont {Toda}(1989)}]{Todahisenkeihadou}%
  \BibitemOpen
  \bibfield  {author} {\bibinfo {author} {\bibfnamefont {M.}~\bibnamefont
  {Toda}},\ }\href@noop {} {\emph {\bibinfo {title} {Nonlinear Waves and
  Solitons}}}\ (\bibinfo  {publisher} {KTK Scientific Publishers},\ \bibinfo
  {address} {Tokyo},\ \bibinfo {year} {1989})\ \bibinfo {note} {[original
  Japanese version: Hisenkei hadou to soriton (Nippon Hyoron Sha, Tokyo,
  1983)]}\BibitemShut {NoStop}%
\bibitem [{\citenamefont {Takahashi}\ and\ \citenamefont
  {Nitta}(2014)}]{takahashinittaJLTP}%
  \BibitemOpen
  \bibfield  {author} {\bibinfo {author} {\bibfnamefont {D.~A.}\ \bibnamefont
  {Takahashi}}\ and\ \bibinfo {author} {\bibfnamefont {M.}~\bibnamefont
  {Nitta}},\ }\href {\doibase 10.1007/s10909-013-0912-8} {\bibfield  {journal}
  {\bibinfo  {journal} {J. Low Temp. Phys.}\ }\textbf {\bibinfo {volume}
  {175}},\ \bibinfo {pages} {250} (\bibinfo {year} {2014}),\ arXiv:1307.3897 }\BibitemShut
  {NoStop}%
\bibitem [{\citenamefont {Gaunt}\ \emph {et~al.}(2013)\citenamefont {Gaunt},
  \citenamefont {Schmidutz}, \citenamefont {Gotlibovych}, \citenamefont
  {Smith},\ and\ \citenamefont {Hadzibabic}}]{PhysRevLett.110.200406}%
  \BibitemOpen
  \bibfield  {author} {\bibinfo {author} {\bibfnamefont {A.~L.}\ \bibnamefont
  {Gaunt}}, \bibinfo {author} {\bibfnamefont {T.~F.}\ \bibnamefont
  {Schmidutz}}, \bibinfo {author} {\bibfnamefont {I.}~\bibnamefont
  {Gotlibovych}}, \bibinfo {author} {\bibfnamefont {R.~P.}\ \bibnamefont
  {Smith}}, \ and\ \bibinfo {author} {\bibfnamefont {Z.}~\bibnamefont
  {Hadzibabic}},\ }\href {\doibase 10.1103/PhysRevLett.110.200406} {\bibfield
  {journal} {\bibinfo  {journal} {Phys. Rev. Lett.}\ }\textbf {\bibinfo
  {volume} {110}},\ \bibinfo {pages} {200406} (\bibinfo {year}
  {2013})}\BibitemShut {NoStop}%
\bibitem [{\citenamefont {Takahashi}(2013)}]{arxiv1304.7567}%
  \BibitemOpen
  \bibfield  {author} {\bibinfo {author} {\bibfnamefont {D.~A.}\ \bibnamefont
  {Takahashi}},\ }\href@noop {} {\bibfield  {journal} {\bibinfo  {journal}
  {arXiv:1304.7567}\ } (\bibinfo {year} {2013})}\BibitemShut {NoStop}%
\bibitem [{\citenamefont {Yamada}\ and\ \citenamefont
  {Ohmi}(1995)}]{YamadaOhmi}%
  \BibitemOpen
  \bibfield  {author} {\bibinfo {author} {\bibfnamefont {K.}~\bibnamefont
  {Yamada}}\ and\ \bibinfo {author} {\bibfnamefont {T.}~\bibnamefont {Ohmi}},\
  }\href@noop {} {\emph {\bibinfo {title} {Chouryuudou (Superfluidity)}}}\
  (\bibinfo  {publisher} {Baifukan},\ \bibinfo {address} {Tokyo},\ \bibinfo
  {year} {1995})\ \bibinfo {note} {[written in Japanese]}\BibitemShut {NoStop}%
\bibitem [{\citenamefont {Takahashi}\ \emph {et~al.}(2006)\citenamefont
  {Takahashi}, \citenamefont {Mizushima}, \citenamefont {Ichioka},\ and\
  \citenamefont {Machida}}]{PhysRevLett.97.180407}%
  \BibitemOpen
  \bibfield  {author} {\bibinfo {author} {\bibfnamefont {M.}~\bibnamefont
  {Takahashi}}, \bibinfo {author} {\bibfnamefont {T.}~\bibnamefont
  {Mizushima}}, \bibinfo {author} {\bibfnamefont {M.}~\bibnamefont {Ichioka}},
  \ and\ \bibinfo {author} {\bibfnamefont {K.}~\bibnamefont {Machida}},\ }\href
  {\doibase 10.1103/PhysRevLett.97.180407} {\bibfield  {journal} {\bibinfo
  {journal} {Phys. Rev. Lett.}\ }\textbf {\bibinfo {volume} {97}},\ \bibinfo
  {pages} {180407} (\bibinfo {year} {2006})}\BibitemShut {NoStop}%
\bibitem [{\citenamefont {Andreev}(1964)}]{Andreev1964}%
  \BibitemOpen
  \bibfield  {author} {\bibinfo {author} {\bibfnamefont {A.~F.}\ \bibnamefont
  {Andreev}},\ }\href@noop {} {\bibfield  {journal} {\bibinfo  {journal} {Sov.
  Phys. JETP}\ }\textbf {\bibinfo {volume} {19}},\ \bibinfo {pages} {1228}
  (\bibinfo {year} {1964})}\BibitemShut {NoStop}%
\bibitem [{\citenamefont {Altland}\ and\ \citenamefont
  {Zirnbauer}(1997)}]{PhysRevB.55.1142}%
  \BibitemOpen
  \bibfield  {author} {\bibinfo {author} {\bibfnamefont {A.}~\bibnamefont
  {Altland}}\ and\ \bibinfo {author} {\bibfnamefont {M.~R.}\ \bibnamefont
  {Zirnbauer}},\ }\href {\doibase 10.1103/PhysRevB.55.1142} {\bibfield
  {journal} {\bibinfo  {journal} {Phys. Rev. B}\ }\textbf {\bibinfo {volume}
  {55}},\ \bibinfo {pages} {1142} (\bibinfo {year} {1997})}\BibitemShut
  {NoStop}%
\bibitem [{\citenamefont {Schnyder}\ \emph {et~al.}(2008)\citenamefont
  {Schnyder}, \citenamefont {Ryu}, \citenamefont {Furusaki},\ and\
  \citenamefont {Ludwig}}]{PhysRevB.78.195125}%
  \BibitemOpen
  \bibfield  {author} {\bibinfo {author} {\bibfnamefont {A.~P.}\ \bibnamefont
  {Schnyder}}, \bibinfo {author} {\bibfnamefont {S.}~\bibnamefont {Ryu}},
  \bibinfo {author} {\bibfnamefont {A.}~\bibnamefont {Furusaki}}, \ and\
  \bibinfo {author} {\bibfnamefont {A.~W.~W.}\ \bibnamefont {Ludwig}},\ }\href
  {\doibase 10.1103/PhysRevB.78.195125} {\bibfield  {journal} {\bibinfo
  {journal} {Phys. Rev. B}\ }\textbf {\bibinfo {volume} {78}},\ \bibinfo
  {pages} {195125} (\bibinfo {year} {2008})}\BibitemShut {NoStop}%
\bibitem [{\citenamefont {Colpa}(1978)}]{Colpa1978}%
  \BibitemOpen
  \bibfield  {author} {\bibinfo {author} {\bibfnamefont {J.~H.~P.}\
  \bibnamefont {Colpa}},\ }\href@noop {} {\bibfield  {journal} {\bibinfo
  {journal} {Physica}\ }\textbf {\bibinfo {volume} {93A}},\ \bibinfo {pages}
  {327} (\bibinfo {year} {1978})}\BibitemShut {NoStop}%
\bibitem [{\citenamefont {Colpa}(1986)}]{Colpa1986}%
  \BibitemOpen
  \bibfield  {author} {\bibinfo {author} {\bibfnamefont {J.~H.~P.}\
  \bibnamefont {Colpa}},\ }\href@noop {} {\bibfield  {journal} {\bibinfo
  {journal} {Physica}\ }\textbf {\bibinfo {volume} {134A}},\ \bibinfo {pages}
  {377} (\bibinfo {year} {1986})}\BibitemShut {NoStop}%
\bibitem [{\citenamefont {Nitta}\ and\ \citenamefont
  {Takahashi}(2015)}]{PhysRevD.91.025018}%
  \BibitemOpen
  \bibfield  {author} {\bibinfo {author} {\bibfnamefont {M.}~\bibnamefont
  {Nitta}}\ and\ \bibinfo {author} {\bibfnamefont {D.~A.}\ \bibnamefont
  {Takahashi}},\ }\href {\doibase 10.1103/PhysRevD.91.025018} {\bibfield
  {journal} {\bibinfo  {journal} {Phys. Rev. D}\ }\textbf {\bibinfo {volume}
  {91}},\ \bibinfo {pages} {025018} (\bibinfo {year} {2015})}\BibitemShut
  {NoStop}%
\bibitem [{\citenamefont {Takahashi}\ and\ \citenamefont
  {Nitta}(2015)}]{Takahashi2015101}%
  \BibitemOpen
  \bibfield  {author} {\bibinfo {author} {\bibfnamefont {D.~A.}\ \bibnamefont
  {Takahashi}}\ and\ \bibinfo {author} {\bibfnamefont {M.}~\bibnamefont
  {Nitta}},\ }\href {\doibase http://dx.doi.org/10.1016/j.aop.2014.12.009}
  {\bibfield  {journal} {\bibinfo  {journal} {Ann. Phys.}\ }\textbf {\bibinfo
  {volume} {354}},\ \bibinfo {pages} {101 } (\bibinfo {year}
  {2015})}\BibitemShut {NoStop}%
\bibitem [{\citenamefont {Arnold}(1989)}]{Arnold}%
  \BibitemOpen
  \bibfield  {author} {\bibinfo {author} {\bibfnamefont {V.~I.}\ \bibnamefont
  {Arnold}},\ }\href@noop {} {\emph {\bibinfo {title} {Mathematical Methods of
  Classical Mechanics}}},\ \bibinfo {edition} {2nd}\ ed.\ (\bibinfo
  {publisher} {Springer},\ \bibinfo {address} {Berlin, Heidelberg},\ \bibinfo
  {year} {1989})\ \bibinfo {note} {, Appendix 6}\BibitemShut {NoStop}%
\bibitem [{\citenamefont {Bronzan}(1988)}]{PhysRevD.38.1994}%
  \BibitemOpen
  \bibfield  {author} {\bibinfo {author} {\bibfnamefont {J.~B.}\ \bibnamefont
  {Bronzan}},\ }\href {\doibase 10.1103/PhysRevD.38.1994} {\bibfield  {journal}
  {\bibinfo  {journal} {Phys. Rev. D}\ }\textbf {\bibinfo {volume} {38}},\
  \bibinfo {pages} {1994} (\bibinfo {year} {1988})}\BibitemShut {NoStop}%
\bibitem [{Tsu()}]{Tsuchidapc}%
  \BibitemOpen
  \href@noop {} {}\bibinfo {note} {T. Tsuchida, private
  communication.}\BibitemShut {Stop}%
\bibitem [{\citenamefont {Anker}\ and\ \citenamefont
  {Freeman}(1978)}]{Anker529}%
  \BibitemOpen
  \bibfield  {author} {\bibinfo {author} {\bibfnamefont {D.}~\bibnamefont
  {Anker}}\ and\ \bibinfo {author} {\bibfnamefont {N.~C.}\ \bibnamefont
  {Freeman}},\ }\href {\doibase 10.1098/rspa.1978.0083} {\bibfield  {journal}
  {\bibinfo  {journal} {Proc. R. Soc. Lond. A}\ }\textbf {\bibinfo {volume}
  {360}},\ \bibinfo {pages} {529} (\bibinfo {year} {1978})}\BibitemShut
  {NoStop}%
\bibitem [{\citenamefont {Barton}\ and\ \citenamefont
  {Moore}(1974)}]{Barton:1974ae}%
  \BibitemOpen
  \bibfield  {author} {\bibinfo {author} {\bibfnamefont {G.}~\bibnamefont
  {Barton}}\ and\ \bibinfo {author} {\bibfnamefont {M.~A.}\ \bibnamefont
  {Moore}},\ }\href@noop {} {\bibfield  {journal} {\bibinfo  {journal} {J.
  Phys.}\ }\textbf {\bibinfo {volume} {C7}},\ \bibinfo {pages} {4220} (\bibinfo
  {year} {1974})}\BibitemShut {NoStop}%
\bibitem [{\citenamefont {Bruder}\ and\ \citenamefont
  {Vollhardt}(1986)}]{PhysRevB.34.131}%
  \BibitemOpen
  \bibfield  {author} {\bibinfo {author} {\bibfnamefont {C.}~\bibnamefont
  {Bruder}}\ and\ \bibinfo {author} {\bibfnamefont {D.}~\bibnamefont
  {Vollhardt}},\ }\href {\doibase 10.1103/PhysRevB.34.131} {\bibfield
  {journal} {\bibinfo  {journal} {Phys. Rev. B}\ }\textbf {\bibinfo {volume}
  {34}},\ \bibinfo {pages} {131} (\bibinfo {year} {1986})}\BibitemShut
  {NoStop}%
\bibitem [{\citenamefont {Ho}(1998)}]{Ho:1998zz}%
  \BibitemOpen
  \bibfield  {author} {\bibinfo {author} {\bibfnamefont {T.-L.}\ \bibnamefont
  {Ho}},\ }\href {\doibase 10.1103/PhysRevLett.81.742} {\bibfield  {journal}
  {\bibinfo  {journal} {Phys. Rev. Lett.}\ }\textbf {\bibinfo {volume} {81}},\
  \bibinfo {pages} {742} (\bibinfo {year} {1998})}\BibitemShut {NoStop}%
\bibitem [{\citenamefont {Ohmi}\ and\ \citenamefont
  {Machida}(1998)}]{JPSJ.67.1822}%
  \BibitemOpen
  \bibfield  {author} {\bibinfo {author} {\bibfnamefont {T.}~\bibnamefont
  {Ohmi}}\ and\ \bibinfo {author} {\bibfnamefont {K.}~\bibnamefont {Machida}},\
  }\href {\doibase 10.1143/JPSJ.67.1822} {\bibfield  {journal} {\bibinfo
  {journal} {J. Phys. Soc. Jpn.}\ }\textbf {\bibinfo {volume} {67}},\ \bibinfo
  {pages} {1822} (\bibinfo {year} {1998})}\BibitemShut {NoStop}%
\end{thebibliography}
%
\end{document}